\documentclass[12pt,a4paper]{article}
\usepackage{times}
\usepackage{amssymb}
\usepackage{lineno}
\usepackage{amsmath}
\usepackage{soul}
\usepackage{etoolbox}
\usepackage{graphicx}
\usepackage{txfonts} 
\usepackage{bbm}
\usepackage[english]{babel}
\usepackage{empheq} 
\usepackage{mathrsfs} 
\usepackage{calc} 
\usepackage{caption} 
\usepackage{subcaption}
\usepackage{enumitem}                                  
\usepackage{amsthm} 
\usepackage{cases}
\usepackage{siunitx}
\usepackage{upgreek}
\DeclareMathAlphabet{\mathdutchcal}{U}{dutchcal}{m}{n}
\usepackage{hyperref}
\hypersetup{colorlinks=true,linkcolor=red,citecolor=blue}
\usepackage[hyperref]{xcolor}
\usepackage[normalem]{ulem}

\usepackage[authoryear,comma,compress]{natbib}
\theoremstyle{remark}

\newcommand{\toVect}[1]{{\boldsymbol{#1}}} 
\newcommand{\toMat}[1]{{\mathbf{#1}}}      

\newcommand{\E}{{\toVect{E}}}
\newcommand{\D}{{\toVect{D}}}
\newcommand{\F}{{\toMat{F}}}
\newcommand{\ff}{{\widetilde{F}}}
\newcommand{\FF}{{\toMat{\ff}}}

\newcommand{\C}{{\toMat{C}}}
\newcommand{\B}{{\toMat{B}}}
\newcommand{\CC}{{\widetilde{\toMat{C}}}}
\newcommand{\cc}{{\widetilde{C}}}

\newcommand{\Pstar}{{\toVect{p}}^{\textbf{r}}}
\newcommand{\pstar}{{p^\text{r}}}
\renewcommand{\P}{{\toVect{P}}}
\newcommand{\p}{{\toVect{p}}}
\newcommand{\e}{{\toVect{e}}}
\renewcommand{\d}{{\toVect{d}}}

\renewcommand{\S}{{\toMat{S}}}
\newcommand{\SMW}{{\toMat{S}^\text{Maxwell}}}

\newcommand{\Smw}{{{S}^\text{Maxwell}}}
\newcommand{\smw}{{\boldsymbol{\sigma}^\text{Maxwell}}}
\newcommand{\SP}{{\widehat{\toMat{S}}}}
\renewcommand{\sp}{{\widehat{S}}}
\renewcommand{\SS}{{\widetilde{\toMat{S}}}}
\renewcommand{\ss}{{\widetilde{S}}}

\DeclareMathOperator*{\SYMM}{symm}
\newcommand{\symm}[2][]{\ifstrempty{#1}
	{\SYMM\left(#2\right)}
	{\SYMM_{#1}\left(#2\right)}}
\newcommand{\id}{\updelta}

\newcommand{\dd}{\text{\,}\mathrm{d}}

\newcommand{\x}{\toVect{x}}
\newcommand{\X}{\toVect{X}}

\newcommand{\Stress}{\toVect{\stress}}
\newcommand{\stress}{\sigma}

\newcommand{\Projector}{\toMat{\projector}}
\newcommand{\projector}{\mathbb{P}}
\newcommand{\curvatureProjector}{\widetilde{N}}
\newcommand{\CurvatureProjector}{\toMat{\curvatureProjector}}

\newcommand{\Flexo}{\toVect{\flexo}}
\newcommand{\flexo}{\mu}
\newcommand{\Flexocoup}{\toVect{\flexocoup}}
\newcommand{\flexocoup}{f}

\newcommand{\StrGr}{\toMat{\strGr}}
\newcommand{\strGr}{h}

\newcommand{\divergence}{\nabla\text{\!}\cdot\text{\!}}
\newcommand{\gradient}{\nabla}

\newcommand{\jump}[1]{\left\llbracket#1\right\rrbracket}
\DeclareMathOperator{\Trace}{Tr}
\newcommand{\trace}[1]{\Trace(\,#1\,)}
\newcommand{\Log}[1]{\log(\,#1\,)}
\newcommand{\curl}{\nabla\times}
\DeclareMathOperator*{\argmin}{\arg\!\min}

\DeclareMathOperator*{\argmax}{\arg\!\max}

\newcommand{\norm}[1]{{\left\lVert #1 \right\rVert}}

\newcommand{\eq}{Eq.~}
\newcommand{\eqs}{Eqs.~}
\newcommand{\fig}{Fig.~}
\newcommand{\figs}{Figs.~}

\newcommand{\ie}{i.e.~}
\newcommand{\eg}{e.g.~}
\newcommand{\cf}{cf.~}

\DeclareSIUnit \uJ  { \micro \joule }
\DeclareSIUnit \uC  { \micro \coulomb }
\newcommand{\um}{~\si{\um}}

\makeatletter
\def\appendixname{Appendix }
\renewcommand\appendix{\par
  \setcounter{section}{0}%
  \setcounter{subsection}{0}%
  \setcounter{equation}{0}
  \gdef\thefigure{\@Alph\c@section.\arabic{figure}}%
  \gdef\thetable{\@Alph\c@section.\arabic{table}}%
  \gdef\thesection{\appendixname~\@Alph\c@section}%
  \@addtoreset{equation}{section}%
  \gdef\theequation{\@Alph\c@section.\arabic{equation}}%
  \addtocontents{toc}{\string\let\string\numberline\string\tmptocnumberline}{}{}
}
\newdimen\appnamewidth
\def\tmptocnumberline#1{%
   \setbox0=\hbox{\appendixname}
   \appnamewidth=\wd0
   \addtolength\appnamewidth{2.5pc}
   \hb@xt@\appnamewidth{#1\hfill}
}
\makeatother

\title{Modeling flexoelectricity in soft dielectrics at finite deformation}
\author
{D. Codony$^1$, P. Gupta$^1$,O. Marco$^1$, I. Arias$^{1,2,\ast}$\\
	\\
	\small{
	$^1$ Laboratori de C\`{a}lcul Num\`{e}ric (LaC\`{a}N), Universitat Polit\`{e}cnica de Catalunya (UPC),\vspace{-.4em}}
	\\
	\small{Campus Nord UPC-C2, E-08034 Barcelona, Spain}
	\\
    \small{$^2$ Centre Internacional de M{\`e}todes Num{\`e}rics en Enginyeria (CIMNE),\vspace{-.4em}}
	\\
	\small{08034 Barcelona, Spain}
	\\
	\small{$^\ast$ Corresponding author; E-mail:  irene.arias@upc.edu.}
}
\date{}
\begin{document}
\maketitle
\newcommand{\sep}{,~}
\begin{abstract} 
This paper develops the equilibrium equations describing the flexoelectric effect in soft dielectrics under large deformations. Previous works have developed related theories using a flexoelectric coupling tensor of mixed material-spatial character. Here, we formulate the model in terms of a flexoelectric tensor completely defined in the material frame, with the same symmetries of the small-strain flexocoupling tensor and leading naturally to objective flexoelectric polarization fields. The energy potential and equilibrium equations are first expressed in terms of deformation and polarization, and then rewritten in terms of deformation and electric potential, yielding an unconstrained system of fourth order partial differential equations (PDEs).  We further develop a theory of geometrically nonlinear extensible flexoelectric rods under open and closed circuit conditions, with which we examine analytically cantilever bending and buckling under mechanical and electrical actuation. Besides being a simple and explicit model pertinent to slender structures, this rod theory also allows us to test our general theory and its numerical implementation using B-splines. This numerical implementation is robust as it handles the electromechanical instabilities in soft flexoelectric materials.
\\

\end{abstract}

\section{Introduction \label{sec_01}}

Flexoelectricity is a two-way coupling between electric polarization and strain gradient, present in any dielectric material. The direct flexoelectric
effect is understood as the material polarization due to inhomogeneous deformation (\eg bending, twisting), and the converse flexoelectric effect consists on the generation of stress due to the presence of an inhomogeneous electric field. The flexoelectric effect is size dependent due to its intrinsic scaling with strain-gradients, and therefore it is only relevant at the micro- and nanoscale.

Flexoelectric effects have been observed and widely studied in hard materials \citep{tolpygo1963long,kogan1964piezoelectric,hong2011first,resta2010towards,maranganti2006electromechanical}, mainly crystalline ceramics such as 
ferroelectric perovskites \citep{zubko2007strain,ma2001large,ma2002flexoelectric,fu2006experimental,ma2001observation,ma2003strain,ma2005flexoelectric,ma2006flexoelectricity}.
However, they are also present in soft materials, such as 
liquid crystals \citep{meyer1969piezoelectric,petrov1975flexoelectric,prost1977microscopic,Pikin1978,Lagerwall1984,Barbero1986,Cepic2000,Kuczynski2005,harden2006giant,trabi2008interferometric},
cellular membranes \citep{petrov1989curvature,todorov1991electrical,todorov1994first,sun1997toward,petrov2002flexoelectricity,Gao2008,Jewell2011} and 
polymers \citep{breger1976bending,marvan1998,baskaran2011,baskaran2012,Deng2014flexoelectricity,zhang2016experimental,zhou2017flexoelectric}.

The mechanism of flexoelectricity in hard materials can be intuitively understood by the ionic crystal model under bending, in which a non-zero net dipole moment arises due to a shift between the centers of gravity of the negative and the positive ions \citep{Zubko2013}. In soft materials such as liquid crystals or lipid bilayers, flexoelectricity results from the reorientation of irregularly shaped polarized molecules under strain gradients, see \eg\cite{meyer1969piezoelectric,petrov1999,Rey2006,Ahmadpoor2013,Liu2013,Mohammadi2014,Ahmadpoor2015,Mozorovska2018}. In these materials, flexoelectricity has been mechanistically linked to the arrangement of not only dipolar but also quadrupolar constituents \citep{prost1977microscopic,Marcerou1990,Derzhanski1990,Gennes1993}, and theories accounting for thermal fluctuations have been proposed \citep{Osipov1995,Liu2013}. The mechanisms leading to flexoelectricity in polymers, however, are not known \citep{Krichen2016}, although they likely involve rearrangements of glassy and crystalline components \citep{baskaran2011,baskaran2012}. Of note is the conceptual model by \cite{marvan1998}, in which flexoelectric polarization results from strain gradient-induced asymmetry of the free-volume of a fluctuating dipole. This or other mechanisms, however, have not been demonstrated.
We refer to \citet{Yudin2013,Nguyen2013,Zubko2013,Krichen2016,wang2019flexoelectricity} for excellent and comprehensive reviews of flexoelectricity in solids.

In recent years, several reasons justify an increasing interest in flexoelectricity in polymer materials. On the one hand, a large flexoelectric response is expected. Experiments suggest that the flexoelectric coefficients of polymers are at least the same order of magnitude as those of hard crystalline materials \citep{chu2012,baskaran2011,baskaran2012}, but being much more deformable, much larger flexoelectric polarization is possible. On the other hand, electromechanical actuation of polymers by flexoelectricity overcomes the current limitations of traditional actuation based on electrostriction, which are: (i) one-way coupling, \ie mechanical deformation does not produce an electric field, (ii) very large electric fields are required (which may lead to dielectric breakdown), and (iii) reversal of electric field does not reverse the direction of the deformation \citep{pelrine1998electrostriction,o2008review,Krichen2016,rosset2016}. Furthermore, only a few polymers exhibit significant piezoelectricity \citep{bauer2008piezoelectric}. Thus, quantifying flexoelectricity at large deformations may enable the design of efficient electromechanical elastomeric devices, such as sensors, actuators and energy harvesters, based on the flexoelectric effect \citep{jiang2013flexoelectric,huang2018flexoelectricity,wang2019flexoelectricity}.

The literature about continuum theories of flexoelectricity in bulk solids ranges from the early works by \citet{Mashkevich1957,tolpygo1963long,kogan1964piezoelectric,indenbom1981,indenbom1981_,Tagantsev1985,Tagantsev1986,sahin1988strain,tagantsev1991electric} to the more recent developments by \citet{maranganti2006electromechanical,Shen2010,hu2010,hadjesfandiari2013size,Liu2014,anqing2015flexoelectric}, to name a few. However, most of these works assume infinitesimal deformations, and are therefore suitable to model crystalline ceramics only. Efforts have been recently made to extend the theory to polymers or elastomers undergoing large deformations, but the literature is still scarce \citep{Liu2014,Yvonnet2017,Thai2018,poya2019family,mcbride2019modelling,zhuang2019meshfree,nguyen2019nurbs}. Some of these works model flexoelectricity as a linear coupling between strain gradients and the electric displacement \citep{poya2019family} or the electric field \citep{mcbride2019modelling,nguyen2019nurbs,zhuang2019meshfree} instead of the electric polarization, which however is the most natural choice \citep{Toupin1956,LandauLifshitz1951,Devonshire1949,Devonshire1951,Devonshire1954,Lines1979}. 
Furthermore, works modeling flexoelectricity as a coupling between strain gradients and electric polarization consider a coupling tensor of mixed material-spatial character  \citep{Liu2014,Yvonnet2017,Thai2018}, leading in general to a lack of objectivity in the resulting polarization as argued in Section \ref{sec_bbp}. 

The equations of flexoelectricity can only be solved analytically in very simple settings, such as simplified Euler-Bernoulli (E-B) \citep{liang2014effects,deng2014nanoscale} and Timoshenko beam \citep{zhang2016timoshenko} models. Such models have been extended to large deformations but moderate rotations \emph{\`a la} von Karmann \citep{baroudi2019dynamic}.  Otherwise, it is necessary to resort to computational flexoelectricity \citep{zhuang2020computational}. 
The major challenge is to handle the $C^1$ continuity of the state variables required by the fourth-order PDE system. To address this, several numerical alternatives have been proposed, such as mesh-free approximations \citep{Abdollahi2014,Abdollahi2015a,Abdollahi2015b,abdollahi2015constructive,zhuang2019meshfree}, isogeometric analysis \citep{ghasemi2017level,nanthakumar2017topology,Thai2018,hamdia2018sensitivity,ghasemi2018multi,nguyen2019nurbs}, $C^1$ Argyris triangular element approximation \citep{Yvonnet2017} and the B-spline-based immersed boundary method \citep{codony2019immersed}.
Another family of numerical methods are those circumventing the $C^1$ continuity requirement by introducing additional variables, such as mixed formulations \citep{mao2016mixed,deng2017mixed,deng2018three}, or those based on micromorphic theories of continua \citep{poya2019family,mcbride2019modelling}. Recently, a few works report the application of these methods to large deformation flexoelectricity \citep{Thai2018,poya2019family,mcbride2019modelling,Yvonnet2017,zhuang2019meshfree,nguyen2019nurbs} but the continuum formulation at finite deformation is still open, see previous paragraph, and there is a need for validation of the computational results. 

To provide a general tool to assess flexoelectricity under large deformations, we propose a formulation with a fully material flexoelectric coupling between strain gradient and electric polarization, leading by construction to objective polarization fields. To facilitate the solution of the associated boundary value problem, we reformulate the balance equations in terms of displacements and electric potential as primal unknowns, yielding an unconstrained system of fourth-order PDE. We solve this system computationally with open uniform B-spline basis in body-fitted Cartesian meshes. We further derive large deformation models for geometrically nonlinear extensible flexoelectric rods under open and closed circuit conditions and derive closed-form solutions for cantilever bending and buckling. We report excellent agreement well into the nonlinear regime between numerical and analytical solutions in conditions mimicking the assumptions of the analytical models, which serves as validation. We then explore general flexoelectric problems beyond the simplifying assumptions of the analytical models and analyze the role of the flexoelectric material parameters in the electromechanical response of the rod. 

The paper is organized as follows. In Section \ref{sec_02} the free energy density and corresponding balance equations of a flexoelectric body are reviewed, the mathematical expression of the flexoelectric coupling is discussed, and the boundary value problem is stated. The numerical implementation used to solve the boundary value problem is presented in Section \ref{sec_03}, and the analytical solutions for one-dimensional geometrically nonlinear flexoelectric rods are derived in Section \ref{sec_04}. In Section \ref{sec_05} the numerical and analytical results of bending and buckling of rods under open/closed circuit are shown. The paper is concluded in Section \ref{sec_06}.

\section{Variational formulation of flexoelectricity in material form}\label{sec_02}

\subsection{Background and balance laws in spatial and material forms}

Consider a deformable dielectric body described by $\Omega_0$ in the reference (or undeformed) configuration, and by $\Omega$ in the current (or deformed) configuration. The deformation map $\boldsymbol{\chi}:\Omega_0 \rightarrow \Omega$ maps every material point $\X\in\Omega_0$ to the spatial point $\x=\boldsymbol{\chi}(\X)\in\Omega$. Whenever index notations are used, uppercase and lowercase indexes refer to quantities in the reference and the current configurations, respectively. The deformation gradient $\F$, the Jacobian determinant $J$, and the right and left Cauchy-Green deformation tensors $\C,\B$ are defined as
\begin{align}
F_{iI}(\X)&\coloneqq\frac{\partial \chi_i(\X)}{\partial X_I},&&
J\coloneqq \det(\F),&&
C_{IJ}\coloneqq F_{kI}F_{kJ},&&
B_{ij}\coloneqq F_{iK}F_{jK}\label{C}.
\end{align}
Standard strain measures in the reference and the current configurations are the Green-Lagrangian $\toMat{\mathfrak{E}}$ and the Almansi-Eulerian $\toMat{\mathfrak{e}}$ strain tensors given by
\begin{align}\label{strains}
\mathfrak{E}_{IJ}\coloneqq\frac{1}{2}\left(C_{IJ}-\id_{IJ}\right),&&
\mathfrak{e}_{ij}\coloneqq\frac{1}{2}\left(\id_{ij}-B_{ij}^{-1}\right)=\mathfrak{E}_{IJ}F^{-1}_{Ii}F^{-1}_{Jj}.
\end{align}
Since the flexoelectricity theory involves high-order derivatives, let us define the gradient of the deformation gradient $\FF$, the gradient of the Cauchy-Green deformation tensor $\CC$ and the Green-Lagrangian strain gradient $\toMat{\widetilde{\mathfrak{E}}}$ as
\begin{align}
\ff_{iJK}\coloneqq\frac{\partial F_{iJ}}{\partial X_K}=\frac{\partial^2 x_i}{\partial X_J \partial X_K},&&
\cc_{IJK}\coloneqq\frac{\partial C_{IJ}}{\partial X_K}=2\symm[IJ]{\ff_{kIK} F_{kJ}},&&
\widetilde{\mathfrak{E}}_{IJK}\coloneqq\frac{\partial\mathfrak{E}_{IJ}}{\partial X_K}=\frac{1}{2}\cc_{IJK}\label{eqfce};
\end{align}
where $\symm[IJ]{A_{IJ}}:=\left(A_{IJ}+A_{JI}\right)/2$. Note that the relation $\toMat{\widetilde{\mathfrak{E}}}(~\FF~)$ in \eq\eqref{eqfce} is inverted as 
\begin{align}\label{straininv}
    \ff_{iJK} = \left(
     \widetilde{\mathfrak{E}}_{IJK}
    +\widetilde{\mathfrak{E}}_{KIJ}
    -\widetilde{\mathfrak{E}}_{KJI}
    \right) F_{Ii}^{-1},
\end{align}
analogously to the relation between second derivative of displacement and strain gradients in the limit of infinitesimal deformation \citep{schiaffino2019metric}.

This body in equilibrium necessarily satisfies mechanical balance laws of linear and angular momentum, and Maxwell equations. In the absence of a magnetic field, they can be expressed in an Eulerian frame as
\begin{subequations}\begin{align}
\divergence \Stress +\toVect{b}&=\toVect{0},\label{bal1}\\
\Stress &= \Stress^T,\\
\curl \e &= \toVect{0} \label{curl},\\
\divergence \d - q &= 0;\label{bal4}
\end{align}\end{subequations}
where $\Stress$ is the physical stress, $\e$ is the the electric field, $\d$ is the electric displacement, and $\toVect{b}$ and $q$ are the body force and electric charge per unit volume. Equation \eqref{curl} implies the existence of an electric potential $\phi$ such that $\e=-\gradient\phi$. The linear constitutive law for $\d$ for a dielectric material is
\begin{gather}
	\d(\p,\e) = \epsilon_0 \e + \p \qquad \text{ or, equivalently, }\qquad \d(\p,\phi) = -\epsilon_0 \gradient\phi + \p,\label{constelec}
\end{gather}
where $\p$ is the electric polarization, which is work-conjugate to $\e$, and $\epsilon_0$ is the electric permittivity of vacuum.

To formulate the problem in a material frame, the Lagrangian second Piola-Kirchhoff physical stress tensor $\S$ is defined from the work-conjugacy relation 
\begin{align}
\stress_{ij} \mathfrak{e}_{ij} = \frac{1}{J} S_{IJ}\mathfrak{E}_{IJ},
\end{align}
where $\stress_{ij} \mathfrak{e}_{ij}$ is a mechanical work density per unit physical volume and $S_{IJ}\mathfrak{E}_{IJ}$ a mechanical work density per unit reference volume, leading to 
\begin{align}
S_{IJ} =& JF^{-1}_{Ii}F^{-1}_{Jj}\stress_{ij},\label{Sa1}
\end{align}
where strictly speaking we should write $S_{IJ}\circ\chi^{-1} = JF^{-1}_{Ii}F^{-1}_{Jj}\stress_{ij}$ to account for the fact that some of these fields are over $\Omega_0$ and others are over $\Omega$. 
To follow an analogous procedure with the electric displacement \citep{Lax1976,Dorfmann2005,Vu2007,dorfmann2014nonlinear,dorfmann2017nonlinear,Steinmann2017}, we first identify the nominal or material electric field. The electric potential can be expressed in the material frame as $\Phi(\X) = \phi(\chi(\X))$, and the nominal electric field $\E$ defined as the negative of its material gradient. By the chain rule, we thus find that 
\begin{align}\label{Sa2}
E_I = -\frac{\partial \Phi}{\partial X_I} = -\frac{\partial \phi}{\partial x_i}\frac{\partial \chi_i}{\partial X_I} = e_i F_{iI}.
\end{align}
Then, from the work-conjugacy relation
\begin{align}
d_i e_i = \frac{1}{J}  D_I E_I,
\end{align}
we identify the nominal electric displacement as
\begin{align}\label{Sa3}
D_I =& JF^{-1}_{Ii}d_i.
\end{align}
Since electric displacement and polarization are physically equivalent quantities, we analogously find
\begin{align}\label{Sa4}
P_I =& JF^{-1}_{Ii}p_i.
\end{align}
Using \eq \eqref{C}, \eqref{Sa1}, \eqref{Sa2}, \eqref{Sa3}, \eqref{Sa4},
the balance equations in \eq \eqref{bal1}-\eqref{bal4}
and the constitutive law for dielectrics in \eq\eqref{constelec} 
are written in material form as
\begin{subequations} \label{SFMat}\begin{align}
	\left(F_{iI}S_{IJ}\right)_{,J} + B_i &= 0_i,\\
	S_{IJ} &= S_{JI},\\
	E_L + \Phi_{,L} &= 0,\label{curlfree}\\
	D_K &= \epsilon_0 JC^{-1}_{KL}E_L + P_K,\label{claw}\\
	D_{K,K} - Q &= 0,
\end{align}\end{subequations}
with $\toVect{B}=J\toVect{b}$ and $Q=Jq$.

\subsection{Constitutive relations and thermodynamic potentials in material form}

We define the Lagrangian internal energy density per unit reference volume of the flexoelectric solid as 
\begin{align}
\Psi^{\rm Int}(\mathfrak{E}, \widetilde{\mathfrak{E}}, \P) = \Psi^{\rm Mech}(\mathfrak{E}, \widetilde{\mathfrak{E}}) + \Psi^{\rm Diele}(\mathfrak{E},\P) + \Psi^{\rm Flexo}(\P, \widetilde{\mathfrak{E}}).
\end{align}
We allow $\Psi^{\rm Mech}$ to depend on Lagrangian strain and strain gradient as required for stability \citep{Liu2014}. The isotropic dielectric energy per unit reference volume follows by transforming the spatial expression per unit physical volume $\psi^\text{Diele}(\p) = \dfrac{1}{2 (\epsilon-\epsilon_0)}p_ip_i$ \citep{Liu2014} by recalling \eq\eqref{Sa4}, resulting in 
\begin{align}
\Psi^{\rm Diele}(\mathfrak{E},\P) = \frac{1}{2J(\epsilon-\epsilon_0)}P_IC_{IJ}P_J,
\end{align}
where $\epsilon$ denotes the electric permittivity of the material. The flexoelectric coupling linking polarization and strain gradient is encoded by $\Psi^{\rm Flexo}$, which for simplicity we assume to be independent on strain.

The spatial expression of the electrostatic energy density $\psi^\text{Elec}(\e)=\frac{1}{2}\epsilon_0e_ie_i$ \citep{Liu2014} can also be expressed in the material frame by recalling  \eq\eqref{Sa2}, resulting in the energy density per unit reference volume
\begin{align}
\Psi^{\rm Elec}(\mathfrak{E}, \E) = \frac{J\epsilon_0}{2}E_IC_{IJ}^{-1}E_J.
\end{align}

To formulate a unified potential self-consistently accounting for the material electromechanics and for electrostatics, $\Psi^{\rm Int}(\mathfrak{E}, \widetilde{\mathfrak{E}}, \P)$ and $\Psi^{\rm Elec}(\mathfrak{E},\E)$ must be expressed in terms of the same variables. To accomplish this, we resort to a partial Legendre transform and define the following internal dual potential
 \begin{align}\label{LT}
\bar{\Psi}^{\rm Int}(\mathfrak{E}, \widetilde{\mathfrak{E}}, \E) = \min_{\P} \left(
\Psi^{\rm Int}(\mathfrak{E}, \widetilde{\mathfrak{E}}, \P) - \P\cdot \E \right).
\end{align}
The stationarity condition of the minimization results in 
\begin{align}
\E (\mathfrak{E}, \widetilde{\mathfrak{E}}, \P)= \frac{\partial \Psi^{\rm Int}}{\partial \P}.
\end{align}
In principle, this expression can be inverted to find $\P (\E,\mathfrak{E}, \widetilde{\mathfrak{E}})$, which plugged into $\Psi^{\rm Int}(\mathfrak{E}, \widetilde{\mathfrak{E}}, \P) - \P\cdot \E$ results in the dual potential $\bar{\Psi}^{\rm Int}(\mathfrak{E}, \widetilde{\mathfrak{E}}, \E)$.

If we  postulate the following flexoelectric coupling 
\begin{align}
\Psi^{\rm Flexo}(\P, \widetilde{\mathfrak{E}}) = -P_Lf_{LIJK}\widetilde{\mathfrak{E}}_{IJK}, \label{flexopot}
\end{align}
where $f_{LIJK}$ is a purely Lagrangian tensor as further discussed later, this inversion can be made explicit yielding
\begin{align}\label{eq:P}
E_L &= \frac{1}{J(\epsilon-\epsilon_0)}C_{LM}P_M - f_{LIJK}\widetilde{\mathfrak{E}}_{IJK} \; \Rightarrow \;
\\
P_M &= J(\epsilon-\epsilon_0) C^{-1}_{ML}\left(E_L+f_{LIJK}\widetilde{\mathfrak{E}}_{IJK}\right) = J(\epsilon-\epsilon_0) C^{-1}_{ML}\left(E_L+E_L^{\rm Flexo}\right),
\end{align}
where we have defined $E_L^{\rm Flexo} = f_{LIJK}\widetilde{\mathfrak{E}}_{IJK}$ for convenience. Replacing this expression for $\P$ in \eq\eqref{LT} and rearranging terms, we find
\begin{align}\label{IntPot}
\bar{\Psi}^{\rm Int}(\mathfrak{E}, \widetilde{\mathfrak{E}}, \E) = \Psi^{\rm Mech}(\mathfrak{E}, \widetilde{\mathfrak{E}}) - \frac{J}{2}(\epsilon-\epsilon_0) E_I^{\rm Flexo}  C^{-1}_{IJ} E_J^{\rm Flexo}  - \frac{J}{2}(\epsilon-\epsilon_0) E_I  C^{-1}_{IJ} E_J - J(\epsilon-\epsilon_0)E_I  C^{-1}_{IJ} E_J^{\rm Flexo}.
\end{align}
Now, the total electromechanical enthalpy accounting for electrostatics $\bar{\Psi}^{\rm Enth} = \bar{\Psi}^{\rm Int}-\Psi^{\rm Elec}$ \citep{Liu2014,dorfmann2014nonlinear,dorfmann2017nonlinear} can be written as
\begin{align}
\bar{\Psi}^{\rm Enth} (\mathfrak{E}, \widetilde{\mathfrak{E}}, \E) = \bar{\Psi}^{\rm Mech}(\mathfrak{E}, \widetilde{\mathfrak{E}}) + \bar{\Psi}^{\rm Diele}(\mathfrak{E},\E) + \bar{\Psi}^{\rm Flexo}(\mathfrak{E}, \widetilde{\mathfrak{E}}, \E),
\end{align}
with 
\begin{align} 
\label{DielePot}
 \bar{\Psi}^{\rm Diele}(\mathfrak{E},\E) =& - \frac{1}{2}J  \epsilon      E_{M}C^{-1}_{ML}E_{L},\\ \label{FlexoPot}
\bar{\Psi}^{\rm Flexo}(\mathfrak{E}, \widetilde{\mathfrak{E}}, \E) = & -JC^{-1}_{ML} E_{M} \flexo_{LIJK} \widetilde{\mathfrak{E}}_{IJK};
\end{align}
where $\Flexo =(\epsilon-\epsilon_0)\Flexocoup$ is the flexoelectricity tensor \citep{Zubko2013,wang2019flexoelectricity}, described in \eq\eqref{flexotensor}. The \emph{effective} mechanical energy density of the system \citep{wang2019flexoelectricity} is
\begin{align} \label{mecheff}
\bar{\Psi}^{\rm Mech}(\mathfrak{E}, \widetilde{\mathfrak{E}}) = & \Psi^{\rm Mech}(\mathfrak{E}, \widetilde{\mathfrak{E}}) - \frac{J}{2}(\epsilon-\epsilon_0) E_M^{\rm Flexo}  C^{-1}_{ML} E_L^{\rm Flexo}, \nonumber \\ = & \Psi^\text{Mech}(\mathfrak{E}, \widetilde{\mathfrak{E}})
-\frac{1}{2}\widetilde{\mathfrak{E}}_{IJK}\left(\frac{\flexo_{AIJK}JC^{-1}_{AB}\flexo_{BLMN}}{\epsilon-\epsilon_0}\right)\widetilde{\mathfrak{E}}_{LMN}.
\end{align}
The standard mechanical contribution accounting for strain gradient elasticity can be written as
\begin{align}
\Psi^\textnormal{Mech}(\mathfrak{E},\widetilde{\mathfrak{E}})=\Psi^\textnormal{Elast}(\mathfrak{E})+\frac{1}{2}\widetilde{\mathfrak{E}}_{IJK}\strGr_{IJKLMN}\widetilde{\mathfrak{E}}_{LMN},
\end{align}
where $\Psi^\textnormal{Elast}$ can be any classical hyperelastic potential, \eg Saint-Venant–Kirchhoff, \cf \eq\eqref{sv}, or Neo-Hookean, \cf \eq\eqref{nh}, constitutive models, and $\StrGr$ is the sixth-order strain gradient elasticity tensor. Upon inspection, it is clear that the second contribution in \eq\eqref{mecheff}, \ie the flexoelectricity-induced mechanical energy, has the same structure as the strain gradient elasticity potential. For convenience, we thus define 
\begin{align} \label{mecheff2}
\bar{\Psi}^{\rm Mech}(\mathfrak{E}, \widetilde{\mathfrak{E}}) = & \Psi^\textnormal{Elast}(\mathfrak{E}) +
\frac{1}{2}\widetilde{\mathfrak{E}}_{IJK}\bar\strGr_{IJKLMN}\widetilde{\mathfrak{E}}_{LMN}, 
\end{align}
where 
\begin{align}\label{effSG}
\bar\strGr_{IJKLMN} = & \strGr_{IJKLMN} - \frac{\flexo_{AIJK}JC^{-1}_{AB}\flexo_{BLMN}}{\epsilon-\epsilon_0}
\end{align}
is the \emph{effective} strain gradient elasticity tensor as described in \eq\eqref{strgrtensor}. To preserve the positive definiteness of $\bar\Psi^\textnormal{Mech}$, it is clear from \eq\eqref{mecheff2} that $\bar\StrGr$ has to be semidefinite positive and thus a stability condition can be derived from \eq\eqref{effSG} 
depending on both $\StrGr$ and $\Flexo$ \citep{Yudin2014,Yudin2015,Morozovska2016}.


\subsection{Variational formulation in material form}
\label{sec_bbp}
The boundary of the reference body, $\partial\Omega_0$, is split in several disjoint Dirichlet and Neumann sets as follows:
\begin{gather}
\partial\Omega_0
= \partial\Omega_0^\chi\cup\partial\Omega_0^T
= \partial\Omega_0^V\cup\partial\Omega_0^R
= \partial\Omega_0^\Phi\cup\partial\Omega_0^W.
\end{gather}
On the Dirichlet boundaries $\partial\Omega_0^\chi$, $\partial\Omega_0^V$ and $\partial\Omega_0^\Phi$, the deformation map $\boldsymbol{\chi}$, normal derivatives of the deformation map $\partial_{0}^N\boldsymbol{\chi}$, and electric potential $\Phi$ are prescribed, respectively.
On the Neumann boundaries $\partial\Omega_0^T$, $\partial\Omega_0^R$ and $\partial\Omega_0^W$, their respective work conjugate quantities (per unit reference volume) are prescribed, \ie the surface traction $\toVect{T}(\boldsymbol{\chi},\Phi)=\overline{\toVect{T}}$, the surface double traction $\toVect{R}(\boldsymbol{\chi},\Phi)=\overline{\toVect{R}}$ and the surface charge $W(\boldsymbol{\chi},\Phi)=\overline{W}$.
As a result of the strain-gradient elasticity potential \citep{Mindlin1964,Mindlin1968a}, additional loads arise in non-smooth regions of $\partial\Omega_0$, \ie edges $C_0$ in a three-dimensional domain \citep{Mao2014,codony2019immersed}. We also split them in Dirichlet in Neumann sets as
\begin{gather}
C_0 = C_0^\chi\cup C_0^J,
\end{gather}
depending on whether the deformation map $\boldsymbol{\chi}$ or edge forces (per unit reference volume) $\toVect{J}(\boldsymbol{\chi},\Phi)=\overline{\toVect{J}}$ are prescribed.
For simplicity, dead loads are considered.

The enthalpy functional governing the physics of a flexoelectric body is written as
\begin{equation}\begin{split}\label{energyfunctional}
	\Pi [\boldsymbol{\chi},\Phi]=&\int_{\Omega_0}\left(\bar\Psi^\text{Enth}(\mathfrak{E}, \widetilde{\mathfrak{E}}, -\nabla_0\Phi) - B_i \chi_i  + Q\Phi 
	\right)\dd\Omega_0
	\\
	&- \int_{\partial\Omega_0^T} \overline{T}_i \chi_i  \dd\Gamma_0
	- \int_{\partial\Omega_0^R} \overline{R}_i\partial_{0}^N \chi_i  \dd\Gamma_0
	- \int_{C_0^J} \overline{J}_i \chi_i  \dd \text{s}_0
	+ \int_{\partial\Omega_0^W}  \overline{W}\Phi \dd\Gamma_0,
\end{split}\end{equation}
where we have used $\E = - \nabla_0 \Phi$ from \eq\eqref{curlfree}. Equilibrium states $\{\boldsymbol{\chi}^*,\Phi^*\}$ are its saddle points satisfying
\begin{gather}\label{argmm}
\{\boldsymbol{\chi}^*,\Phi^*\}=\arg\min_{\chi\in\mathcal{X}} \max_{\Phi\in\mathcal{P}}  \Pi[\boldsymbol{\chi},\Phi] ,
\end{gather}
where $\mathcal{X}$ and $\mathcal{P}$ are the functional spaces for $\boldsymbol{\chi}$ and $\Phi$ with sufficient regularity fulfilling Dirichlet boundary conditions.

A necessary condition for equilibrium is the vanishing of the first variation of $\Pi[\boldsymbol{\chi},\Phi]$
\begin{align}\label{res}	
0=	\delta \Pi[\boldsymbol{\chi},\Phi;\delta\boldsymbol{\chi},\delta\Phi]=&
	\int_{\Omega_0}\left(
	\frac{\partial \bar\Psi^\text{Enth} }{\partial \mathfrak{E}_{IJ}} \delta \mathfrak{E}_{IJ} + 
	\frac{\partial \bar\Psi^\text{Enth} }{\partial \widetilde{\mathfrak{E}}_{IJK}} \delta \widetilde{\mathfrak{E}}_{IJK} + 
	\frac{\partial \bar\Psi^\text{Enth} }{\partial E_{L}} \delta E_L
	- B_i\delta \chi_i  + Q\delta \Phi 
	\right)\dd\Omega_0
	\nonumber\\&
	- \int_{\partial\Omega_0^T} \overline{T}_i\delta \chi_i  \dd\Gamma_0
	- \int_{\partial\Omega_0^R} \overline{R}_i\partial_{0}^N \delta \chi_i  \dd\Gamma_0
	- \int_{C_0^J}  \overline{J}_i\delta \chi_i  \dd \text{S}_0
	+ \int_{\partial\Omega_0^\Phi}  \overline{W}\delta \Phi \dd\Gamma_0
	\nonumber\\
	{}=&
	\int_{\Omega_0}\left(
	\sp_{IJ}\delta \mathfrak{E}_{IJ}
	+\ss_{MJK}\delta\widetilde{\mathfrak{E}}_{MJK}
	-D_L \delta E_L
	- B_i\delta \chi_i  + Q\delta \Phi 
	\right)\dd\Omega_0
	\nonumber\\&
	- \int_{\partial\Omega_0^T}  \overline{T}_i\delta \chi_i \dd\Gamma_0
	- \int_{\partial\Omega_0^R} \overline{R}_i\partial_{0}^N \delta \chi_i  \dd\Gamma_0
	- \int_{C_0^J}  \overline{J}_i\delta \chi_i \dd \text{S}_0
	+ \int_{\partial\Omega_0^\Phi} \overline{W}\delta \Phi  \dd\Gamma_0,
\end{align}
for all admissible variations $\delta\boldsymbol{\chi}$ and $\delta\Phi$, and where
\begin{align}
&\delta E_L\coloneqq -\frac{\partial(\delta\Phi)}{\partial X_L},
\qquad
\delta F_{iI}\coloneqq\frac{\partial(\delta \chi_i)}{\partial X_I},
\qquad
\delta \ff_{iIJ}\coloneqq\frac{\partial^2(\delta \chi_i)}{\partial X_I\partial X_J},
&\\&
\delta\mathfrak{E}_{IJ}=\frac{1}{2}\delta C_{IJ}\coloneqq \symm[IJ]{\delta F_{kI} F_{kJ}},
\qquad
\delta\widetilde{\mathfrak{E}}_{IJK}=\frac{1}{2}\delta \cc_{IJK}\coloneqq \symm[IJ]{\delta F_{kI} \ff_{kJK} + F_{kI} \delta\ff_{kJK}}
.
&\end{align}
We have introduced the local second Piola-Kirchhoff stress $\SP$, the second Piola-Kirchhoff double stress $\SS$ and the electric displacement $\D$ defined as follows:
\begin{align}
	\sp_{IJ}(\boldsymbol{\chi},\Phi)&= \frac{\partial\bar\Psi^\text{Enth}}{\partial \mathfrak{E}_{IJ}}=
	2\frac{\partial\Psi^\text{Elast}(\C)}{\partial C_{IJ}}+J \mathscr{C}_{MLIJ}E_M
	\left( \frac{1}{2}\epsilon E_L + \mu_{LABK}\widetilde{\mathfrak{E}}_{ABK} \right),\label{S_cons}
	\\
	\ss_{IJK}(\boldsymbol{\chi},\Phi)&= \frac{\partial\bar\Psi^\text{Enth}}{\partial \widetilde{\mathfrak{E}}_{IJK}}
	=
	\bar\strGr_{IJKLMN}\widetilde{\mathfrak{E}}_{LMN}-JC^{-1}_{LM}E_M \mu_{LIJK},\label{SS_cons}
	\\
	D_L(\boldsymbol{\chi},\Phi)&= -\frac{\partial\bar\Psi^\text{Enth}}{\partial E_L}
	=
	J C^{-1}_{KL} \left( \epsilon E_K + \mu_{KIJM}\widetilde{\mathfrak{E}}_{IJM} \right), \label{D_cons}
\end{align}
with
\begin{align}
\mathscr{C}_{ABCD}
=
\frac{2}{J}\frac{\partial\left(-JC_{AB}^{-1}\right)}{\partial C_{CD}}
=\left(C^{-1}_{AC}C^{-1}_{BD}+C^{-1}_{BC}C^{-1}_{AD}-C^{-1}_{AB}C^{-1}_{CD}\right).
\end{align}

Analogously to the infinitesimal strain theory of flexoelectricity \citep{Mao2014,codony2019immersed}, \eq\eqref{res} can be integrated by parts and, by invoking the divergence and surface divergence theorems, the strong form in \eq\eqref{SFMat} is recovered along with the following definitions of the physical second Piola-Kirchhoff stress $\S$, the surface traction ${\bf T}$, the double traction   ${\bf R}$, the surface charge density $W$ and the edge forces  ${\bf J}$:\begin{subequations}\begin{align}
S_{IJ}(\boldsymbol{\chi},\Phi)\coloneqq&\sp_{IJ}(\boldsymbol{\chi},\Phi)-\ss_{IJK,K}(\boldsymbol{\chi},\Phi)\nonumber\\{}=&
	2\frac{\partial\Psi^\text{Elast}(\C)}{\partial C_{IJ}}
	-\bar\strGr_{IJKLMN}\widetilde{\mathfrak{E}}_{LMN,K}
	+\frac{J}{2}\mathscr{C}_{MLIJ}E_M\epsilon E_L
	+JC^{-1}_{LM}E_{M,K} \mu_{LIJK}
	&& \text{ in }\Omega_0,\label{eq:S}
	\\
	T_i(\boldsymbol{\chi},\Phi) \coloneqq& F_{iI}\left[\left(S_{IJ}(\boldsymbol{\chi},\Phi)-\ss_{IKJ,N}\projector_{NK}\right)N_J+\ss_{IJK}\curvatureProjector_{JK}\right]
	-\ff_{iIN}\projector_{NK}\ss_{IKJ}N_J
	&& \text{ on }\partial\Omega_0 \label{eq_Traction},\\
	R_i(\boldsymbol{\chi},\Phi) \coloneqq& F_{iI}\ss_{IJK}N_JN_K
	&& \text{ on } \partial\Omega_0, \\
	W(\boldsymbol{\chi},\Phi) \coloneqq& -D_LN_L
	&& \text{ on }\partial\Omega_0, \\
	J_i(\boldsymbol{\chi},\Phi) \coloneqq& \jump{F_{iI}\ss_{IJK}M_JN_K}
	&& \text{ on } C_0;
\end{align}\end{subequations}
where $\toVect{N}$ is the outward unit normal vector on $\partial\Omega_0$, $\toVect{M}$ is the outward unit co-normal vector on $C_0$, $\Projector=\toMat{I}-\toVect{N}\times\toVect{N}$ is the projection operator on $\partial\Omega_0$, $\CurvatureProjector=\gradient_0\toVect{N}:\Projector(\toVect{N}\times\toVect{N})-\gradient_0\toVect{N}\cdot\Projector$ is the second-order geometry tensor on $\partial\Omega_0$ and $\jump{~}$ is the jump operator defined on $C$ as the sum of its argument evaluated at each boundary adjacent to $C$ (we refer to \citet{codony2019immersed} for a detailed definition of the quantities involved here).

Upon inspection, the second Piola-Kirchhoff stress tensor $\S$ in \eq\eqref{eq:S} is composed by four terms. The first two terms correspond to the classical and high-order mechanical stresses, respectively. The third one corresponds to the total second Piola-Maxwell stress tensor $\SMW$. This becomes evident by expanding it as
\begin{equation}
\Smw_{IJ}\coloneqq\frac{J}{2}\mathscr{C}_{MLIJ}E_M\epsilon E_L=JF^{-1}_{Ii}F^{-1}_{Jj}\epsilon\left[
\left(E_MF^{-1}_{Mi}\right)\left(E_LF^{-1}_{Lj}\right)-\frac{1}{2}\left(E_MF^{-1}_{Ma}\right)\left(E_LF^{-1}_{La}\right)\id_{ij}\right],
\end{equation}
and obtaining its spatial counterpart by using \eq\eqref{Sa1} and \eqref{Sa2} as
\begin{equation}
\smw\coloneqq\epsilon\left(\toVect{e}\otimes\toVect{e}-\frac{1}{2}|\toVect{e}|^2\toMat{I}\right).
\end{equation}
The last term in \eq\eqref{eq:S} corresponds to the total flexoelectricity-induced stress, and is analogous to the term appearing in the linear theory of flexoelectricity, \cf\eqs (31-33) in \citet{codony2019immersed}.

Equations \eqref{eq:P} and \eqref{D_cons} show that the Lagrangian flexoelectric polarization in the present theory is
$ P_M = JC^{-1}_{ML}(\epsilon-\epsilon_0)E_L
+
J C^{-1}_{ML}\flexo_{LIJK}\widetilde{\mathfrak{E}}_{IJK}$
, and hence its spatial counterpart is derived with \eq\eqref{Sa2} and \eqref{Sa4} as
\begin{equation}
p_m = (\epsilon-\epsilon_0)e_m + F^{-1}_{Lm} \mu_{LIJK} \widetilde{\mathfrak{E}}_{IJK}.
\end{equation}
In the present formulation, $\mu_{LIJK}$ is a purely Lagrangian tensor, and hence it is meaningful to view it as a material constant with the same material symmetries and intrinsic symmetry ($\mu_{LIJK} = \mu_{LJIK}$) as the small strain flexoelectric tensor \citep{Majdoub2008,Zubko2013,Krichen2016}. We note, however, that in previous literature a distinct notion of polarization per unit undeformed volume is introduced as $\Pstar = J\p$, i.e.~a volume-normalized spatial polarization related to our material or nominal polarization by $\pstar_i=F_{iI}P_I$ \citep{Liu2014,dorfmann2014nonlinear,dorfmann2017nonlinear}. The polarization $\Pstar$ is not work-conjugate to the Lagrangian electric field $\E$. Furthermore, when it is used to formulate flexoelectric models it can be problematic. Indeed, the flexoelectric coupling has been defined in terms of $\Pstar$ \citep{Liu2014,Deng2014electrets,Deng2014flexoelectricity,Yvonnet2017,Thai2018} as
\begin{equation}\label{flexopotwrong}
\Psi^\text{Flexo}(\FF,\Pstar)=-\pstar_l \mathdutchcal{F}_{liJK}\ff_{iJK},
\end{equation}
with $\mathdutchcal{F}$ a mixed spatial-material flexoelectric tensor, which unlike the infinitesimal flexoelectric tensor is intrinsically symmetric with respect to its last two indices ($\mathdutchcal{F}_{liJK}=\mathdutchcal{F}_{liKJ}$). By comparing \eq\eqref{flexopotwrong} and \eqref{flexopot}, using \eq\eqref{eqfce} and \eqref{straininv}, the relation $\pstar_i=F_{iI}P_I$ and the chain rule, we find the relation between $\boldsymbol{f}$ and $\boldsymbol{\mathdutchcal{F}}$  as
\begin{subequations}\begin{align} \label{mixedFCTensor}
    \mathdutchcal{F}_{liJK}&=
    -\frac{\partial^2\Psi^\text{Flexo}}{\partial\pstar_l\partial\ff_{iJK}}
    =\symm[JK]{f_{LIJK}}F_{iI}F^{-1}_{Ll},\\ \label{matFCTensor}
    f_{LIJK}&=
    -\frac{\partial^2\Psi^\text{Flexo}}{\partial P_L\partial\widetilde{\mathfrak{E}}_{IJK}}
    =\left(
    \mathdutchcal{F}_{liJK}F^{-1}_{Ii}
   +\mathdutchcal{F}_{ljIK}F^{-1}_{Jj}
   -\mathdutchcal{F}_{lkIJ}F^{-1}_{Kk}
   \right)F_{lL}.
\end{align}\end{subequations}
In the limit of infinitesimal deformation, $\mathdutchcal{F}$ and $\boldsymbol{f}$ correspond to the so-called type-I ($\boldsymbol{f}^\text{I}$) and type-II ($\boldsymbol{f}^\text{II}$) flexocoupling tensors, respectively, and choosing one or the other is just a matter of convenience \citep{schiaffino2019metric}. However this equivalence does not hold anymore in a finite deformation framework, since $\boldsymbol{f}$ is purely Lagrangian whereas $\mathdutchcal{F}$ is not.

Equation \eqref{matFCTensor} clearly shows that taking $\boldsymbol{\mathdutchcal{F}}$ as a material constant, as done in \citet{Yvonnet2017} and \citet{Thai2018}, directly implies a very particular dependence of the Lagrangian flexoelectric tensor $\boldsymbol{f}$ on deformation. Thus, formulating flexoelectricity as in \eq\eqref{flexopotwrong} leads implicitly to a material flexoelectric tensor whose magnitude and symmetry depend on deformation in a way that is unphysical. To illustrate this assertion, consider a particular case in which $\boldsymbol{\chi}$ corresponds to a rigid body deformation map, and thus $\toMat{F}$ is a rotation matrix $\toMat{R} \neq\toMat{I}$. Then, \eq\eqref{matFCTensor} leads to
\begin{equation}
    f_{LIJK}=
    \left(
    \mathdutchcal{F}_{liJK}R_{Ii}
   +\mathdutchcal{F}_{ljIK}R_{Jj}
   -\mathdutchcal{F}_{lkIJ}R_{Kk}
   \right)R_{lL}
\end{equation}
showing that, if \eq\eqref{flexopotwrong} is used to model flexoelectricity, then the Lagrangian flexoelectric material tensor, and hence the enthalpy functional $\Pi [\boldsymbol{\chi},\Phi]$, are not invariant with respect to a superimposed rigid body motion and hence not objective.

\section{Numerical implementation}\label{sec_03}
\begin{figure}[!htb]
\centering\minipage{0.7\textwidth}
  \includegraphics[width=\textwidth]{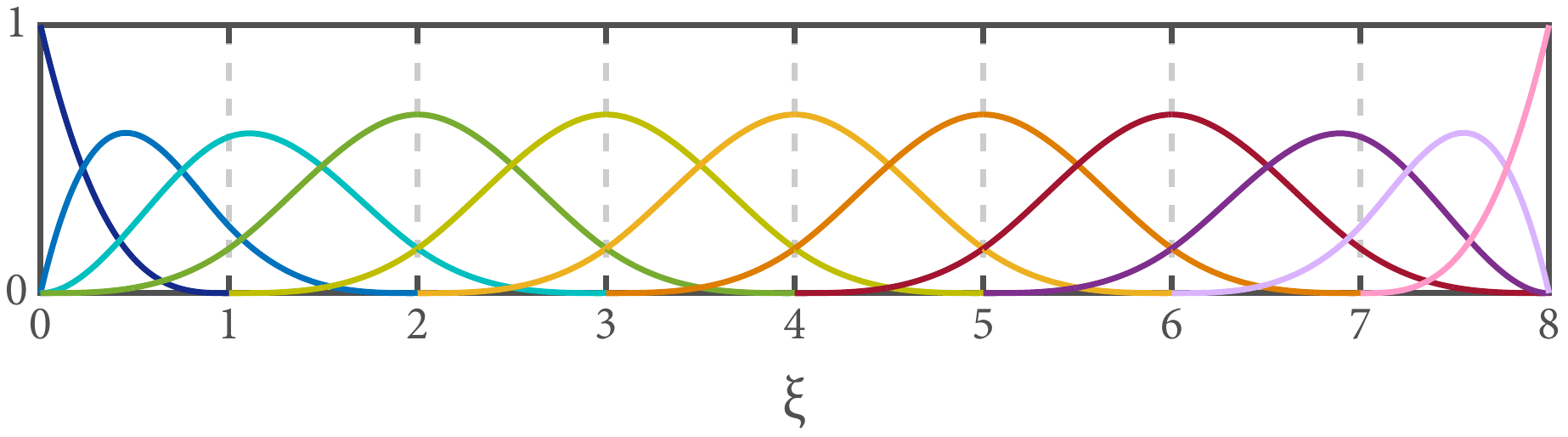}
\endminipage\qquad
\caption{\label{fig:bspline}Univariate open uniform B-spline basis of degree $p=3$. Each basis function is a smooth ($C^{p-1}$) piece-wise polynomial on a compact ($\leq p+1$) support. Multivariate B-spline bases are constructed by means of the tensor product of multiple univariate bases.}
\end{figure}

In this Section, we develop a direct numerical approach to solve the boundary value problem in Section \ref{sec_bbp}.
We restrict ourselves to 2D rod-like geometries, which can be easily discretized by Cartesian grids.
The state variables $\{\boldsymbol{\chi},\Phi\}$ are approximated by an open uniform B-spline basis \citep{deBoor2001,Rogers2001,Piegl2012} of degree $p\geq2$ in order to provide the smoothness required by the high-order model (see \fig\ref{fig:bspline}). Since the basis is interpolant at the boundaries of the reference domain, Dirichlet boundary conditions are strongly enforced. Domain and boundary integrals are approximated by standard Gaussian quadrature rules.

The discretization of \eq\eqref{res} yields a nonlinear system of equations (for the sake of brevity, we keep the same notation to denote discretized quantities). In order to solve it, we consider a modified-step Newton-Raphson algorithm. At the $k$-th iteration, an increment of the solution $\{\Delta\boldsymbol{\chi},\Delta\Phi\}^{(k)}$ is found by vanishing the first order Taylor expansion of the residual $\mathcal{R}$ in \eq\eqref{res} around the previous solution $\{\boldsymbol{\chi},\Phi\}^{(k-1)}$:
\begin{align}\label{system}
	\mathcal{R}[\boldsymbol{\chi}^{(k)},\Phi^{(k)};\delta\boldsymbol{\chi},\delta\Phi]
	\approx{}&\ 
	\mathcal{R}[\boldsymbol{\chi}^{(k-1)},\Phi^{(k-1)};\delta\boldsymbol{\chi},\delta\Phi]
	+
	\frac{\partial \mathcal{R}[\boldsymbol{\chi}^{(k-1)},\Phi^{(k-1)};\delta\boldsymbol{\chi},\delta\Phi]}{\partial\boldsymbol{\chi}}\Delta\boldsymbol{\chi}^{(k)}
	\nonumber\\&
	+
	\frac{\partial \mathcal{R}[\boldsymbol{\chi}^{(k-1)},\Phi^{(k-1)};\delta\boldsymbol{\chi},\delta\Phi]}{\partial\Phi}\Delta\Phi^{(k)}
	=0,
\end{align}
leading to an algebraic system of equations for $\{\Delta\boldsymbol{\chi},\Delta\Phi\}^{(k)}$ of the form
\begin{equation}
{\begin{bmatrix}
\toMat{H}_{\boldsymbol{\chi}\boldsymbol{\chi}} & \toMat{H}_{\boldsymbol{\chi}\Phi} \\
\toMat{H}_{\Phi\boldsymbol{\chi}} & \toMat{H}_{\Phi\Phi}
\end{bmatrix}}^{(k-1)}
\cdot
{\begin{bmatrix}
\Delta\boldsymbol{\chi}\\
\Delta\Phi
\end{bmatrix}}^{(k)}
=
-{\begin{bmatrix}
\toMat{R}_{\boldsymbol{\chi}}\\
\toMat{R}_\Phi
\end{bmatrix}}^{(k-1)}
,
\end{equation}
given $\{\boldsymbol{\chi},\Phi\}^{(k-1)}$ at the previous iteration. The explicit form of the variations of the residual $\mathcal{R}$ can be found in \ref{App02}.

Once $\{\Delta\boldsymbol{\chi},\Delta\Phi\}^{(k)}$ are found, we compute the \emph{modified} increments of the solution at the $k$-th iteration, namely $\{\overline{\Delta\boldsymbol{\chi}},\overline{\Delta\Phi}\}^{(k)}$, by ensuring that the total increment \emph{i)} leads to an enthalpy decrease along $\boldsymbol{\chi}$, \emph{ii)} leads to an enthalpy increase along $\Phi$, and \emph{iii)} has a predefined maximum norm $\gamma_\text{max}\in\mathbb{R}^{+}$. The first two conditions are required in accordance to the variational principle in \eq\eqref{argmm}, whereas the latter is just a numerical requirement to avoid too large increments of the solution at each iteration. To formulate those conditions mathematically, let us recast the variational principle in \eq\eqref{argmm} as
\begin{subequations}\label{recast}
\begin{align}
\widehat\Phi(\boldsymbol{\chi}):=&\argmax_{\Phi\in\mathcal{P}} \Big( \Pi[\boldsymbol{\chi},\Phi] \Big);\\
\boldsymbol{\chi}^*=&\argmin_{\boldsymbol{\chi}\in\mathcal{X}} \Big( \widehat\Pi[\boldsymbol{\chi}] \Big),\quad\text{ with }\quad\widehat\Pi[\boldsymbol{\chi}]:=\Pi[\boldsymbol{\chi},\widehat\Phi(\boldsymbol{\chi})];\\
\Phi^*=&\,\widehat\Phi(\boldsymbol{\chi}^*).
\end{align}
\end{subequations}
Numerically, \eq\eqref{recast} is equivalent to solving two linear systems consecutively, constructed from \eq\eqref{system} by writing $\Delta\Phi^{(k)}$ as a function of $\Delta\boldsymbol{\chi}^{(k)}$, as follows:
\begin{subequations}\label{recast2}
\begin{align}
\left.\toMat{\widehat{H}}_{\boldsymbol{\chi}\boldsymbol{\chi}}\right.^{\!(k-1)}
\cdot
\left.\Delta\boldsymbol{\chi}\right.^{\!(k)}
=&
-\left.\toMat{\widehat{R}}_{\boldsymbol{\chi}}\right.^{\!(k-1)}
\quad\text{with}\quad
\left\{\begin{aligned}
\left.\toMat{\widehat{H}}_{\boldsymbol{\chi}\boldsymbol{\chi}}\right.^{\!(k-1)}&:=\left.\toMat{H}_{\boldsymbol{\chi}\boldsymbol{\chi}}\right.^{\!(k-1)}-\left.\toMat{H}_{\boldsymbol{\chi}\Phi}\right.^{\!(k-1)}\cdot{\left.\toMat{H}_{\Phi\Phi}^{-1}\right.^{\!(k-1)}}\cdot\left.\toMat{H}_{\Phi\boldsymbol{\chi}}\right.^{\!(k-1)}\\
\left.\toMat{\widehat{R}}_{\boldsymbol{\chi}}\right.^{\!(k-1)}&:=\left.\toMat{R}_{\boldsymbol{\chi}}\right.^{\!(k-1)}-\left.\toMat{H}_{\boldsymbol{\chi}\Phi}\right.^{\!(k-1)}\cdot\left.\toMat{H}_{\Phi\Phi}^{-1}\right.^{\!(k-1)}\cdot\left.\toMat{R}_{\Phi}\right.^{\!(k-1)}
\end{aligned}\right.
;
\\
\left.\toMat{H}_{\Phi\Phi}\right.^{\!(k-1)}
\cdot
\left.\Delta\Phi\right.^{\!(k)}
=&
-\left.\toMat{\widehat{R}}_{\Phi}\right.^{\!(k-1)}
\quad\text{with}\quad
\begin{aligned}
\left.\toMat{\widehat{R}}_{\Phi}\right.^{\!(k-1)} &:= \left.\toMat{R}_{\Phi}\right.^{\!(k-1)}+\left.\toMat{H}_{\Phi\boldsymbol{\chi}}\right.^{\!(k-1)}\cdot\left.\Delta\boldsymbol{\chi}\right.^{\!(k)}
\end{aligned}
.
\end{align}
\end{subequations}
From \eq\eqref{recast2} it is clear that the descent and ascent directions are respectively identified by
$\left.\toMat{\widehat{R}}_{\boldsymbol{\chi}}\right.^{\!(k-1)}$
and
$\left.\toMat{\widehat{R}}_{\Phi}\right.^{\!(k-1)}$, \ie the \emph{modified} residuals which take into account the coupled nature of the enthalpy potential. Therefore, the modified increments are computed as follows:
\begin{subequations}\begin{align}
{\alpha_{\boldsymbol{\chi}}}^{(k)} &=
	\begin{cases}
	-1 & \text{if } \left.\toMat{\widehat{R}}_{\boldsymbol{\chi}}\right.^{\!(k-1)}\cdot\Delta\boldsymbol{\chi}^{(k)} > 0,\\
	+1 & \text{otherwise};
	\end{cases}
\\{\alpha_\Phi}^{(k)} &=
\begin{cases}
-1 & \text{if } \left.\toMat{\widehat{R}}_{\Phi}\right.^{\!(k-1)}\cdot\Delta\Phi^{(k)} < 0,\\
+1 & \text{otherwise};
\end{cases}
\\
{\beta}^{(k)} &=\min\left\{+1,\gamma_\text{max}/\sqrt{\norm{
\frac{\Delta\boldsymbol{\chi}^{(k)}}{{\chi}_0}}^2 + \norm{\frac{\Delta\Phi^{(k)}}{\Phi_0}}^2}\right\}; 
\\
\overline{\Delta\boldsymbol{\chi}}^{(k)}&={\alpha_{\boldsymbol{\chi}} }^{(k)}{\beta}^{(k)}\Delta\boldsymbol{\chi}^{(k)};
\\
\overline{\Delta\Phi}^{(k)}&={\alpha_\Phi}^{(k)}{\beta}^{(k)}\Delta\Phi^{(k)};
\end{align}\end{subequations}
with ${\chi}_0$ and $\Phi_0$ characteristic factors of the problem for displacement and potential. In practice, $\gamma_\text{max}$ is treated as an adaptive heuristic parameter, tunable for proper convergence.

Finally, the solution at the $k$-th iteration is updated with
\begin{align}
\{\boldsymbol{\chi},\Phi\}^{(k)} = \{\boldsymbol{\chi},\Phi\}^{(k-1)}+\{\overline{\Delta\boldsymbol{\chi}},\overline{\Delta\Phi}\}^{(k)}.
\end{align}

The external loads are applied incrementally in a sequence of load steps, and the modified-step Newton-Raphson algorithm presented here is used to obtain converged solutions at every load step. Once convergence is reached, the stability of the solution is checked by assuring  $\{\boldsymbol{\chi},\Phi\}^{(k)}$ is a saddle point in the enthalpy functional $\Pi[\boldsymbol{\chi},\Phi]$ in accordance to the variational principle in \eq\eqref{argmm}. By means of \eq\eqref{recast}, 
stability of $\{\boldsymbol{\chi},\Phi\}^{(k)}$ is given by
\begin{subequations}\label{stabcond}\begin{align}
\delta_{\boldsymbol{\chi}}^2\widehat\Pi[\boldsymbol{\chi}^{(k)};\Delta\boldsymbol{\chi};\Delta\boldsymbol{\chi}]&>0\quad\forall\Delta\boldsymbol{\chi}\in\mathcal{X},\\
\delta_\phi^2\Pi[\boldsymbol{\chi}^{(k)},\Phi^{(k)};\Delta\Phi;\Delta\Phi]&<0\quad\forall\Delta\Phi\in\mathcal{P}.
\end{align}\end{subequations}
Numerically, \eq\eqref{stabcond} is met by checking the sign of the extremal eigenvalues $\lambda$ of $\left.\toMat{\widehat{H}}_{\boldsymbol{\chi}\boldsymbol{\chi}}\right.^{\!(k)}$ and $\left.\toMat{H}_{\Phi\Phi}\right.^{\!(k)}$ as follows:
\begin{equation}\label{stabcond2}
\lambda_\text{min}\left[\left.\toMat{\widehat{H}}_{\boldsymbol{\chi}\boldsymbol{\chi}}\right.^{\!(k)}\right]>0,\qquad
\lambda_\text{max}\left[\left.\toMat{H}_{\Phi\Phi}\right.^{\!(k)}\right]<0.
\end{equation}
We recognize convergence to unstable solutions by the violation of \eq\eqref{stabcond2}. In such case, the solution $\{\boldsymbol{\chi},\Phi\}^{(k)}$ is slightly perturbed and the iterative algorithm is run again until a stable solution is found.
In practice, we found that $\lambda_\text{max}\left[\left.\toMat{H}_{\Phi\Phi}\right.^{\!(k)}\right]$ remains always negative, and therefore the encountered instabilities are given by $\lambda_\text{min}\left[\left.\toMat{\widehat{H}}_{\boldsymbol{\chi}\boldsymbol{\chi}}\right.^{\!(k)}\right]$ becoming negative only (\ie \emph{geometrical} instabilities). The eigenvector associated to $\lambda_\text{min}\left[\left.\toMat{\widehat{H}}_{\boldsymbol{\chi}\boldsymbol{\chi}}\right.^{\!(k)}\right]$ is an appropriate direction for numerical perturbations on $\boldsymbol{\chi}^{(k)}$ to reach stable solutions.

\section{One-dimensional analytical models for flexoelectric rods undergoing large displacements and rotations}\label{sec_04}
\begin{figure}[!htb]
    \centering
    \includegraphics[width=0.6\textwidth]{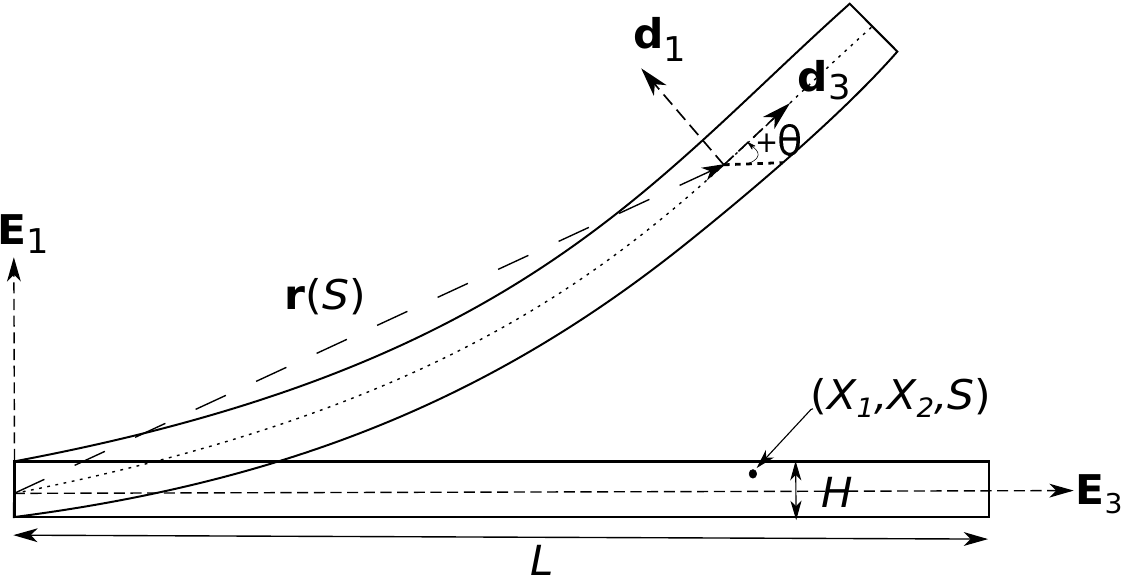}
    \caption{A typical schematic of deformed planar rod of length $L$ and height $H$ from its reference straight configuration. For upward bending, $\theta >0$.  }
    \label{fig:deformedrod}
\end{figure}

In this Section, we derive simplified closed-form solutions for planar bending and buckling of flexoelectric slender uniform rods  undergoing large displacements and rotations under open circuit and close circuit conditions. A material point in the reference configuration is denoted by $\X = X_{1} \mathbf{E}_1 +X_2\mathbf{E}_3+S\mathbf{E}_3$, where $\{{\mathbf{E}_1,\mathbf{E}_2,\mathbf{E}_3}\}$ is a global right-handed orthonormal basis of $\mathbb{R}^3$, ($X_1,X_2$) denotes the coordinates of the undeformed cross-section and $X_3=S$ is the Lagrangian coordinate along the undeformed arc-length, see \fig\ref{fig:deformedrod}. The rod is assumed to be extensible, and thus $S$ is not arc-length of the deformed centerline, but unshearable,  following the special Cosserat rod kinematics \citep{Antman1995}. We further assume that the cross-sections of the rod remain plane and rigid during the deformation. The corresponding deformation map can be defined as $\boldsymbol{\chi}(X_1, X_2, S) = \mathbf{r}(S) +X_1 \mathbf{d}_{1} + X_2 \mathbf{d}_2$, where $\mathbf{r}(S)$ is the deformed position of the centerline at $S \mathbf{E}_3$  and ($\mathbf{d}_{1},\mathbf{d}_2$)  are the director vectors associated with the cross section. For planar bending (\fig\ref{fig:deformedrod}),
\begin{eqnarray}
\mathbf{d}_1 &=& -\sin{\theta} \ \mathbf{E}_3 + \cos{\theta} \ \mathbf{E}_1 ,\\ 
\mathbf{d}_2 &=& \mathbf{E}_2,
  \end{eqnarray}
where $\theta$ is the angle of deflection.
The deformation gradient $F_{iI} = \dfrac{\partial x_i}{\partial X_I}$ is obtained as
\begin{align}
\dfrac{\partial \boldsymbol{\chi}}{\partial X_1} &=  \mathbf{d}_1,  \label{def_grad_1}\\
\dfrac{\partial \boldsymbol{\chi}}{\partial X_2} &=  \mathbf{E_2},  \label{def_grad_2}\\
\dfrac{\partial \boldsymbol{\chi}}{\partial S}   &=   \mathbf{r}' +X_1 \mathbf{d}_1'.\label{def_grad_3} \end{align}
where $\dfrac{\dd}{\dd S} = ()'$.  Now, following \cite{Antman1995} and \cite{gupta2017effect} we have  
\begin{eqnarray}
\mathbf{r}' &=&  \nu_3 \mathbf{d}_3, \label{t-vector}\\
 \mathbf{d}_1' &=& -\theta' \mathbf{d_3},
\end{eqnarray}
where $\nu_3$ is the stretch, and 
\begin{equation} \label{tangent}
\mathbf{d}_3 = \mathbf{d}_1 \times \mathbf{d}_2 = \cos{\theta} \ \mathbf{E}_3 + \sin{\theta} \ \mathbf{E}_1.
\end{equation}
Thus, \eq \eqref{def_grad_3} becomes 
\begin{eqnarray}\label{dxX3}
\frac{\partial \boldsymbol{\chi}}{\partial S}  &=  (\nu_3 - X_1\theta') ~\mathbf{d}_3,
\end{eqnarray} 
the deformation gradient tensor can written as 
\begin{eqnarray}
\mathbf{F} &= \left [
\begin{array}{ccc}
\cos{\theta} & 0  & (\nu_3 - X_1\theta') \sin{\theta}\\
0 & 1 & 0 \\
-\sin{\theta} & 0 & (\nu_3 - X_1\theta') \cos{\theta}
\end{array}
\right ],
\end{eqnarray}
and the Green-Lagrange strain tensor as
\begin{eqnarray}\label{Green-Lagrange} 
\mathfrak{E} &=  \dfrac{1}{2} \left ( \mathbf{F}^T \mathbf{F}  - \mathbf{I}\right ) = \dfrac{1}{2} \left [
\begin{array}{ccc}
0 & 0  & 0\\
0 & 0 & 0 \\
0 & 0 & (\nu_3-X_1 \theta')^2 -1
\end{array}
\right ].
\end{eqnarray}
We rewrite the only non-vanishing component of  $\toMat{\mathfrak{E}}$ as
\begin{equation}
2\toMat{\mathfrak{E}}_{33} =  (\nu_3-X_1 \theta')^2 -1 \approx  \nu_3^2 -  2 X_1 \nu_3 \theta' -1, 
\end{equation}
where the term $X_{1}^{2} \theta'^{2}$ has been neglected for thin rods. Expanding $\nu_3$  around $\nu_3 = 1$ in a Taylor series and neglecting higher order terms, since for a thin rod stretches are expected to be small, yields
\begin{equation} \label{strain-rod}
\toMat{\mathfrak{E}}_{33} =  \zeta-  X_1 \theta',
\end{equation}
where $\zeta = \nu_3 -1$ is the axial strain. Retaining the above thin rod approximations and further assuming that the stretch and the curvature vary slowly along $S$, the dominant strain gradient component is
\begin{equation} \label{strain-grad-rod}
\widetilde{\toMat{\mathfrak{E}}}_{331} =   - \theta'.
\end{equation}

In the absence of body forces, neglecting strain gradient elasticity and  the effect of $E_3$, the equilibrium  condition  \eq \eqref{res}, reduces to 
\begin{eqnarray}\label{rod-functional}
\int^L_0 \left[\int_A \sp_{33}\delta \mathfrak{E}_{33} \dd A + \int_A \ss_{331}\delta\widetilde{\toMat{\mathfrak{E}}}_{331} \dd A - \int_A D_1 \delta E_1 \dd A\right] ~ \dd X_{3} -\delta \hat{T} + \delta \hat{W}=0.
\end{eqnarray}
where $L$ is the undeformed length of the rod, $A$ is area of the cross-section, and $\delta \hat{T}$ and $\delta \hat{W}$ are the variations of the external work done by mechanical tractions and surface charges. 
Since  bending of slender rods can involve large displacements but typically small Lagrangian strains, all isotropic constitutive models are very close. For convenience, we consider the isotropic Kirchhoff-Saint-Venant model, requiring two elastic constants, here  Young's modulus $Y$ and Poisson's ratio $\nu$, see \eq \eqref{sv}.  The flexoelectric tensor $\Flexo$ is assumed to have cubic symmetry with three independent constants $\flexo_\text{L}$, $\flexo_\text{T}$ and $\flexo_\text{S}$, namely the longitudinal, transversal and shear coefficients (\eq\eqref{flexotensor}).  We assume that all material properties are homogeneous in the cross-section. 

Using \eqs \eqref{strain-rod} and \eqref{strain-grad-rod} in \eq\eqref{rod-functional}, the corresponding local stress, higher order stress and electric displacement relations in \eq\eqref{S_cons}-\eqref{D_cons} reduce to

\begin{eqnarray}
\sp_{33} 
\label{rod-stress}
&=& \bar{Y}(\zeta-X_1\theta') 
	 -(1-\zeta+X_1\theta')\left(-E_1 \mu_{T}\theta' 
	 +  \frac{1}{2}\epsilon E_1^2 \right) 
	, \\  \label{rod-stress-grad}
	\ss_{331} &=& -(1+\zeta-X_1\theta')\ \flexo_\text{T} E_1
	,
	\\  \label{rod-D}
	D_1 &=& \left ( 1+\zeta-X_1\theta' \right ) \left(\epsilon E_1 - \flexo_\text{T} \theta' \right) 
	,
\end{eqnarray}
with $\bar{Y}=Y (1-\nu)/(1+\nu)(1-2\nu)$.                                                                                                                                                          
Note that in this reduced order theory, only transverse flexoelectricity is relevant.
 
\begin{figure}
\begin{subfigure}[b]{.48\textwidth}
  \centering
  \includegraphics[width=0.87\linewidth]{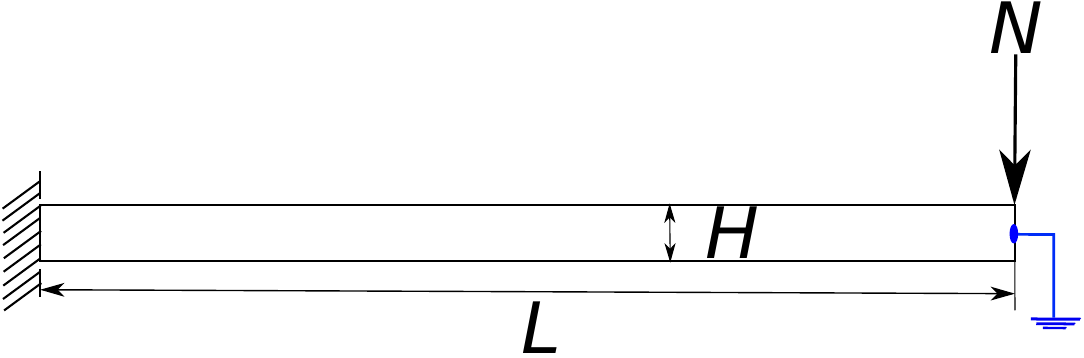}  
  \caption{ A cantilever rod subjected to an endpoint load $N$ at the right end tip, and electrically grounded at the mid-point in the right end cross-section.\vspace{1em}}
  \label{fig:OC-bending}
\end{subfigure}\hfill
\begin{subfigure}[b]{.48\textwidth}
\vspace{1.3cm}
  \centering
  \includegraphics[width=1\linewidth]{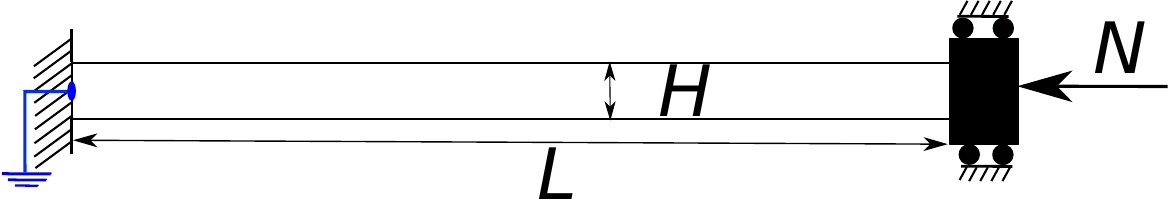}  
  \caption{A clamped-clamped rod subjected to a compressive load $N$ at the right end, and electrically grounded at the mid-point in the left end cross-section. However, the right end is allowed to displace horizontally. }
  \label{fig:OC-buckling}
\end{subfigure}

\begin{subfigure}[b]{.48\textwidth}
  \centering
  \includegraphics[width=.83\linewidth]{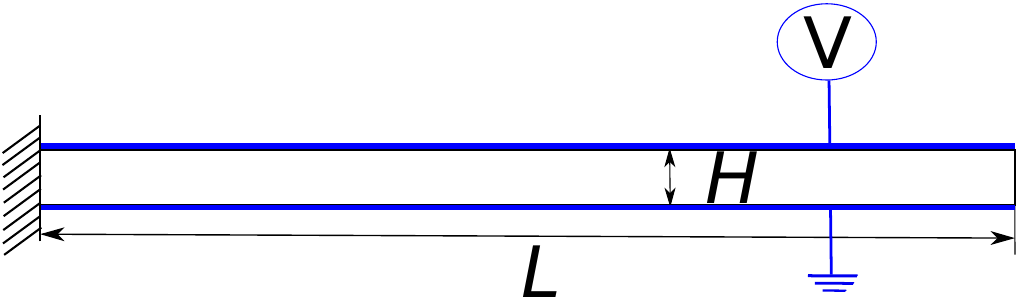}  
  \caption{A cantilever actuator sandwiched between two electrodes (blue in color) under voltage $V$.}
  \label{fig:CC-bending}
\end{subfigure}\hfill
\begin{subfigure}[b]{.48\textwidth}
  \centering
  \hspace{-3em}
  \includegraphics[width=.83\linewidth]{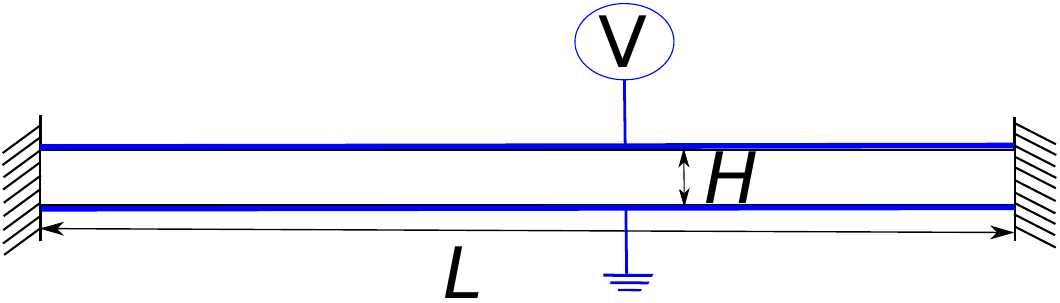}  
  \caption{A clamped-clamped actuator sandwiched between two electrodes (blue in color) under voltage $V$ .}
  \label{fig:CC-buckling}
\end{subfigure}
\caption{A schematic of flexoelectric rod under external mechanical load or external voltage.}
\end{figure}

\subsection{Flexoelectric rod in open circuit under mechanical load}\label{sec_FRS}

We consider now a flexoelectric rod in open circuit conditions, \ie one of the rod end's is grounded and all other boundaries are free of surface charges, \ie they satisfy that $\mathbf{D} \cdot \mathbf{n} =0$, see \figs \ref{fig:OC-bending}, \ref{fig:OC-buckling}. Thus, at the top and bottom surfaces, the vertical electric displacement vanishes, $D_{1}=0$, and for thin rods it can be assumed to vanish within the cross-section as well \citep{Majdoub2008,Majdoub2009,liang2014effects}. In this case, the vertical electric field can be computed from \eq 
\eqref{rod-D} as
\begin{eqnarray}\label{rod-OC-EF}
E_1= \dfrac{\mu_\text{T}}{\epsilon}\theta',
\end{eqnarray}
and then, \eqs \eqref{rod-stress} and \eqref{rod-stress-grad} reduce to
\begin{eqnarray}
\sp_{33} 
 \label{rod-stress-OC}
&=& \bar{Y} (\zeta-X_1\theta') +  (1-\zeta+X_1\theta') \ \dfrac{\flexo_\text{T}^{2} \theta'^{2}}{2\epsilon}
	, \\  \label{rod-stress-grad-OC}
	\ss_{331} &=&  - (1-\zeta+X_1\theta')\ \dfrac{\flexo_\text{T}^2}{\epsilon}\theta'.
\end{eqnarray}
By substituting \eqs \eqref{rod-stress-OC} and \eqref{rod-stress-grad-OC} into \eq \eqref{rod-functional}, and using $\hat{W} =0$ we obtain the following equilibrium condition 
\begin{eqnarray}\label{rod-WF-OC-full}
\int^L_0 \left \{ \bar{Y}A \left  [ \zeta + \dfrac{1}{2} (1-\zeta) \ell_\mu^{2} \theta'^{2}\right ] \delta \zeta + \bar{Y}  \left [ I \left ( 1-\dfrac{1}{2} \ell_\mu^2 \theta'^2 \right ) +  (1-\zeta)\ell_\mu^{2} A
\right ] \theta' \delta \theta'
\right \} \dd X_{3} -\delta \hat{T} =0,
\end{eqnarray}
where $I = \int_A X_1^2 \dd A$ is the moment of inertia of the cross-section and $\ell_\mu = \flexo_\text{T}/\sqrt{\bar{Y} \epsilon}$ is a lengthscale arising from transversal flexoelectricity. Since the stretch in thin rods is expected to be small, even if deformations are not, we can approximate $1-\zeta \approx 1$ in \eq\eqref{rod-WF-OC-full}, which yields 
\begin{eqnarray}\label{rod-WF-OC-full-2}
\int^L_0 \left \{ \bar{Y}A \left  [ \zeta + \dfrac{1}{2} \ell_\mu^{2} \theta'^{2}\right ] \delta \zeta + \bar{Y}  \left [ I \left ( 1-\dfrac{1}{2} \ell_\mu^2 \theta'^2 \right ) +  \ell_\mu^{2} A
\right ] \theta' \delta \theta'
\right \} \dd X_{3} -\delta \hat{T} =0,
\end{eqnarray}
where we identify the axial force and the bending moment as
\begin{eqnarray}
N &=& \bar{Y}A \left  [ \zeta + \dfrac{1}{2} \ell_\mu^{2} \theta'^{2}\right ]  , \label{axial-force-OC} \\
M &=& \bar{Y}  \left [ I -\dfrac{1}{2} \ell_\mu^2 \theta'^2 I +  \ell_\mu^{2} A 
\right ] \theta'. \label{bending-moment-OC} 
\end{eqnarray}
Interestingly, \eq\eqref{rod-WF-OC-full-2} points out the two main size-dependent effects of flexoelectricity. On one hand, flexoelectricity induces a positive size-dependent axial strain in the rod which depends quadratically on the flexural strain $\theta'$. On the other hand, flexoelectricity modifies the effective bending stiffness by two size-dependent contributions of opposite sign. The first is a reduction in the rod's stiffness which depends quadratically on the flexural strain while the second makes the rod stiffer independent of deformation. 

In order to evaluate the relative importance of the different contributions, let us consider a rectangular cross-section of unit width and thickness $H$, \ie $I=H^3/12$ and $A=H$. The second and third contribution to the effective bending stiffness are comparable in magnitude for a radius of curvature $R = \nu_3/\theta' \approx H/5$, which is unphysically small. For reasonable radii of curvature, we expect $\dfrac{1}{2} \ell_\mu^2 \theta'^2 I << \ell_\mu^{2} A$, and thus 
\eq\eqref{rod-WF-OC-full-2} reduces to
\begin{eqnarray}\label{rod-WF-OC}
\int^L_0 \left [ \bar{Y}A  \left  [ \zeta + \dfrac{1}{2} \ell_\mu^{2} \theta'^{2}\right ] \delta \zeta + \bar{Y}  I^\text{eff} \theta' \delta \theta'
\right ] \dd X_{3} -\delta \hat{T} =0,
\end{eqnarray}
with
\begin{equation} \label{Ieff}
I^\text{eff} =  I +  \ell_\mu^{2} A.
\end{equation}
Furthermore, the values of $\ell_\mu$ for typical flexoelectric polymers are in the order of $1-10 \,\si{\nano\metre}$ \citep{chu2012,zhang2015,zhou2017flexoelectric}. The minimum radius of curvature for a rectangular cross-section is $R = H/2$, which implies that the maximum flexoelectrically-induced axial strain is approximately $2 \ell_\mu^{2}/H^2$, in the order of $10^{-3}$ for a $H = 100 \si{\nano\meter} $ thick rod. Thus, we expect the flexoelectrically-induced axial strain to be small, as well as $\zeta$. This is later verified in the numerical examples in Section \ref{Sec:OCBending}, with $\zeta$ in the order of $10^{-4}$ for a $H = 100 \si{\nano\meter} $ thick rod.

Keeping nevertheless the full axial strain $\zeta + \ell_\mu^2 \theta'^{2}/2$, we consider now a flexoelectric cantilever rod subjected to a point load $\mathbf{N}= N_1 \mathbf{E}_1 + N_3 \mathbf{E}_3$ on one of its ends, \figs \ref{fig:OC-bending}, \ref{fig:OC-buckling}. The work done by the external force is
\begin{eqnarray} \label{EW}
\hat{T}= \mathbf{N} \cdot \mathbf{r}(L) =
\mathbf{N}  \cdot \int^L_0 (1+\zeta) \ \mathbf{d}_3(S) \dd S =  \int^L_0 (1+\zeta)  \left (N_1 \sin{\theta}  + N_3 \cos{\theta}  \right ) \dd S, 
\end{eqnarray}
where we have used \eq \eqref{t-vector}. Substituting the first variation of \eq \eqref{EW} in \eq \eqref{rod-WF-OC}, and assuming that the stretch is small, \ie $1+\zeta \approx 1$  yields
\begin{multline}\label{variational-principle-1}
\int^L_0 \left [ \bar{Y}A \left  [ \zeta + \dfrac{1}{2} \ell_\mu^{2} \theta'^{2}\right ] \delta \zeta + \bar{Y}  I^\text{eff} \theta'  \delta \theta' \right ] \dd S  \\ =
   \int^L_0 \left [( N_1 \sin{\theta} + N_3 \cos{\theta})~ \delta \zeta + (N_1 \cos{\theta} - N_3 \sin{\theta})~ \delta \theta \right ] \dd S.
\end{multline} 
Upon integration by parts, \eq \eqref{variational-principle-1} becomes
\begin{multline}\label{variational-principle-2}
\int^L_0 \left [ \bar{Y}A  \delta \left  [ \zeta + \dfrac{1}{2} \ell_\mu^{2} \theta'^{2}\right ] - \bar{Y}  I^\text{eff}  \theta '' \delta \theta \right ] \dd S + \left. \bar{Y}  I^\text{eff} \theta' \delta \theta \right \vert_0^L \\ = 
\int^L_0 \left [( N_1 \sin{\theta} + N_3 \cos{\theta})~ \delta \zeta + (N_1 \cos{\theta} - N_3 \sin{\theta})~ \delta \theta \right ] \dd S,
\end{multline}
from where the Euler-Lagrange equations can be derived for all admissible $\delta \zeta$ and $\delta \theta$ as
\begin{subequations} \label{EL}
\begin{equation}\label{EL-force}
    \bar{Y}A  \zeta + \dfrac{\bar{Y}A}{2} \ell_\mu^{2} \theta'^{2} -  N_1 \sin \theta  - N_3 \cos{\theta}= 0, 
\end{equation}
\begin{equation} \label{EL-moment}
 \bar{Y} I^\text{eff}\theta'' + N_1  \cos{\theta} - N_3 \sin{\theta}= 0,
 \end{equation}
\end{subequations}
where we have assumed that the external force $\mathbf{N}$ is known. Equations \eqref{EL} form a system of two coupled equations for the two unknowns $\zeta$ and $\theta$, where $\theta$ can be obtained from \eq \eqref{EL-moment}  and used in \eq \eqref{EL-force} to compute $\zeta$. Note that 
\eq \eqref{EL-moment} corresponds to bending moment balance of a purely mechanical non-linear Kirchhoff rod with modified (larger) bending rigidity $I^\text{eff}$ \citep{Antman1995}. This effective stiffness coincides with that identified by  \cite{Majdoub2008,Majdoub2009,liang2014effects} for linear flexoelectric rods.
Equation \eqref{EL-moment} can be rewritten in standard form as
\begin{equation} \label{EL-moment-standard}
\theta'' + \bar{\beta}^{2} \, \mathbf{N}\cdot\mathbf{d}_1 =0,
\end{equation}
with $\bar{\beta}^{-2} = \bar{Y} I^\text{eff}$.

We derive next the solution for bending of a cantilever flexoelectric rod under a vertical point load, and buckling of a doubly clamped rod under axial compression, see \fig \ref{fig:OC-bending}.

\subsubsection{Bending of a flexoelectric cantilever under a vertical point load}\label{OC-KH-bending}

We consider a cantilever flexoelectric rod subjected to a vertical force $\mathbf{N} = -N \mathbf{E}_1$, see \fig \ref{fig:OC-bending}. In this case, \eq \eqref{EL-moment-standard} reduces to 
\begin{equation} \label{EL-cantilever-moment}
 \theta'' - \beta^2  \cos{\theta} = 0,
 \end{equation}
with $\beta^{2} = \bar{\beta}^{2} \, N$  and boundary conditions
\begin{subequations}\label{BC-cantilever} 
\begin{eqnarray}
\theta(0) &=& 0,\\
\theta' (L) &=& 0.
\end{eqnarray}
\end{subequations}
The solution to this problem was obtained by \citet{Bisshopp1945}.
As derived in detail in \ref{appendix:bending},  the vertical displacement of the tip is
\begin{eqnarray}\label{tip_deflection}
r_1(L)  = L+\dfrac{2}{\beta} \left[\tilde{E}(p,\psi_0) -\tilde{E}(p)\right],
\end{eqnarray}
where $\tilde{E}(p)$ and $\tilde{E}(p,\psi_0)$ are the complete and incomplete elliptical integrals of the second kind, respectively, see \eq \eqref{elipticalInt_2ndKind}, with $p=\sqrt{(1-\sin{\theta^\text{max}})/2}$, and  $1/\sin{\psi_0} = \sqrt{1-\sin{\theta^\text{max}}}$, with $\theta^\text{max} = \theta(L)$. For a given load $N$, $\theta^\text{max}$ is obtained by the shooting method, using \eq\eqref{eqF}. 
Using \eqs\eqref{EL-force} and \eqref{eqn:dX3-1}, the axial strain can be computed as
\begin{equation}\label{stretch-cantilever-OC}
\zeta(S) =  - \frac{N}{\bar{Y} A} \sin{\theta} - \beta^2 \ell_\mu^{2} \left ( \sin{\theta} - \sin{\theta_\text{max}}\right),
\end{equation}
which attains its maximum value (in magnitude) at the free end
\begin{equation}\label{max-stretch-cantilever-OC}
\zeta(L) =  - \frac{N}{\bar{Y} A} \sin{\theta_\text{max}}.\end{equation}
Note that for $N>0$, the rod bends downwards ($\theta<0$) while for $N<0$, the rod bends upwards ($\theta>0$) and thus in all cases $\zeta >0$.

The electric field at the fixed end is  (see \eq\eqref{rod-OC-maxEF})
\begin{equation}\label{rod-OC-maxEF-main}
E_1(0) = \dfrac{\mu_\text{T} }{\epsilon } \theta'(0)  = - \dfrac{\mu_\text{T} }{\epsilon }\beta \sqrt{2\sin{\vert \theta^\textnormal{max} \vert}}  = -\ell_\flexo\sqrt{\dfrac{2N}{\epsilon I^\text{eff}} \sin{\vert \theta^\textnormal{max} \vert}} 
.
\end{equation}

In the limit case of small deflections, or small $N$, we recover the well-known flexoelectric theory relying on  linear Euler-Bernoulli beams \citep{Majdoub2008,Majdoub2009}, yielding the vertical displacement at the free end and the curvature at the fixed end as
\begin{eqnarray} \label{Euler-Bernoulli_tip_deflection}
	r_1(L)  &=& -\dfrac{NL^3}{3\bar{Y}I^\text{eff}},
	\\
	\theta' (0) &\approx& - \dfrac{NL}{\bar{Y}{I^\text{eff}}}, 
	\label{Euler-Bernoulli_tip_curvature}
\end{eqnarray}
and thus the electric field at the fixed end for the linear Euler-Bernoulli beam is
\begin{eqnarray} 
E_1(0)  &=& \dfrac{\mu_\text{T}}{\epsilon} \theta'(0)  \approx - \dfrac{\mu_\text{T}NL}{\epsilon \bar{Y}{I^\text{eff}}} 
. \label{Euler-Bernoulli_tip_EF}
\end{eqnarray}
Interestingly, for a given rod of length $L$, with cross-section area $A$ and moment of inertia $I$, the vertical electric field at the fixed end in \eq \eqref{Euler-Bernoulli_tip_EF} attains a maximum for \
\begin{eqnarray}\label{muTstar}
\mu^*_T = \sqrt{\epsilon \bar{Y} \dfrac{I}{A}}.
\end{eqnarray}
From a physical point of view, this maximum is a result of two competing effects of flexoelectricity. On one hand, flexoelectricity increases the bending rigidity of the rod by increasing the effective moment of inertia in $\ell_{\mu}^2 A \propto \flexo_\text{T}^2$, see \eq \eqref{Ieff}, and thus reduces the rod deflection, the curvature and the resulting vertical electric field. On the other hand, for a given curvature, the vertical electric  field is proportional to $\flexo_\text{T}$. 
The maximum electric field at the fixed end for this optimum flexoelectric coefficient $\flexo_\text{T}^{*}$ becomes
\begin{eqnarray}\label{E_max-1}
E^{\text{max}}_1(0) = -\dfrac{NL}{2\sqrt{\epsilon \bar{Y}IA}}.
\end{eqnarray}
Similarly, for a given material with properties $Y,\epsilon,$ and $\flexo_\text{T}$, one can find an optimal design that maximizes the flexoelectric response. For instance, considering a rod with square/rectangular cross-section, the optimal thickness is  
\begin{eqnarray} \label{Hstar}
H^* = \mu_\text{T} \sqrt{\dfrac{12}{\bar{Y}\epsilon}} = 2\sqrt{3} \, \ell_\flexo,
\end{eqnarray}
and the corresponding maximum vertical electric field at the fixed end for a rod with a unit width is
\begin{equation} \label{E_max-2}
\E^{\text{max}}_1(0) = -\dfrac{\sqrt{\epsilon \bar{Y}} NL}{4\sqrt{3} \flexo_\text{T}^{2}}. 
\end{equation}

\subsubsection{Buckling of a flexoelectric rod under axial compression}\label{OC-KH-buckling}
We consider next an open-circuit flexoelectric rod clamped at the left end and with vertical displacement and rotation prevented at the right end, subjected to a compressive force $\mathbf{N} = - N \mathbf{E}_{3}$, see \fig\ref{fig:OC-buckling}. We examine the buckling critical load $N_\text{cr}$ and the post-buckling behavior. 
In this case, the vertical reaction at the right end $N_{1}$ is unknown. Thus, the Euler-Lagrange \eqs\eqref{EL} have to be supplemented with the constraint of vanishing vertical displacements at the right end given by
\begin{equation} \label{constraint-OC-buckling}
0= \mathbf{E}_{1} \cdot \mathbf{r}(L)  = 
\mathbf{E}_{1}  \cdot \int^L_0 (1+\zeta) \ \mathbf{d}_3(s) \dd s =  \int^L_0 (1+\zeta) \sin{\theta}  \dd s. 
\end{equation}
The bending moment balance \eq\eqref{EL-moment-standard} is written as
\begin{equation} \label{EL-moment-buckling}
 \theta'' + \beta^{2} \sin{\theta} + N_{1} \bar{\beta}^2 \cos{\theta}= 0,
 \end{equation}
with $\beta^2=N/\bar{Y} I^\text{eff}$, subject to the boundary conditions
\begin{subequations}\label{BC-CC} 
\begin{eqnarray}
\theta(0) &=& 0, \label{BC-CC-1}\\
\theta (L) &=& 0. \label{BC-CC-2}
\end{eqnarray}
\end{subequations}
Since the expected lowest buckling mode is symmetric, the constraint in \eq\eqref{constraint-OC-buckling} is fulfilled by symmetry, and thus the reaction $N_{1}=0$, and \eq\eqref{EL-moment-buckling} reduces to
\begin{equation} \label{EL-moment-buckling-2}
 \theta'' + \beta^{2} \sin{\theta} = 0.
 \end{equation}
The expected lowest mode 
exhibits inflection points at $S=L/4$ and $S=3L/4$, which require special attention \citep{Lin1998}. 
Instead, we invoke symmetry considerations and avoid the inflection points by solving \eq \eqref{EL-moment-buckling-2} over a quarter of the rod and replace \eq \eqref{BC-CC-2} with
\begin{eqnarray}\label{BC:Buckling:OC}
\theta '\left(\frac{L}{4}\right) &=& 0.
\end{eqnarray}
After solving the BVP, see \ref{appendix:buckling} for a detailed derivation, the vertical displacement and the vertical electric field at the center of the rod for upward buckling are, 
\begin{subequations}\label{buckre}\begin{align}
r_1\left(\frac{L}{2}\right) &=  \dfrac{2}{\beta} \sqrt{2\left(1 -\cos \theta^\text{max}\right)},\label{buckrea}\\
E_1\left(\frac{L}{2}\right) &= -\dfrac{\mu_\text{T}}{\epsilon}\sqrt{\dfrac{2N\left(1 -\cos \theta^\text{max}\right)}{\bar{Y}I^\text{eff}}}.\label{buckreb}
\end{align}\end{subequations}
where $\theta^\textnormal{max} = \theta\left({L}/{4}\right)$ is computed for a given load $N$ by the shooting method using \eq\eqref{buckling-CC-OC-solution}. Since the right end of the rod is allowed to move horizontally, its length is assumed to remain unchanged and the stretch is $\nu_3\approx 1$.

The post buckling load can be determined as
\begin{eqnarray} \label{post-buckling-CC}
N = \frac{\beta^{2}}{\bar{\beta}^{2}} = 16 F^2\left(\sin{\frac{\theta^\text{max}}{2}}\right) \dfrac{\bar{Y} I^{\text{eff}}}{L^2},
\end{eqnarray}
where $F$ is the complete elliptical integral of first kind, \cf\eq \eqref{elipticalInt_1stKind}, and we have used \eq\eqref{buckling-CC-OC-solution}. By letting $\theta^\text{max} \rightarrow 0$ in \eq \eqref{post-buckling-CC}, the critical load for buckling is obtained as
\begin{eqnarray}\label{criticalbucklingload}
N_\text{cr} =4\pi^2 \dfrac{\bar{Y}I^\text{eff}}{L^2},
\end{eqnarray}
which coincides with the buckling load for a linear flexoelectrically-stiffened Euler-Bernoulli beam with a modified bending stiffness \citep{timoshenko2009theory}, see \eq \eqref{Ieff}. 

\subsection{Flexoelectric rod actuator in closed circuit}\label{Sec:CC}  
We consider a flexoelectric rod in closed circuit, \ie electrodes are attached to the top and bottom surfaces. Under actuation operation mode, \ie the bottom electrode is grounded ($\phi=0$), while a potential $\phi=V$ is applied to the top electrode,  two setups are studied, bending of a cantilever, \fig \ref{fig:CC-bending}, and buckling of a doubly clamped rod, \fig \ref{fig:CC-buckling}.

In these setups, neglecting the localized boundary effects at the ends of the rod, the non-vanishing electric field component is
\begin{equation}
    E_1 = -\dfrac{V}{H},
    \label{E-field-CC}
\end{equation}
and then, \eqs \eqref{rod-stress} -- \eqref{rod-D} reduce to
\begin{eqnarray}
\sp_{33} 
 \label{rod-stress_{CC}}
&=& \bar{Y}(\zeta-X_1\theta') 
	 -(1-\zeta+X_1\theta')\ \left(\dfrac{V}{H} \mu_{T}\theta' 
	 +  \dfrac{\epsilon V^{2}}{2 H^{2}} \right) , \\  \label{rod-stress-grad-CC}
	\ss_{331} &=& (1+\zeta-X_1\theta')\ \flexo_\text{T} \dfrac{V}{H}, 
	\\  \label{rod-D-CC}
	D_1 &=& \left ( 1+\zeta-X_1\theta' \right ) \left(-\epsilon \dfrac{V}{H} - \flexo_\text{T}\theta'\right).
\end{eqnarray}
Hence, the balance law in \eq \eqref{rod-functional} becomes 
\begin{multline}\label{variational-principle-CC-1}
\int^L_0 \Bigg \{  \left [ \left ( \bar{Y} + \flexo_\text{T} \dfrac{V}{H} \theta' + \dfrac{\epsilon V^{2}}{2H^2} \right ) \zeta - \flexo_\text{T} \dfrac{V}{H} \theta' - \dfrac{\epsilon V^{2}}{2H^2} \right ] A \delta \zeta  
\\+ \left [ \left ( \bar{Y} + \flexo_\text{T} \dfrac{V}{H} \theta' + \dfrac{\epsilon V^{2}}{2H^2} \right ) I \theta' - (1+\zeta) \flexo_\text{T} \dfrac{V}{H} A \right ] \delta \theta' \Bigg \} \dd s - \delta \hat{T}=0,
\end{multline}
where $\delta \hat{T}$ is given by \eq \eqref{EW} for an external force $\mathbf{N}= N_1 \mathbf{E}_1 + N_3 \mathbf{E}_3$ applied at the right end, and we have used $\delta E_{1} =0$, $\delta \hat{W} =0$. Assuming again that the strain is small, integration by parts yields, 
\begin{multline}\label{variational-principle-CC-2}
\int^L_0 \left \{  \left [ \left ( \tilde{Y} + \flexo_\text{T} \dfrac{V}{H} \theta' \right ) \zeta - \flexo_\text{T} \dfrac{V}{H}  \theta' - \dfrac{\epsilon V^{2}}{2H^2} \right ] A \delta \zeta  -  \left [ \left ( \tilde{Y} + \flexo_\text{T} \dfrac{V}{H}  \theta' \right ) I  \theta'  -  (1+\zeta)\flexo_\text{T} \dfrac{V}{H} A \right ]' \delta \theta \right \} \dd s \\- \delta \hat{T} =  -\left. \left [ \left ( \tilde{Y} + \flexo_\text{T} \dfrac{V}{H} \theta' \right ) I  \theta'  - (1+\zeta) \flexo_\text{T} \dfrac{V}{H} A 
\right ] \delta \theta \right \vert_{\theta(0)}^{\theta(L)},
\end{multline}
 where we have defined an effective Young's modulus modified by electrostriction as
\begin{equation} \label{Ybar}
\tilde{Y} = \bar{Y} + \dfrac{\epsilon V^{2}}{2H^2}.
\end{equation}

\subsubsection{Bending of a flexoelectric cantilever under applied voltage}\label{CC-KH-bending}
We consider next a flexoelectric cantilever rod sandwiched between two electrodes as depicted in \fig \ref{fig:CC-bending}. In this case, there are no applied mechanical loads and there is no kinematical constraint at the right  end, and thus $\mathbf{N}=\mathbf{0}$. The Euler-Lagrange equations are identified as  
\begin{subequations}
\begin{equation}
\left (\tilde{Y} + \flexo_\text{T} \dfrac{V}{H} \theta'  \right) \zeta - \flexo_\text{T} \dfrac{V}{H} \theta' - \dfrac{\epsilon V^{2}}{2H^2}  = 0,\label{forcebalance-ff}
\end{equation}
\begin{equation}
 \left [\left ( \tilde{Y} + \flexo_\text{T} \dfrac{V}{H} \theta' \right ) I  \theta'  - (1+\zeta) \flexo_\text{T} \dfrac{V}{H} A \right ]' = 0,
\label{momentbalance-ff}
\end{equation}
\end{subequations} 
Equation \eqref {forcebalance-ff} yields
\begin{equation} \label{zeta-CC-cantilever-0}
\zeta = \frac{\flexo_\text{T} \dfrac{V}{H} \theta' + \dfrac{\epsilon V^{2}}{2H^2}}{\tilde{Y} + \flexo_\text{T} \dfrac{V}{H} \theta'} \approx  \dfrac{\epsilon V^{2}}{2H^2 \tilde{Y}} +  \dfrac{\flexo_\text{T} V}{\tilde{Y} H} \left( 1 -\dfrac{\epsilon V^2}{2 \tilde{Y} H^2} \right ) \theta ', 
\end{equation}
where we have expanded $\zeta$ in a Taylor series around $\theta' =0$ 
and have neglected the higher order terms, thereby assuming that the flexural strain is small. Replacing \eq\eqref{zeta-CC-cantilever-0} in  \eq \eqref{momentbalance-ff} leads to
\begin{equation}
	 \left [ \tilde{Y} \left ( I - \dfrac{\flexo_\text{T}^2 A V^2}{\tilde{Y}^2 H^2} \left ( 1 - \dfrac{\epsilon  V^2}{2 \tilde{Y} H^2} \right )\right )  \theta'  -  \flexo_\text{T} \dfrac{V}{H} A \left ( 1 + \dfrac{\epsilon V^{2}}{2\tilde{Y}H^2} \right )
	\right ]' = 0.\label{momentbalance-ff-2}
\end{equation}
By defining, an effective moment of inertia modified by flexoelectricity and electrostriction as 
\begin{eqnarray}
\tilde{I} =&  I - \dfrac{\flexo_\text{T}^2 A V^2}{\tilde{Y}^2 H^2} \left ( 1 - \dfrac{\epsilon  V^2}{2 \tilde{Y} H^2} \right ),
\end{eqnarray}
and an effective cross-section area modified by electrostriction as
\begin{eqnarray}
\tilde{A} =& A \left ( 1 + \dfrac{\epsilon V^{2}}{2 \tilde{Y}H^2 } \right ),
\end{eqnarray}
\eq \eqref{momentbalance-ff-2} reduces to
\begin{equation}
	 \left [ \tilde{Y}\ \tilde{I}  \theta'  -  \flexo_\text{T} \dfrac{V}{H} \tilde{A} \right ]' = 0.\label{momentbalance-ff-3}
\end{equation}
Equation \eqref{momentbalance-ff-3} implies that the flexural strain $\theta'$ is uniform along $s$.
By \eq\eqref{variational-principle-CC-2}, the corresponding boundary conditions are
\begin{eqnarray}
\theta(0) &=& 0, \label{BC-cantilever-CC-1}\\
 \tilde{Y} \ \tilde{I} \theta'(L)  -  \flexo_\text{T} \dfrac{V}{H} \tilde{A} &=& 0. \label{BC-cantilever-CC-2}
 \end{eqnarray}
The resulting uniform flexural strain in this case is 
\begin{equation} \label{varkappa-CC-cantilever}
  \theta' = \theta'(L) =\flexo_\text{T} \dfrac{V}{H}  \dfrac{\tilde{A}}{\tilde{Y} \ \tilde{I}},  
\end{equation}
which agrees with the expression given in \cite{Bursian1974} for a linear flexoelectric rod by replacing the effective quantities ($\tilde{Y}, \bar{I}, \bar{A}$) by the nominal ones ($\bar{Y}, I, A$). From \eq  \eqref{zeta-CC-cantilever-0} the axial strain is obtained as
\begin{equation} \label{zeta-CC-cantilever}
\zeta = \flexo^2_\text{T} \dfrac{\tilde{A} }{\tilde{Y}^2 \tilde{I}} \dfrac{V^2}{H^2}\left( 1 - \dfrac{\epsilon V^{2}}{2\tilde{Y}H^2}\right)+ \dfrac{\epsilon V^{2}}{2H^2 \tilde{Y}}.  
\end{equation}
 We now examine \eqs\eqref{varkappa-CC-cantilever} and \eqref{zeta-CC-cantilever} by Taylor expansion of these expressions around $V/H = 0$ as

\begin{subequations} \label{Taylor-ActCF-FS-AS}
\begin{align}
\theta' &= \left(\dfrac{V}{H}\right)\frac{A \mu_T}{I \bar{Y}} + \left(\frac{V}{H}\right)^3\frac{A^2 \mu_T^3 }{I^2 \bar{Y}^3} + \mathcal{O}\left (\left(\dfrac{V}{H}\right)^5 \right ) \label{Taylor-ActCF-FS}, \\
\zeta &=
 \frac{1}{2}\left(\frac{V}{H}\right)^2\frac{\epsilon}{\bar{Y}}\left(1+2\frac{A\ell_\flexo^2}{I}\right)
-\frac{1}{4}\left(\frac{V}{H}\right)^4\left(\frac{\epsilon}{\bar{Y}}\right)^2 \left(1+4\frac{A\ell_\flexo^2}{I}-4\left(\frac{A\ell_\flexo^2}{I}\right)^2\right)+\mathcal{O}\left ( \left(\dfrac{V}{H} \right)^6 \right) \nonumber \\
&\approx
 \frac{1}{2}\left(\frac{V}{H}\right)^2\frac{\epsilon}{\bar{Y}}
-\frac{1}{4}\left(\frac{V}{H}\right)^4\left(\frac{\epsilon}{\bar{Y}}\right)^2 +\mathcal{O}\left ( \left(\dfrac{V}{H} \right)^6 \right) \label{Taylor-ActCF-AS}
\end{align}
\end{subequations}
According to \eq\eqref{Taylor-ActCF-FS}, under the application of a voltage $V$, the flexoelectric cantilever bends upwards for $V >0$ and downwards for $V<0$ due to the positive flexoelectric coupling, and elongates regardless of the sign of $V$, due to both flexoelectricity and electrostriction from \eq \eqref{Taylor-ActCF-AS}. However, the contribution of the flexoelectric effect on the axial strain is negligible as for typical flexoelectric elastomers  $A\ell_\flexo^2/I\approx 10^{-2} \ll1$ for a $H = 100 \si{\nano\meter}$ thick rod, as previously argued in Section \ref{sec_FRS}.

Finally,
keeping only the leading order terms, 
the curvature of the rod is obtained from \eqs\eqref{Taylor-ActCF-FS-AS} 
as
\begin{eqnarray}\label{curvature:CC}
\dfrac{1}{R} = \dfrac{\theta'}{(1+\zeta)} \approx \flexo_{\text{T}}\dfrac{A}{I\tilde{Y}} \dfrac{V}{H}. 
\end{eqnarray}
Integrating $\theta$ from \eq \eqref{curvature:CC} and accounting for the clamping condition \eq \eqref{BC-cantilever-CC-1}, we have
\begin{eqnarray}
\theta(S) = (1+\zeta) \flexo_{\text{T}}\dfrac{A}{I\tilde{Y}}\dfrac{V}{H} \ S.
\end{eqnarray}
Finally, the vertical deflection at the free end can be evaluated as
\begin{align}
        r_1(L) &= \int_{0}^{L} (1+\zeta) \sin \theta \dd S =  (1+\zeta) \int_{0}^{L}\sin \left((1+\zeta) \flexo_{\text{T}}\dfrac{A}{I\tilde{Y}}\dfrac{V}{H} \ S\right) \dd S \nonumber\\ &= \dfrac{\tilde{Y}I}{\flexo_{\text{T}} A} \dfrac{H}{V} \left[1 -\cos \left((1+\zeta) \flexo_{\text{T}}\dfrac{A L}{I\tilde{Y}}\dfrac{V}{H}\right)\right] \approx \dfrac{\tilde{Y}I}{\flexo_{\text{T}} A} \dfrac{H}{V} \left[1 -\cos \left(\flexo_{\text{T}}\dfrac{A L}{I\tilde{Y}}\dfrac{V}{H}\right)\right] .
\end{align}

\subsubsection{Buckling of a doubly-clamped flexoelectric rod under applied voltage}
\label{CC-KH-buckling}

We consider now a doubly clamped flexoelectric rod in closed circuit conditions subjected to an external electrical bias $V$, \fig \ref{fig:CC-buckling}. Since as we have seen above, an applied bias leads to an elongation of the rod, if axially constrained this should lead to buckling, and hence here we study the  critical  buckling load $V_\text{cr}$ and the post-buckling behavior. Similarly to Section \ref{OC-KH-buckling}, the kinematic constraints of vanishing vertical and horizontal displacements at the right end, give rise to a reaction force at the right end $\mathbf{N}= N_1 \mathbf{E}_1 + N_3 \mathbf{E}_3$, where now $N_1$ and $N_3$ are unknown quantities. Since the expected lowest buckling mode is symmetric, the vertical displacement at the right end vanishes by symmetry and thus $N_{1}=0$. Hence, \eq\eqref{EW} reduces to
\begin{eqnarray} \label{EW-CC-buckling}
\hat{T}=  \int^L_0 (1+\zeta)  N_3 \cos{\theta} \dd S, 
\end{eqnarray}
and its variation is
\begin{equation} \label{VarEW-CC-buckling}
\delta \hat{T} = \int^L_0 \left [  N_3 \cos{\theta}~ \delta \zeta - (1+\zeta) N_3 \sin{\theta}~ \delta \theta \right ] \dd S.
\end{equation}
Replacing \eq\eqref{VarEW-CC-buckling} in \eq\eqref{variational-principle-CC-2}, the Euler-Lagrange equations are derived as
\begin{subequations}
\begin{equation}
 \left (\tilde{Y} + \flexo_\text{T} \dfrac{V}{H}  \theta'  \right) A \zeta - \flexo_\text{T} A \dfrac{V}{H} \theta' - \dfrac{\epsilon V^{2}}{2H^2} A - N_3 \cos{\theta} =0,   \label{forcebalance-ff-buckling}
\end{equation}
\begin{equation}
 \left [\left ( \tilde{Y} + \flexo_\text{T} \dfrac{V}{H} \theta' \right ) I \theta'  - (1+\zeta) \flexo_\text{T} \dfrac{V}{H} A \right ]' - (1+\zeta) N_3 \sin{\theta} = 0,
\label{momentbalance-ff-buckling}
\end{equation}
\end{subequations}
and the constraint of vanishing horizontal displacement at the right end is 
\begin{equation} \label{constraint-CC-buckling}
\mathbf{E}_{3} \cdot \mathbf{r}(L) - L  = 
\mathbf{E}_{3}  \cdot \int^L_0 (1+\zeta) \ \mathbf{d}_3(S) \dd S -L =  \int^L_0 (1+\zeta) \cos{\theta}  \dd S - L =0.
\end{equation}
The unknown reaction force magnitude $N_{3}$ is calculated by evaluating \eq \eqref{forcebalance-ff-buckling} at the left end, with $\theta(0)=0$, as
\begin{eqnarray}\label{Act-CC-Force}
 N_{3} = \left (\tilde{Y} + \flexo_\text{T} \dfrac{V}{H} \left. \theta'\right|_0  \right)A \left.\zeta \right \vert_{0} - \flexo_\text{T} A \dfrac{V}{H}\left. \theta'\right|_0 -  \dfrac{\epsilon V^{2}}{2H^2} A.
\end{eqnarray}
Furthermore, by assuming that the axial strain and all material parameters are uniform along $S$ and neglecting the nonlinear term  $2\flexo_\text{T} \dfrac{V}{H} \theta'  I \theta''$, \eq \eqref{momentbalance-ff-buckling} reduces to
\begin{eqnarray} \label{torquebalance-CC-buckling}
\tilde{Y}I \theta'' - (1+\zeta)  N_{3} \sin{\theta} = 0. 
\end{eqnarray}
Finally, substituting $N_{3}$ from \eq\eqref{Act-CC-Force}, \eq\eqref{torquebalance-CC-buckling} simplifies to 
\begin{eqnarray}\label{momentbalance-final-cc}
 \theta'' + (1+\zeta) \ \tilde{\beta}~^2 \sin{\theta} = 0, 
\end{eqnarray}
with 
\begin{equation} \label{beta-CC-Buckling}
\tilde{\beta}~^2 = \dfrac{A}{\tilde{Y}I}\left(-\tilde{Y} \zeta + \dfrac{\epsilon V^2}{2 H^2} + \left (1-\zeta \right)  \flexo_\text{T}\dfrac{V}{H}\left.\theta'\right|_0 \right),
\end{equation}
and subject to the boundary conditions
\begin{subequations}\label{BC-CC-buckling} 
\begin{eqnarray}
\theta(0) &=& 0, \label{BC-CC-buckling-1}\\
\theta (L) &=& 0. \label{BC-CC-buckling-2}
\end{eqnarray}
\end{subequations}
Similarly to the problem in Section \ref{OC-KH-buckling}, the expected lowest mode exhibits inflection points at $S = L/4$ and $S = 3L/4$. To avoid having to deal with them, we consider only a quarter of the rod and replace \eq\eqref{BC-CC-buckling-2} with  
\begin{eqnarray}
\theta'\left(\frac{L}{4}\right) &=& 0, \label{BC-CC-buckling-L/4}
\end{eqnarray}
After solution of the above BVP, see \ref{appendix:buckling_CC} for a detailed derivation, the vertical displacement at the center of the rod is obtained as
\begin{eqnarray}\label{r1-CC-postbuckling}
r_1\left(\dfrac{L}{2}\right) = - \dfrac{4 \sqrt{1+\zeta}}{\tilde{\beta}} \sin{\dfrac{\theta^\text{max}}{2}},
\end{eqnarray}
where $\theta^\text{max} = \theta(L/4)$, and $\tilde{\beta}$ and $\zeta$  are computed from \eqs \eqref{eqn:buckling:CC:CC:UL} and \eqref{eqn:strain:CC} in terms of $\theta^{\text{max}}$. The curvature at the left end is
\begin{eqnarray}\label{flexuralstrain0}
 \theta' (0) = \tilde{\beta} \sqrt{2 (1+\zeta)(1-\cos{\theta^\text{max}})}.
\end{eqnarray}
Finally, using \eq \eqref{beta-CC-Buckling} and \eqref{flexuralstrain0}, the postbuckling voltage can be obtained as 
\begin{multline}\label{V-CC-postbuckling}
 V=\dfrac{H \sqrt{2\bar{Y}/\epsilon}}{ (1-\zeta )-\tilde{\beta} ^2 I/A} \Bigg(\sqrt{\left( \tilde{\beta}^2 (\zeta +1) (\zeta -1)^2 \ell_\flexo^2 (1-\cos {\theta^\text{max}})- \left( (\zeta -1)+\tilde{\beta} ^2 I/A\right) \left( \zeta +\tilde{\beta} ^2 I/A\right)\right)} 
 \\
 + (\zeta -1) \tilde{\beta}  \ell_\flexo \sqrt{(\zeta +1) (1-\cos {\theta^\text{max}})}\Bigg)
 \end{multline}
 The critical buckling voltage is determined from \eq\eqref{V-CC-postbuckling} in the limit $\theta^{\text{max}} \rightarrow 0$ as 
\begin{eqnarray}\label{criticalV}
V_\text{cr} = \dfrac{2 \pi  H    }{L} \sqrt{\dfrac{2\bar{Y}}{\epsilon \left(\dfrac{A}{I}-\dfrac{4 \pi ^2}{L^2}\right)}}.
\end{eqnarray}
The critical electric field for a rectangular/square cross section becomes:
\begin{eqnarray}\label{criticalE}
E_\text{cr} = \dfrac{2 \pi }{L} \sqrt{\dfrac{2\bar{Y}}{\epsilon \left(\dfrac{12}{H^2}-\dfrac{4 \pi ^2}{L^2}\right)}} = \left(\dfrac{H}{L}\right) \sqrt{\dfrac{2\bar{Y}}{\epsilon \left(\dfrac{3}{\pi^2}-\left(\dfrac{H}{L}\right)^2\right)}},
\end{eqnarray}
 and for slender rods, the Taylor approximation around $H/L\rightarrow 0$ provides
\begin{eqnarray}\label{criticalE_Taylor}
E_\text{cr} \approx  \left(\dfrac{H}{L}\right) \pi  \sqrt{\dfrac{2\bar{Y}}{3\epsilon }}.
\end{eqnarray}

\section{Numerical examples for general nonlinear flexoelectric rod problems}\label{sec_05}
In this Section, we present numerical results of our general nonlinear model of flexoelectricity for bending and buckling of flexoelectric rods, both in open-circuit and in closed-circuit conditions. We compare these results with the solutions of the 1D nonlinear analytical model for rods developed in Section~\ref{sec_04} and its linearized Euler-Bernoulli (E-B) counterpart, by considering material parameters to match the assumptions of these models. This comparison allows us to validate our computational approach. We then explore more general flexoelectric problems and establish the limits of the simplified 1D flexoelectric rod models.

To model standard elasticity, we consider isotropic hyperelastic potentials, either Saint-Venant– Kirchhoff (\eq\eqref{sv}) or Neo-Hookean (\eq\eqref{nh}) models, requiring two elastic constants, here  Young's modulus $Y$ and  Poisson's ratio $\nu$. Strain-gradient elasticity is modeled by the analogous isotropic hyperelastic Saint-Venant–Kirchhoff law (\eq\eqref{strgrtensor}), which additionally depends on the characteristic length scale $\ell$.
The flexoelectric tensor $\Flexo$ is assumed to have cubic symmetry with three independent constants $\flexo_\text{L}$, $\flexo_\text{T}$ and $\flexo_\text{S}$, namely the longitudinal, transversal and shear coefficients (\eq\eqref{flexotensor}). Isotropic flexoelectricity tensor is just a particular case with only two independent parameters, with $2\flexo_\text{S}=\flexo_\text{L}-\flexo_\text{T}$.

The dielectric strength (\ie maximum electric field magnitude that a dielectric can sustain before electric breakdown occurs) is typically around $1-100\si{\volt\per{\micro\metre}}$ \citep{Liu2014}. Here, for simplicity, electrical breakdown is neglected, \ie we assume an infinite dielectric strength in all the examples. In all simulations, we consider a cubic ($p=3$) spline mesh with square cells of size $h=H/10$, being $H$ the thickness of the rod. 

\subsection{Bending of open-circuit flexoelectric cantilever under a vertical point load}
\label{Sec:OCBending}

We consider here a flexoelectric cantilever rod under bending by a vertical point load in an open circuit configuration with the mechanically free end electrically grounded, \cf Fig.~\ref{fig:OC-bending}. Young's modulus is chosen as $Y = 1.725 \si{\giga\pascal}$ and the dielectric permittivity as $\epsilon = 0.092 \si{\nano\joule / \square\volt\meter}$, which correspond to polyvinylidene fluoride (PVDF) \citep{chu2012,zhang2016experimental,zhou2017flexoelectric}.

\subsubsection{Validation}
\label{Sec:ValidationBending}

\begin{figure}[p!]\vspace{-5em}\centering

\begin{subfigure}[b]{0.75\textwidth}\centering\vspace{-1em}
\includegraphics[width=0.93\textwidth]{mu14.pdf}
\caption{\label{fig:deformedshapebending}Deformed shape and electric potential [\si{\volt}] distribution of cantilever rods of slenderness $L/H=20$ under a point load of {\SI{20}{\nano\newton}}, for different transversal flexoelectric coefficients $\flexo_\text{T}$.}
\end{subfigure}

\begin{subfigure}[b]{0.75\textwidth}\centering
\centering
 \includegraphics[width=0.46\textwidth]{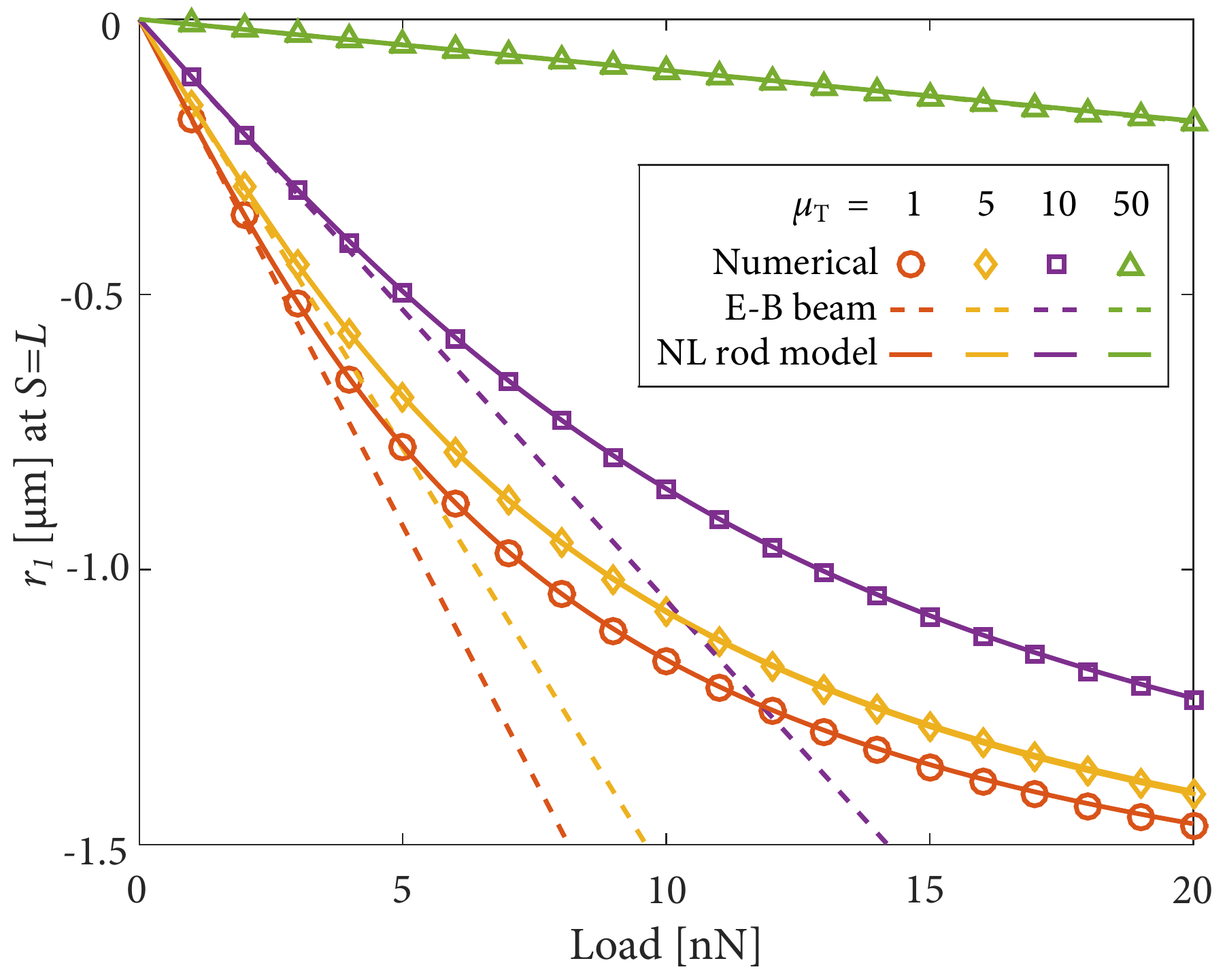}
\quad
\includegraphics[width=0.46\textwidth]{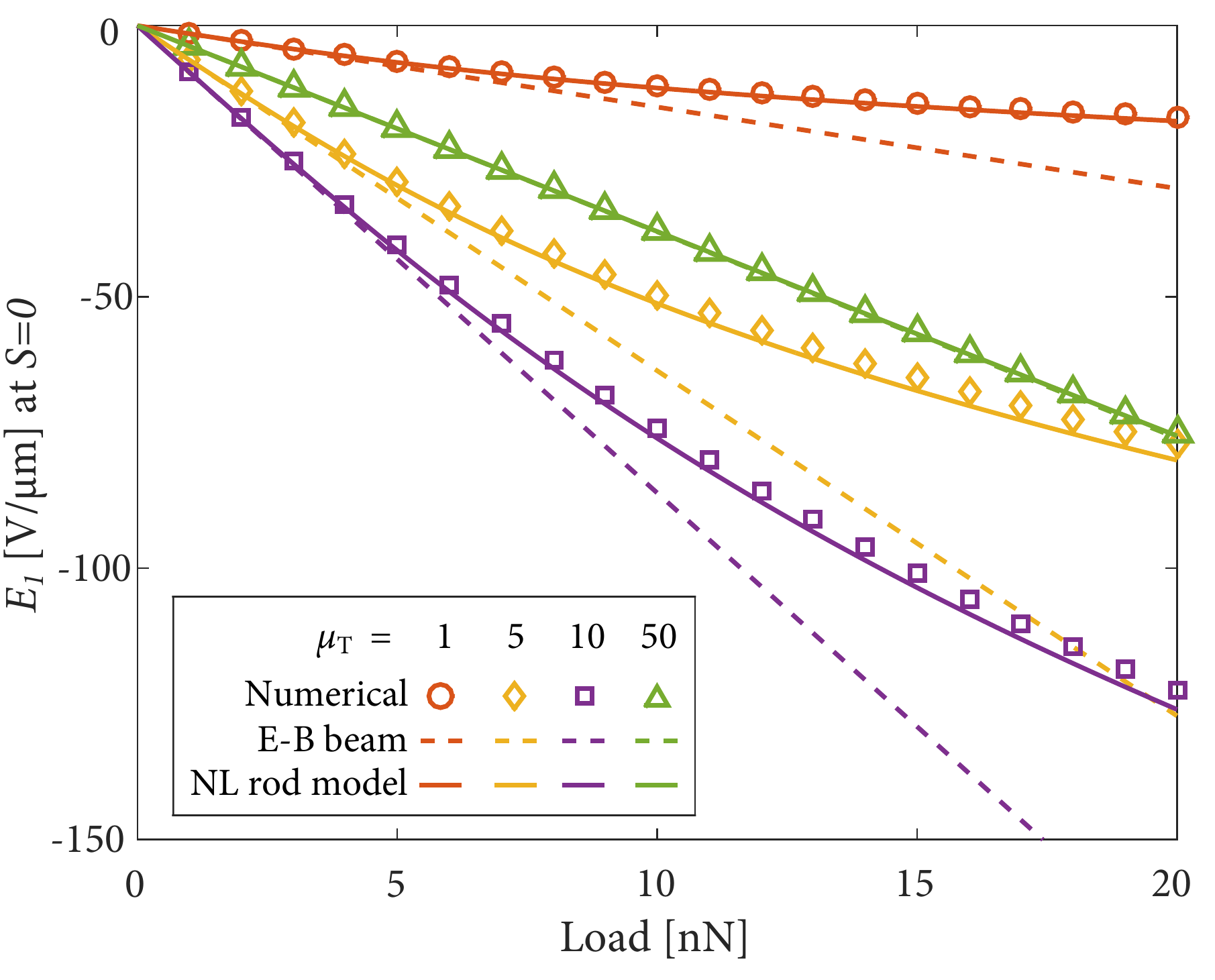}
\caption{\label{fig:flexo_rod}Bending of a cantilever rod of slenderness $L/H=20$ with varying transversal flexoelectric coefficient $\flexo_\text{T}$. The left plot shows the vertical displacement at the loaded end, and the right one shows the vertical electric field at the fixed end.}
\end{subfigure}

\begin{subfigure}[h]{0.75\textwidth}\centering
\includegraphics[width=.45\textwidth]{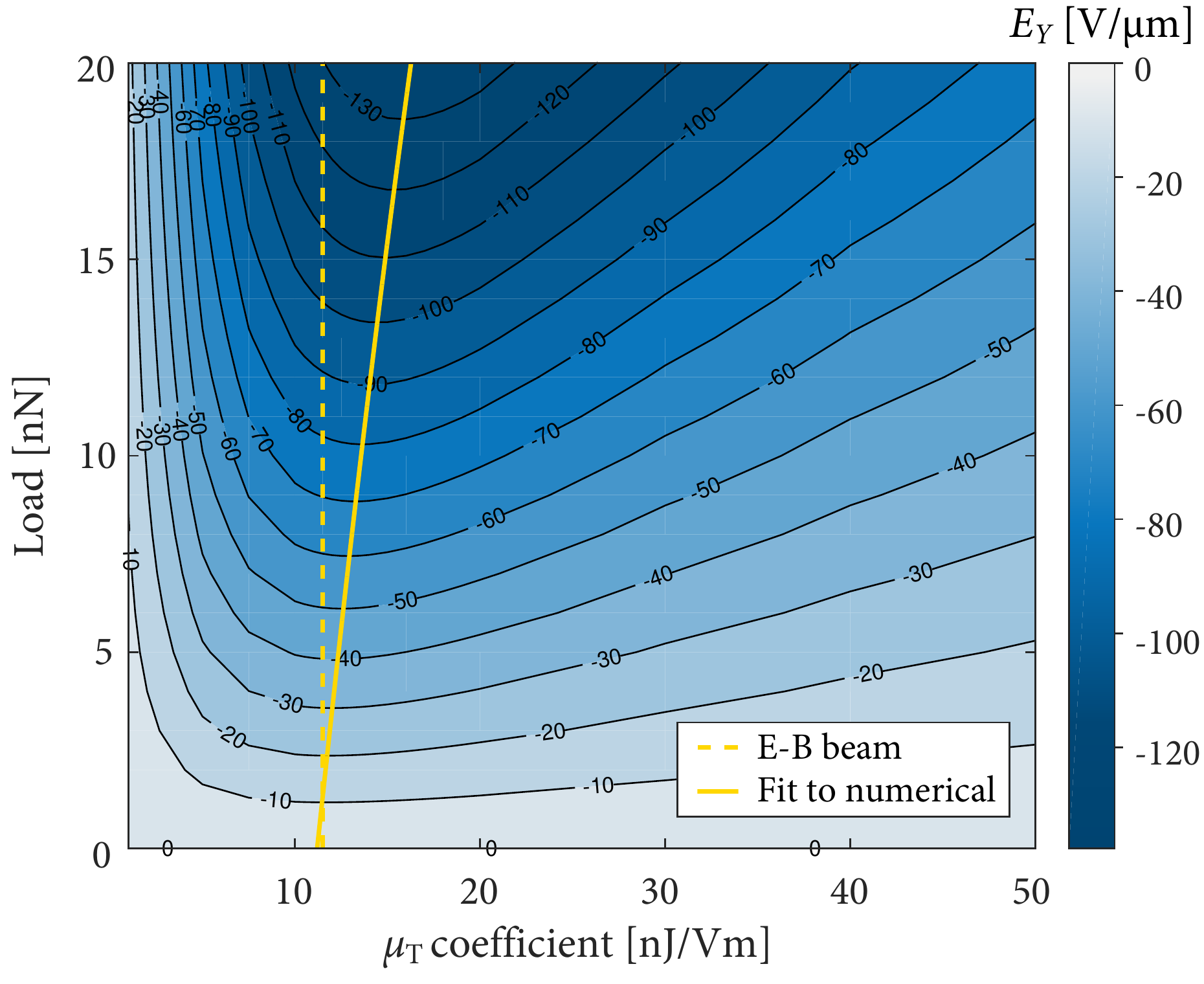}
\caption{\label{fig:flexo_rod3} Countour plot the vertical electric field $E_1$ at the fixed end of the rod of $L/H=20$ as a function of the applied load and the transversal flexoelectric coefficient $\flexo_\text{T}$.}
\end{subfigure}

\begin{subfigure}[h]{0.75\textwidth}\centering
\includegraphics[width=.42\textwidth]{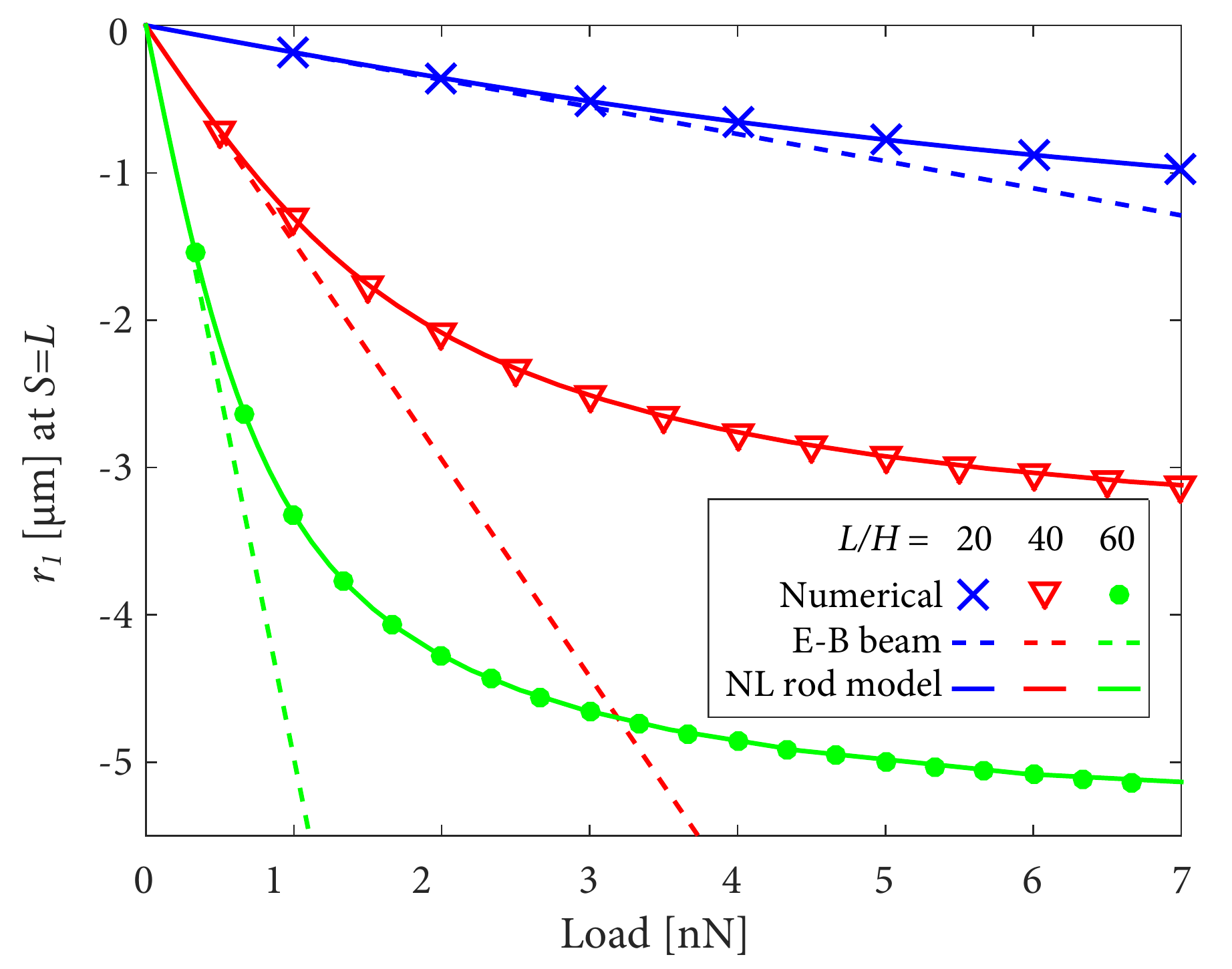}
\quad
\includegraphics[width=.45\textwidth]{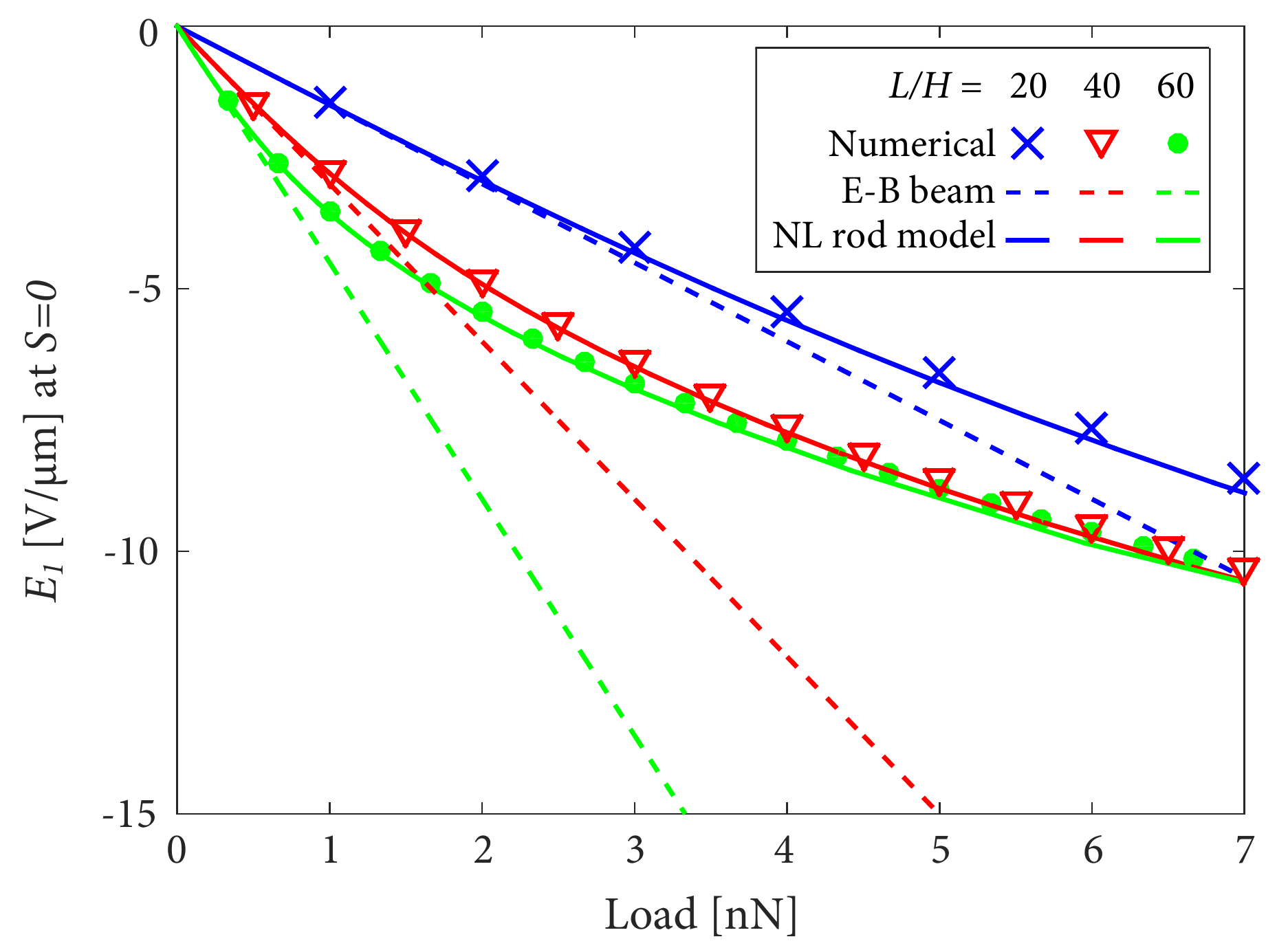}
\caption{\label{fig:slenderness}Bending of a cantilever rod of { $\flexo_\text{T}=1\si{\nano\joule \per{\volt\metre}}$} with varying slenderness. The left panel shows the vertical displacement at the loaded end, and the right one shows the vertical electric field at the fixed end.}
\end{subfigure}

\caption{\label{OC-bending-validation}Validation results for bending of open-circuit flexoelectric cantilever in sensor mode. The transversal flexoelectric coefficient $\flexo_\text{T}$ in the legends is expressed in
    \si{\nano\joule \per{\volt\metre}} = \si{\nano\coulomb \per{\metre}}.
}
\end{figure}

We first validate the full computational model in Section \ref{sec_02} and \ref{sec_03} against the 1D nonlinear model for flexoelectric rods presented in Section \ref{OC-KH-bending}, and its linearized Euler-Bernoulli counterpart. For this, we choose a Saint-Venant–Kirchhoff mechanical constitutive law {with $\nu=0$ and} material parameters consistent with the assumptions of the 1D reduced model, namely $\flexo_\text{L} = \flexo_\text{S} = 0$, $\ell=0$. {We consider a thickness $H=100\si{\nano\meter}$ and a slenderness of $L/H \geq 20$}. \fig\ref{OC-bending-validation} collects all the validation results. Typical computational solutions are shown in \fig\ref{fig:deformedshapebending}, where the electric potential $\phi$ is plotted on the deformed configuration. These simulations highlight the very large deformations attained. In this figure, we show  numerical calculations for a given force at the tip, and for several values of $\flexo_\text{T}$. As in the linear case \citep{Majdoub2008,Majdoub2009} and as expected by the expression $I^{\rm eff}$ in the reduced theory, \cf \eq\eqref{Ieff}, flexoelectricity leads to an effective stiffening of the system even though the elastic constants are kept fixed. As anticipated in Section \ref{OC-KH-bending} for the linearized Euler-Bernoulli beam, \cf \eq\eqref{muTstar}, we find that also for the non-linear rod the maximum electric field generated at the clamping cross-section exhibits a maximum for an intermediate value of the flexoelectric constant. The existence of an optimal value of $\flexo_\text{T}$, for which the flexoelectric response is maximized results from the competition of the two conflicting effects of $\flexo_\text{T}$: (1) the stiffening and (2) the flexoelectric coupling. For small values of $\flexo_\text{T}$  the structure is very compliant and larger strain gradients are attained but the generated field is small due to the small coupling, whereas for very large values of $\flexo_\text{T}$ the flexoelectric coupling is large but the stiffer beam attains smaller deformations and thus smaller strain gradients. 

To further analyze these effects, we present in \fig\ref{fig:flexo_rod} the dependence of the cantilever rod vertical displacement at the tip on the endpoint load, and the vertical electric field on the clamped edge, for different values of transversal flexoelectric coefficient $\flexo_\text{T}$. The results for the tip displacement show \emph{i)} the stiffening as $\flexo_\text{T}$ increases, \emph{ii)} the nonlinearity in the response of the system (particularly for the most deformable systems), \emph{iii)} an excellent quantitative agreement with the nonlinear flexoelectric rod model given by the analytical expression in \eq\eqref{tip_deflection}, and \emph{iv)} an agreement with the linearized E-B  model for small deformations, \ie smaller loads or stiffer cantilevers (large values of $\flexo_\text{T}$). 
Similarly, we find an excellent agreement between the numerical simulations and the nonlinear rod model in the vertical electric field on the clamped end. Its behavior is nonlinear for large loads since the electric field is directly proportional to the curvature, \cf\eq\eqref{rod-OC-EF}. The non-monotonicity in the maximum electric field as a function of $\mu_\text{T}$ discussed above is apparent from this plot. To further examine this point, we represent in \fig\ref{fig:flexo_rod3} a contour plot showing the dependence of the vertical electric field at the clamped cross-section on $\mu_\text{T}$ and on the load. We find that the load for maximum electrical output depends on the value of the flexoelectric coupling in the nonlinear model, whereas it is independent of it according to the linearized E-B model, see \eq\eqref{E_max-1}.  

Finally, we examine the effect of the slenderness on the load vs.~deflection and the load vs.~electric field curves for a given $\mu_\text{T}$, see \fig\ref{fig:slenderness}. As the slenderness ratio increases, the rod becomes more flexible and therefore nonlinearity is stronger and manifests for smaller loads, with a larger overestimation of the vertical displacement by the linear E-B model. In contrast, the nonlinear 1D rod model closely follows our simulations even deep into the nonlinear regime.

\subsubsection{General flexoelectric problem}

\begin{figure}[tb!]
\centering
\begin{subfigure}[b]{0.45\textwidth}
\includegraphics[width=0.8\textwidth]{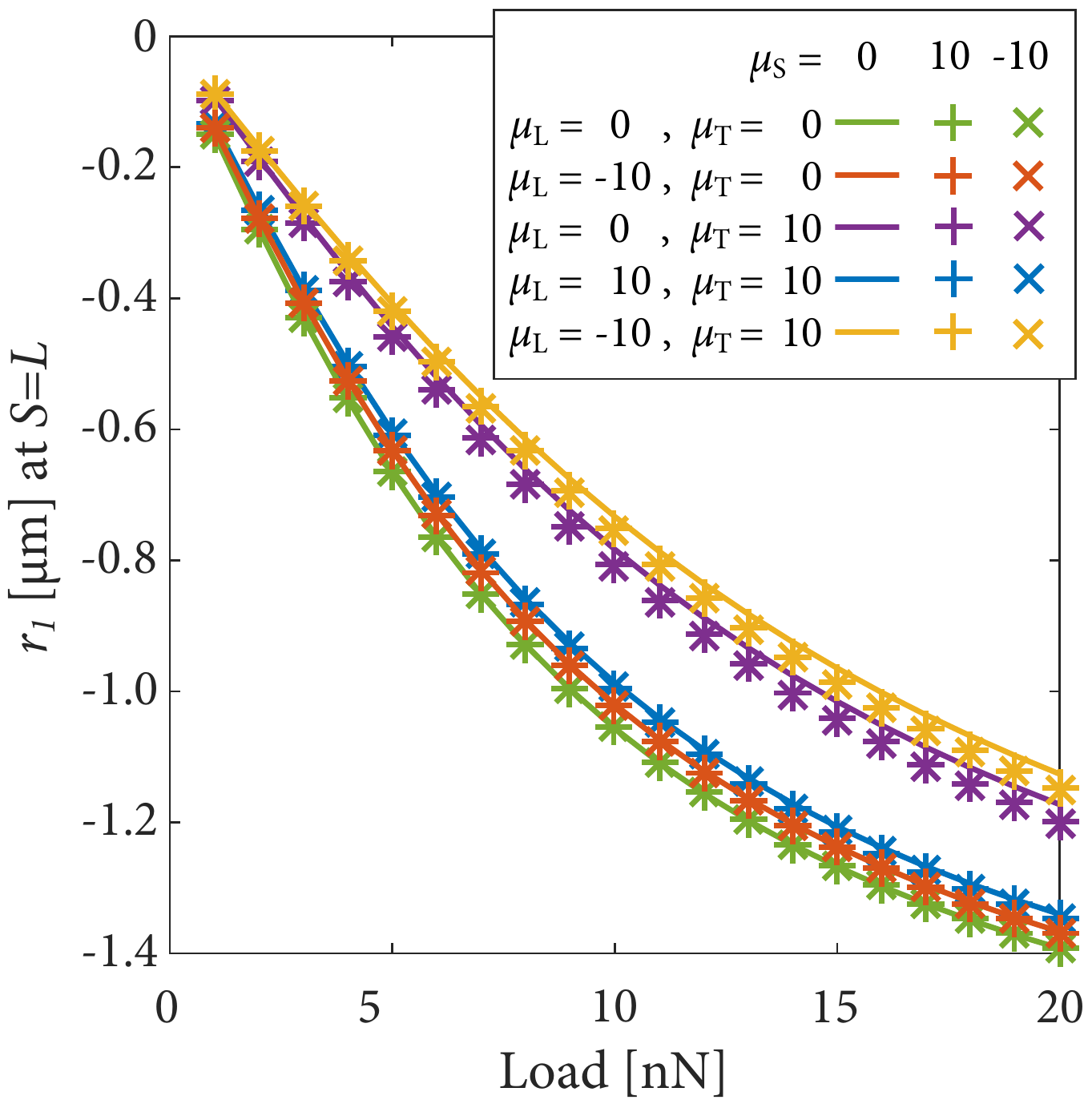}
\caption{\label{fig:flexo_rodaNH}Vertical displacement at the loaded end.}
\end{subfigure}%
\begin{subfigure}[b]{0.45\textwidth}
\includegraphics[width=0.8\textwidth]{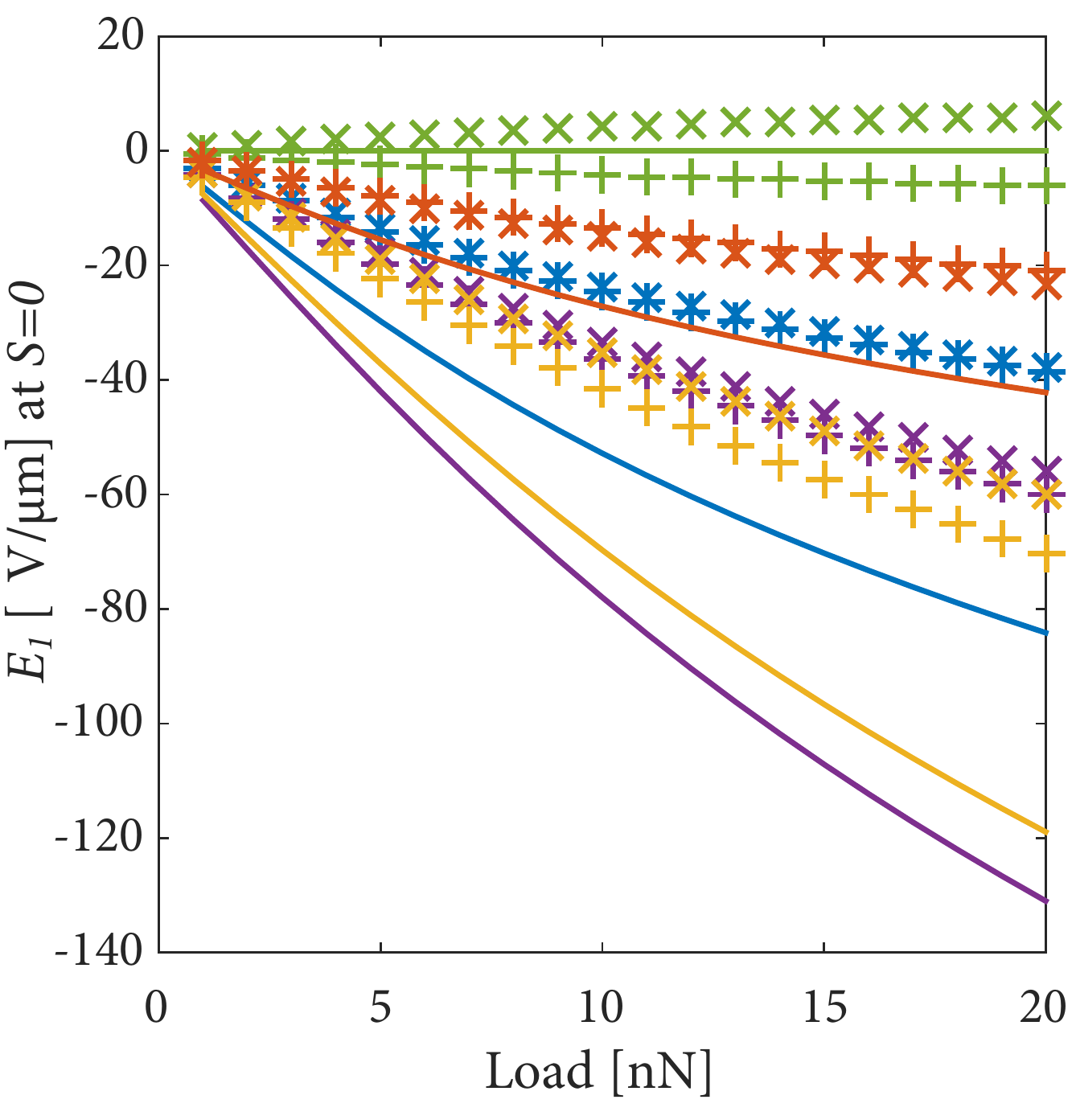}
\caption{\label{fig:flexo_rodbNH}Vertical electric field at the fixed end.}
\end{subfigure}
\caption{\label{fig:flexo_rodNH}Electromechanical response of Neo-Hookean cantilever flexoelectric sensor under bending, with different flexoelectric tensors (expressed in \si{\nano\joule\per{\volt\metre}}). }
\end{figure}

We investigate now more general flexoelectric conditions beyond the restrictive assumptions of the reduced model in Section \ref{OC-KH-bending}. We consider an
    $L=2\si{\micro\meter}$ by $H=100\si{\nano\meter}$
   
isotropic Neo-Hookean hyperelastic rod, \cf\eq\eqref{nh}, augmented with strain gradient elasticity, with $\nu = 0.3$, $\ell = 0.1$ \si{\micro\meter} and varying flexoelectric constants.

\fig\ref{fig:flexo_rodNH} represents the electromechanical response of the open circuit cantilever rod under point load for varying flexoelectric constants 
    $\flexo_\text{L},\flexo_\text{T},\flexo_\text{S}=\{ -10,0,10 \} \si{\nano\joule \per{\volt\metre}}.$
\fig\ref{fig:flexo_rodaNH} shows the deflection $r_1$ of the loaded end, whereas \fig\ref{fig:flexo_rodbNH} shows the vertical electric field $E_1$ at the clamped end.
For the sake of brevity, some combinations of flexoelectric tensors are omitted, since we found that the responses are analogous to the ones of other combinations as follows:
\begin{subequations}\begin{gather}
r_1|_{\Flexo}=r_1|_{-\Flexo};\\
E_1|_{\Flexo}=-E_1|_{-\Flexo}.
\end{gather}\end{subequations}

From \fig\ref{fig:flexo_rodaNH}, it is clear that flexoelectricity is always increasing the bending stiffness of the rod. 
The largest stiffening is found with opposite $\flexo_\text{T}$ and $\flexo_\text{L}$, followed by the case of vanishing $\flexo_\text{L}$. On the contrary, the simulations with $\flexo_\text{L}\sim\flexo_\text{T}$ and the ones with vanishing $\flexo_\text{T}$ present a smaller stiffening.
In all cases, the effect of the shear flexoelectric coefficient $\flexo_\text{S}$ on bending stiffness is much smaller, and therefore less relevant.

\fig\ref{fig:flexo_rodbNH} shows the electric response of the rod at the clamped tip, revealing that all three flexoelectric coefficients are relevant here. Within the studied range, a larger flexoelectricity-induced bending stiffness leads also to a larger electric field. However, in addition, the shear flexoelectric effect $\flexo_\text{S}$ has a large influence on the electric field. In most cases, a non-vanishing $\flexo_\text{S}$ leads to a substantial decrease in the reported electric field, which slightly depends also on the sign of $\flexo_\text{S}$. The only case in which a non-vanishing $\flexo_\text{S}$ increases the electric field is the one where $\flexo_\text{S}$ is the \emph{only} non-vanishing flexoelectric coefficient.

\subsection{Buckling of open-circuit flexoelectric rod under mechanical load}
\label{Sec:ValidationBuckling}

\begin{figure}[p]\vspace{-3em}\centering
\minipage{0.42\textwidth}
  \includegraphics[width=\textwidth]{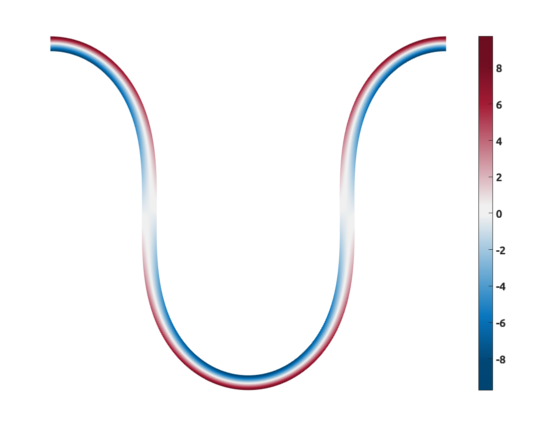}\subcaption{\label{buck-d}Buckled rod geometry and resulting electric potential [\si{\volt }].}
\endminipage\qquad
\minipage{0.42\textwidth}
\includegraphics[width=\textwidth]{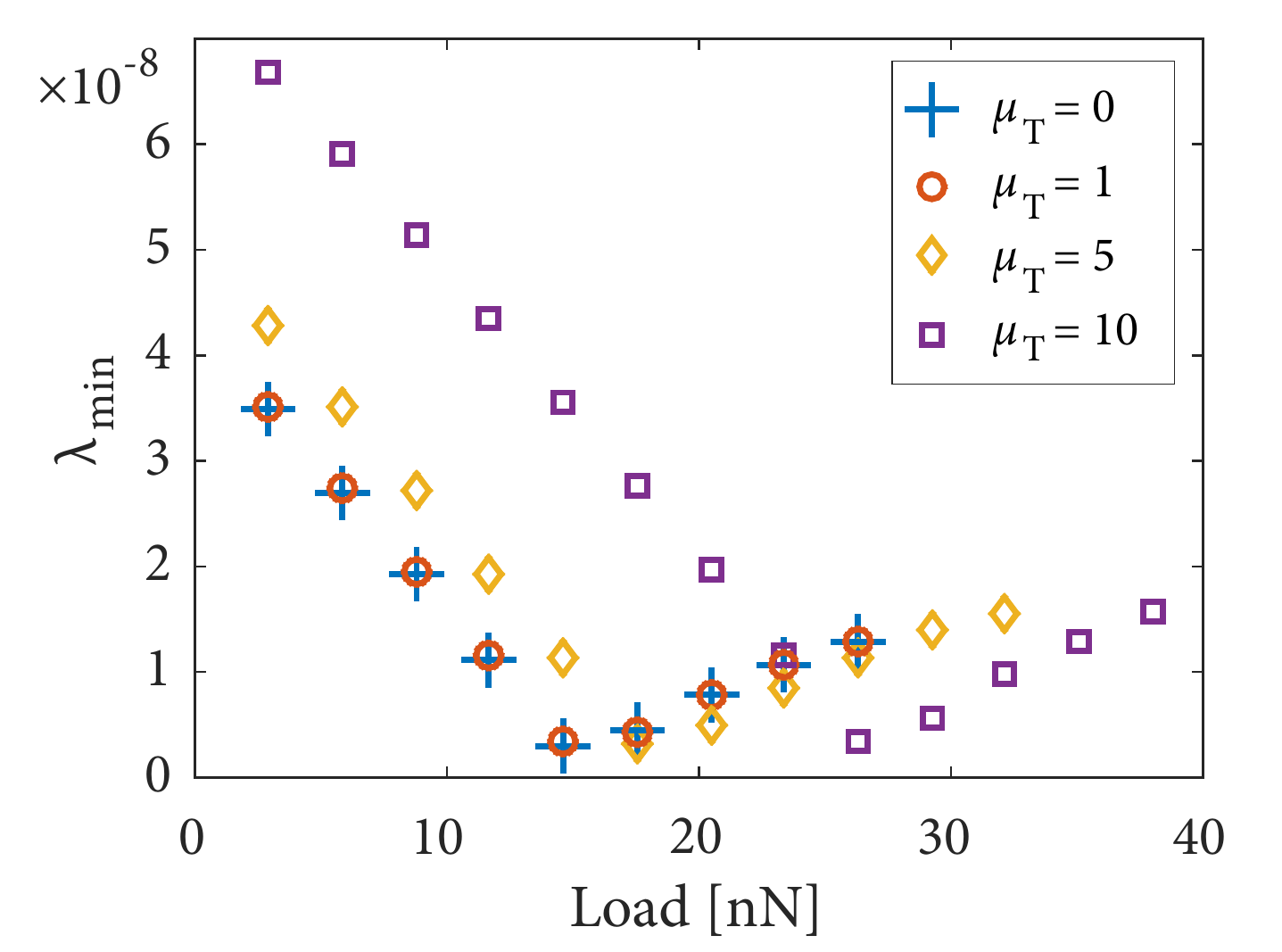}\subcaption{\label{buck-l}Minimum eigenvalue
$\lambda_\text{min}\left[\left.\toMat{\widehat{H}}_{\boldsymbol{\chi}\boldsymbol{\chi}}\right.^{\!(k)}\right]$.}
\endminipage\qquad
  \vspace{1em}
 \\
\minipage{0.42\textwidth}
  \includegraphics[width=\textwidth]{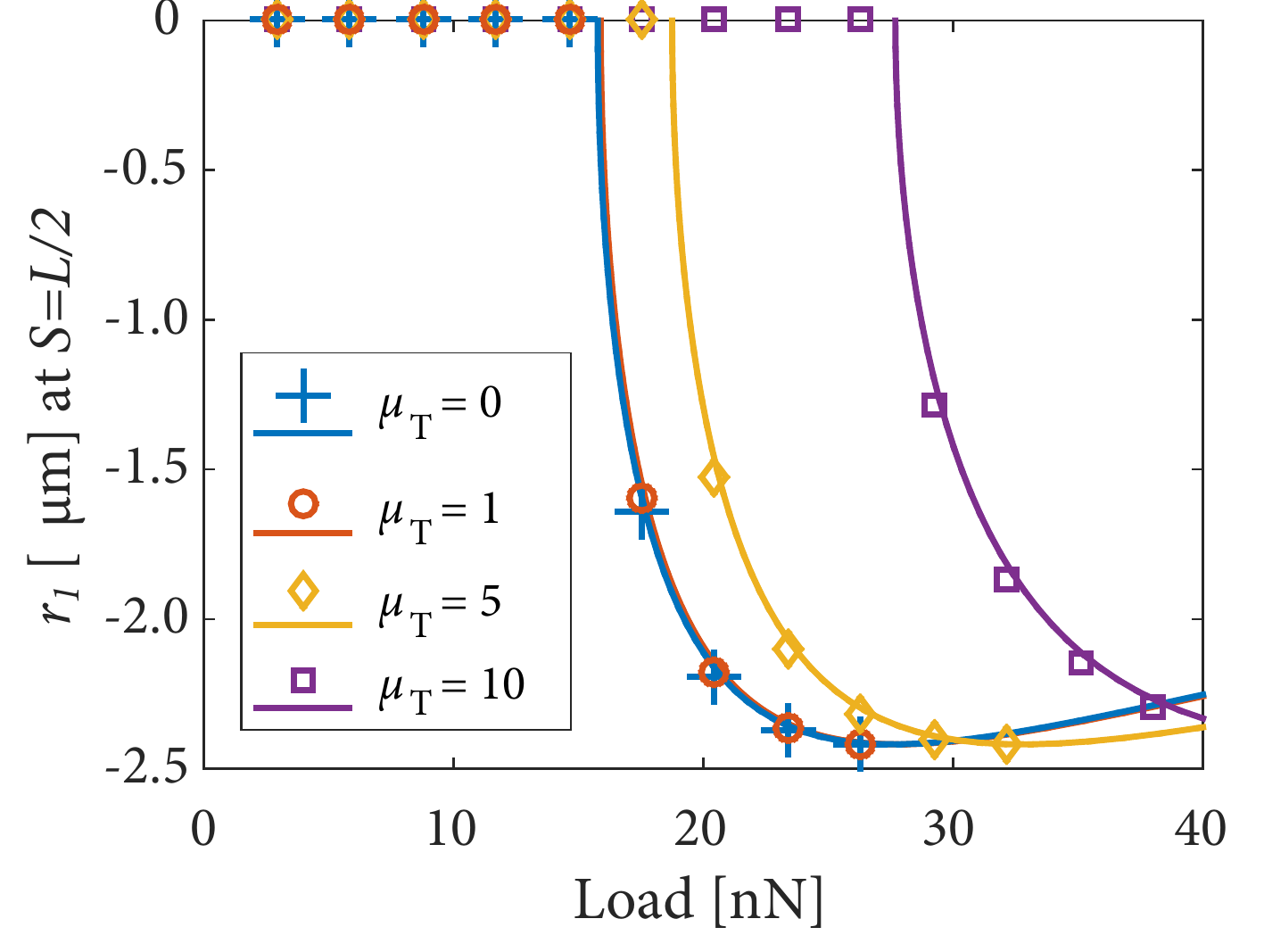}\subcaption{\label{buck-y}Vertical displacement at the middle cross-section of the rod.}
\endminipage\qquad
\minipage{0.42\textwidth}
 \includegraphics[width=\textwidth]{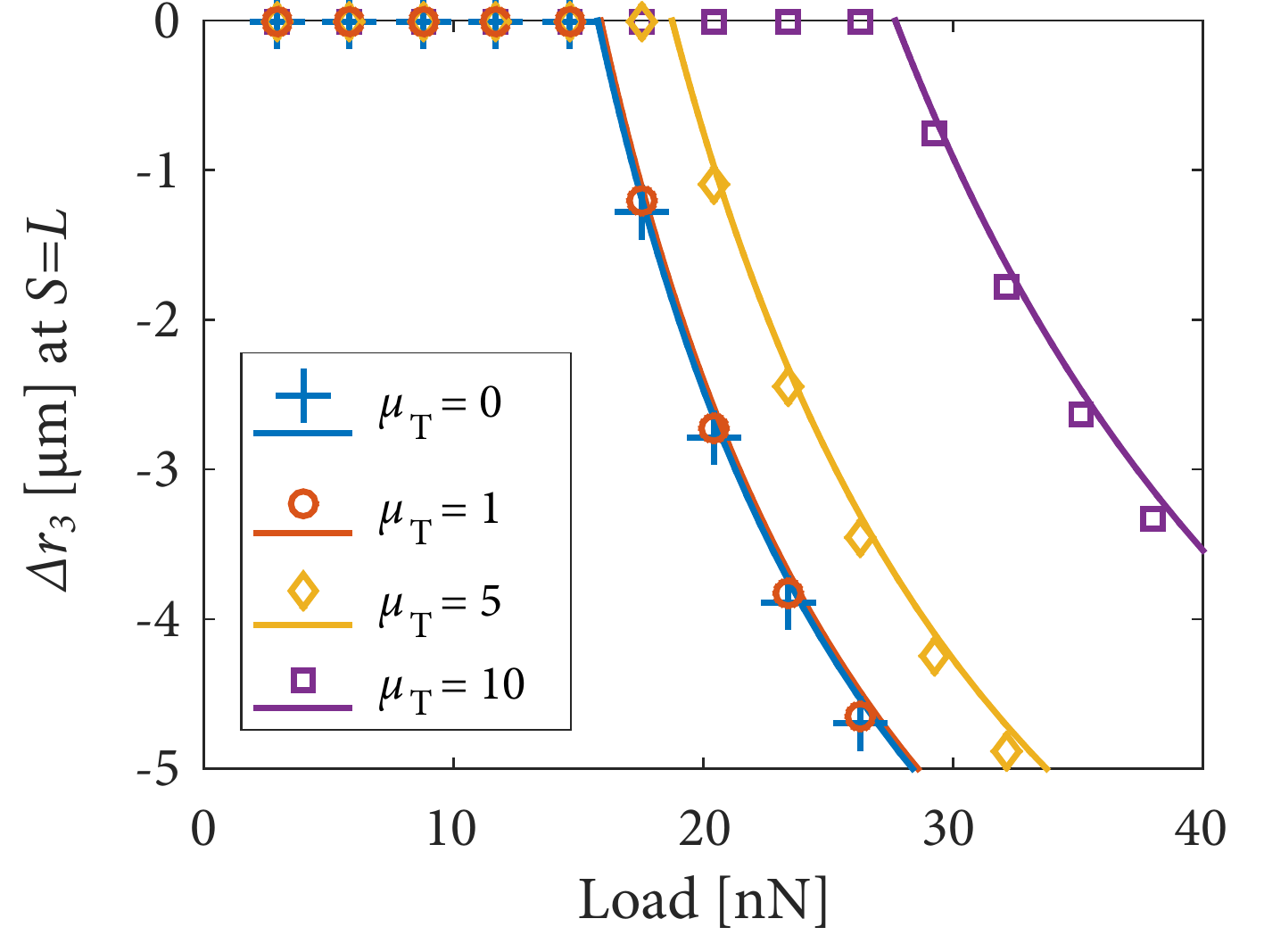}\subcaption{\label{buck-x}Horizontal displacement at the right end of the rod.}
\endminipage\qquad
  \vspace{1em}
 \\
\minipage{0.42\textwidth}
 \includegraphics[width=\textwidth]{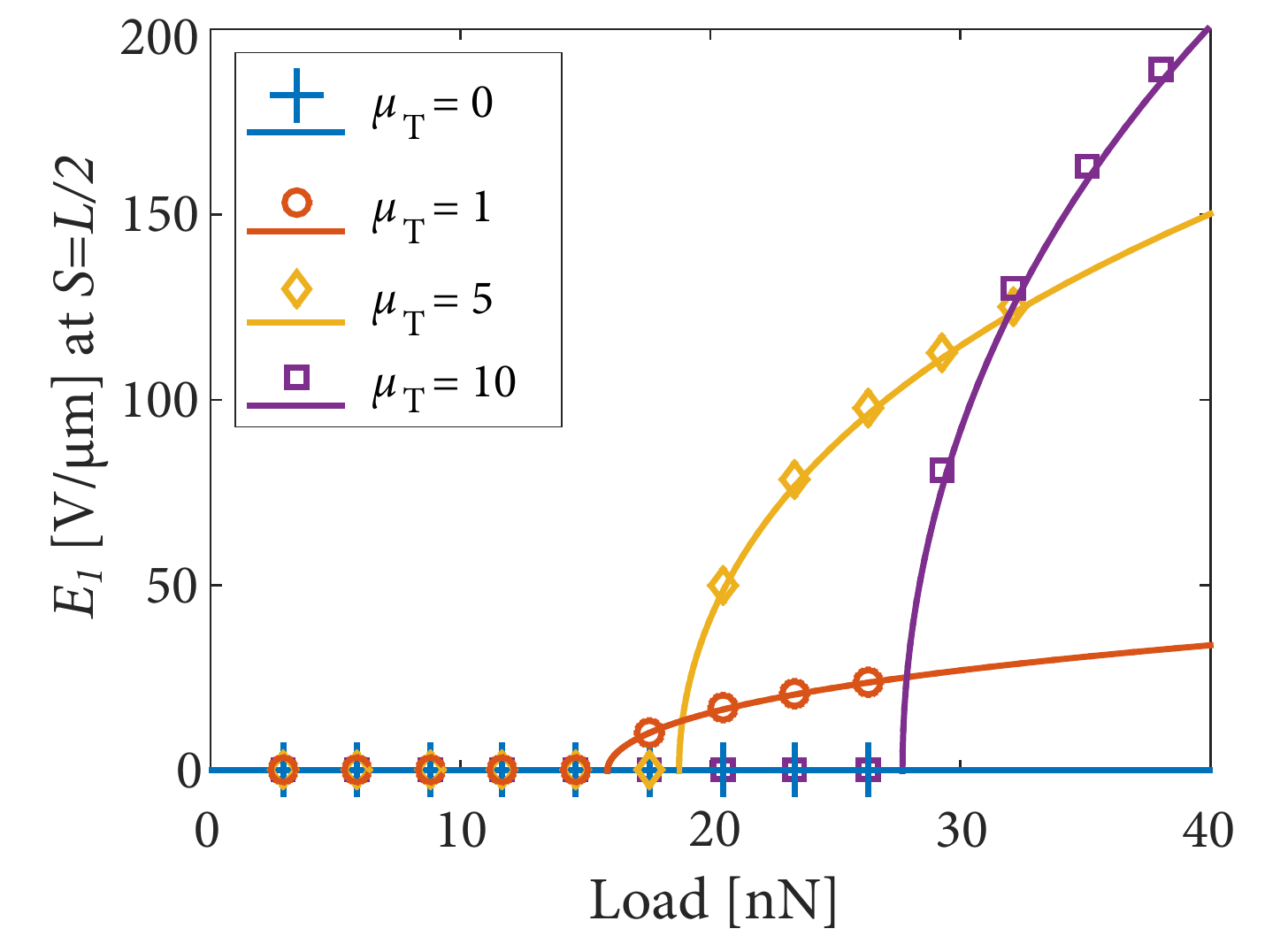}\subcaption{\label{buck-e}Vertical electric field at the middle cross-section of the rod.}\vfill
\endminipage\qquad
\minipage{0.418\textwidth}\hspace{-1.5em}
\includegraphics[width=\textwidth]{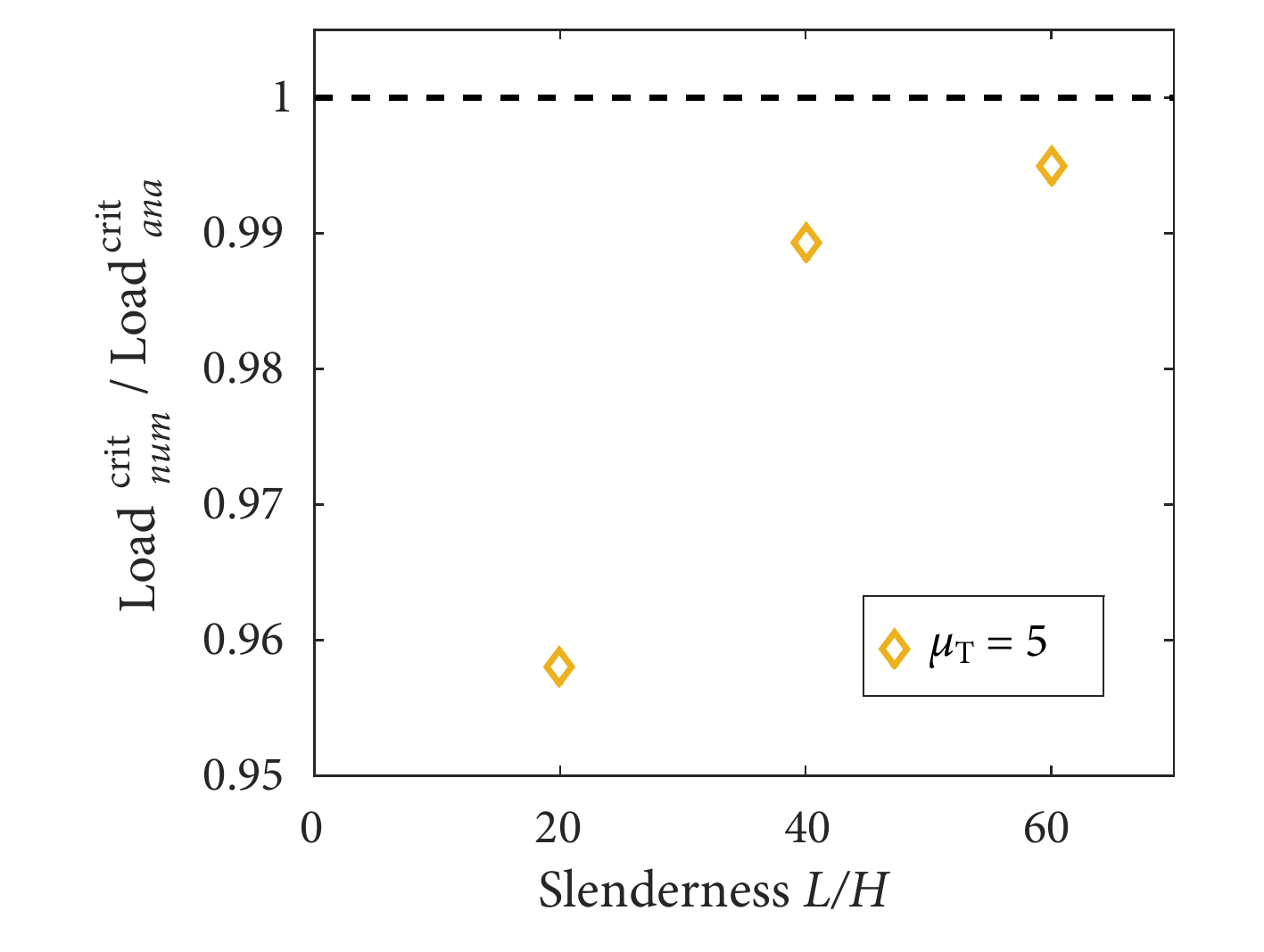}\subcaption{\label{buck-s}Convergence of the buckling critical load with respect to $L/H$ ratio.}
\endminipage
  \caption{\label{buck}Force-controlled buckling of a flexoelectric rod of $L/H=60$. Markers refer to the numerical implementation and solid lines refer to the analytical nonlinear model for rods. The transversal flexoelectric coefficient $\flexo_\text{T}$ is expressed in  \si{\nano\joule \per{\volt\metre}}.}
\end{figure}

We now compress a slender flexoelectric rod
    ($L=\SI{6}{\micro\metre}$, $H=\SI{100}{\nano\metre}$ )
in open-circuit until buckling occurs, and also during the post-buckling stage. The left tip is clamped and a uniform horizontal load is applied on the right cross-section, which can only move uniformly in axial direction, \ie vertical displacement and rotation of the right end are prevented (see \fig\ref{fig:OC-buckling}). We consider an isotropic Saint-Venant–Kirchhoff model with Young's modulus $Y = 1.725 \si{\giga\pascal}$, dielectric permittivity
    $\epsilon = 0.092 \si{\nano\joule / \square\volt\meter}$
and different transversal flexoelectric coefficients:
    $\flexo_\text{T}=\{0,1,5,10\}~\si{\nano\joule \per{\volt\metre}}$.
The other material parameters are set to zero ($\nu=\flexo_\text{L}=\flexo_\text{S}=\ell$ =0 ).

As shown in \fig\ref{buck}, the numerical simulations and the analytical 1D model agree remarkably well.
The highly nonlinear nature of the electromechanical system is clear in the responses reported in the post-buckling regime.
Before buckling, the system is uniformly compressed and the flexoelectric effect is not present yet since the rod is not bent, and hence the electric response is zero. Once the rod has buckled (see \fig\ref{buck-d}), the vertical displacement at $s=L/2$ (\fig\ref{buck-y}) and the horizontal displacement at $s=L$ (\fig\ref{buck-x}) suddenly deviate from zero and evolve nonlinearly with respect to the applied load. The flexoelectric effect arises due to the curvature induced by buckling, leading to a measurable electric field at $s=L/2$, which also evolves nonlinearly with applied load (\fig\ref{buck-e}).

The role of the magnitude of the flexoelectric coefficient $\flexo_\text{T}$ is  twofold. On the one hand, the critical buckling load becomes larger with a larger $\flexo_\text{T}$ coefficient, as suggested by the nonlinear rod model, \cf \eq\eqref{criticalbucklingload}, for an effectively stiffer structure.
Numerically, the precise value of the critical buckling load is identified by the load at which the eigenvalue $\lambda_\text{min}\left[\left.\toMat{\widehat{H}}_{\boldsymbol{\chi}\boldsymbol{\chi}}\right.^{\!(k)}\right]$ vanishes, as reported in \fig\ref{buck-l}.
On the other hand, the electric field at the post-buckling stage grows faster with a larger $\flexo_\text{T}$ coefficient, which is also predicted by the nonlinear rod model, \cf \eq\eqref{buckreb}. Thus, the buckling-induced flexoelectric response is delayed but stronger when $\flexo_\text{T}$ is larger.

We expect the agreement of the simplified rod model and the computational model to deteriorate for thicker rods, and thus the assumptions of the rod model loose validity. In \fig\ref{buck-s} we show the effect of the finite thickness of the rod on the buckling critical load by plotting the value predicted by the computational model normalized by that estimated by the nonlinear rod model for different values of slenderness $L/H$. For all $L/H$ values, the 1D nonlinear rod model overestimates the buckling load, as it provides a more constrained model. As expected, \fig\ref{buck-s} shows that the buckling critical load computed with the 2D computational model converges towards the approximated value given by the 1D nonlinear rod model as the slenderness $L/H$ increases and thus the 1D assumption is approached.

\subsection{Bending of closed-circuit flexoelectric cantilever under electric actuation}\label{actuator}
\label{sec:CC-bending-actuator}

We now consider a closed-circuit flexoelectric cantilever rod with Young's modulus $Y = 1.0 \si{\giga\pascal}$, dielectric permittivity  $\epsilon = 0.11 \si{\nano\joule / \square\volt\meter}$, and dimensions $L=20\si{\micro\meter}$, $H=1\si{\micro\meter}$, which rolls up into a circle upon electrical stimulus. The geometry and boundary conditions are depicted in \fig\ref{fig:CC-bending}. The left tip cross-section of the rod is clamped, while all other boundaries are traction-free. The electric potential at the top boundary is set to a certain non-zero value $\phi=V$, and the bottom boundary is grounded ($\phi=0$). The voltage difference $\Delta \phi =V$ induces a transverse electric field across the rod thickness, \cf \eq\eqref{E-field-CC}, which triggers the flexoelectric and electrostrictive effects, thereby generating a non-uniform strain that bends the rod, as shown in \fig\ref{roll-up}. Depending on the sign of the applied electric field the cantilever will bend upwards or downwards. This bending actuator was first used by \cite{bursian1968} to experimentally demonstrate for the first time the flexoelectric effect, which had been predicted theoretically by \cite{Mashkevich1957}.

\subsubsection{Validation}

\begin{figure}[!b]\centering
\minipage{0.33\textwidth}
\includegraphics[scale=.62]{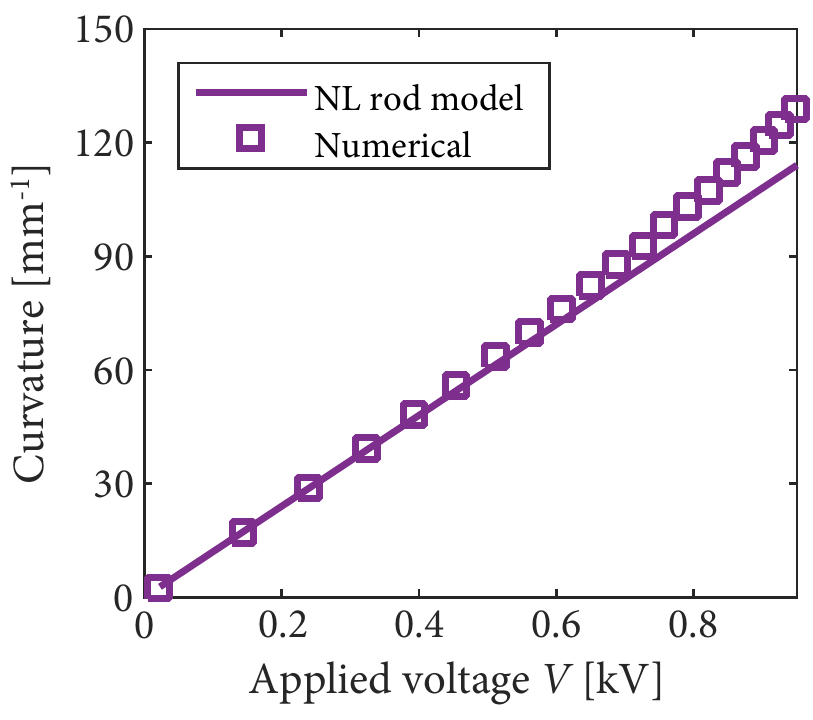}\subcaption{\label{actuator:bSV}Curvature $R^{-1}(V)$.}
\endminipage ~
\minipage{0.33\textwidth}
\includegraphics[scale=.63]{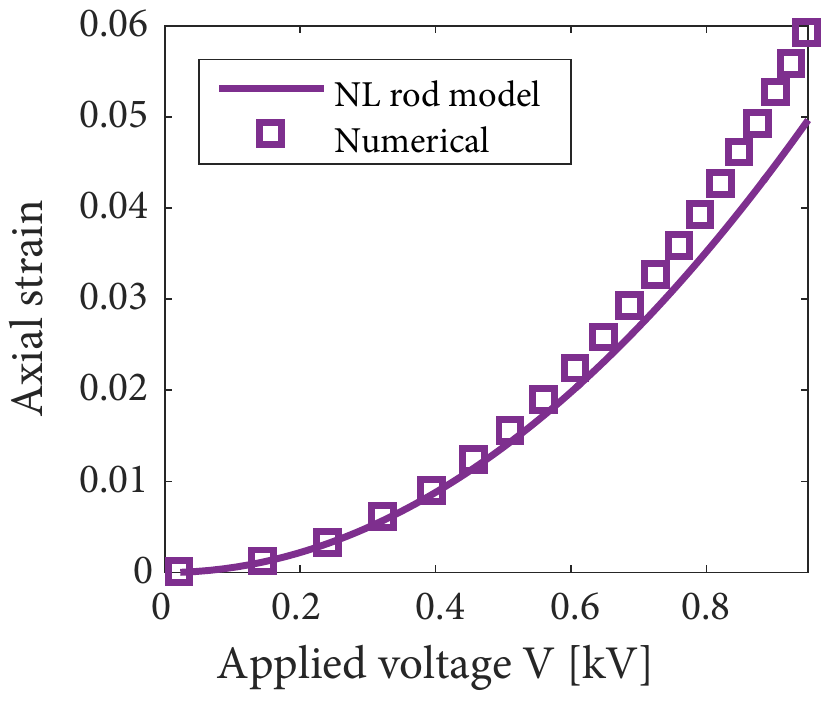}\subcaption{\label{actuator:aSV}Axial strain $\zeta(V)$.}
\endminipage ~
\caption{\label{fig:actuatorSV}Actuation of Saint-Venant–Kirchhoff cantilever rod with transversal flexoelectric coefficient $\flexo_\text{T}=10\si{\nano\joule \per{\volt\metre}}$. Numerically, the axial strain corresponds to the axial component of the Green-Lagrangian strain tensor ($\mathfrak{E}_{33}$), whereas the value from the 1D model corresponds to its Taylor approximation in \eq\eqref{strain-rod}, evaluated at $X_1=0$.}
\end{figure}

Figure \ref{fig:actuatorSV} shows the electromechanical response of an elastically isotropic Saint-Venant–Kirchhoff flexoelectric rod ($\nu = l = 0$) with the flexoelectric constants
$\flexo_\text{T}=10\si{\nano\joule \per{\volt\metre}}, 
\flexo_\text{L}=\flexo_\text{S}=0$
.
The curvature $1/R$ (\fig\ref{actuator:bSV}) and the axial strain $\zeta$ (\fig\ref{actuator:aSV}) are captured very well by the closed-circuit flexoelectric rod model, where we have considered only the leading term in the expansions in \eq\eqref{Taylor-ActCF-FS-AS}, up to a relatively large value of applied voltage $V$. Beyond this limit, the small strains assumption of the 1D non-linear model loose validity. According to \eq\eqref{Taylor-ActCF-FS-AS}, the rod bends thanks to the flexoelectric coupling, and elongates mainly due to electrostriction, cf Section \ref{CC-KH-bending}.

\subsubsection{General flexoelectric problem}

\begin{figure}[!t]\centering
\minipage{0.33\textwidth}
\includegraphics[scale=.62]{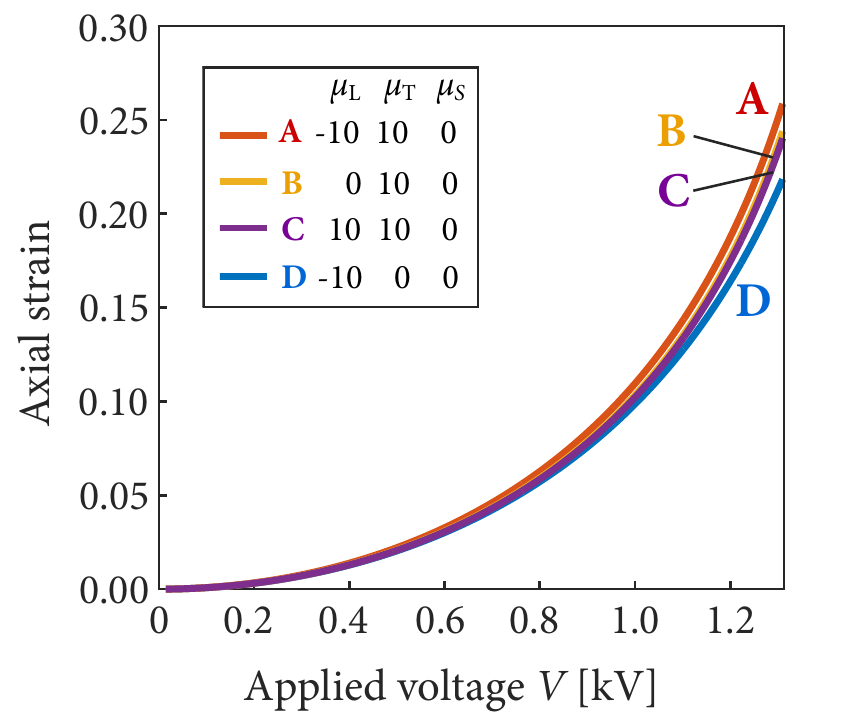}\subcaption{\label{actuator:a}Axial strain $\zeta(V)$}
\endminipage ~
\minipage{0.33\textwidth}
  \includegraphics[scale=.62]{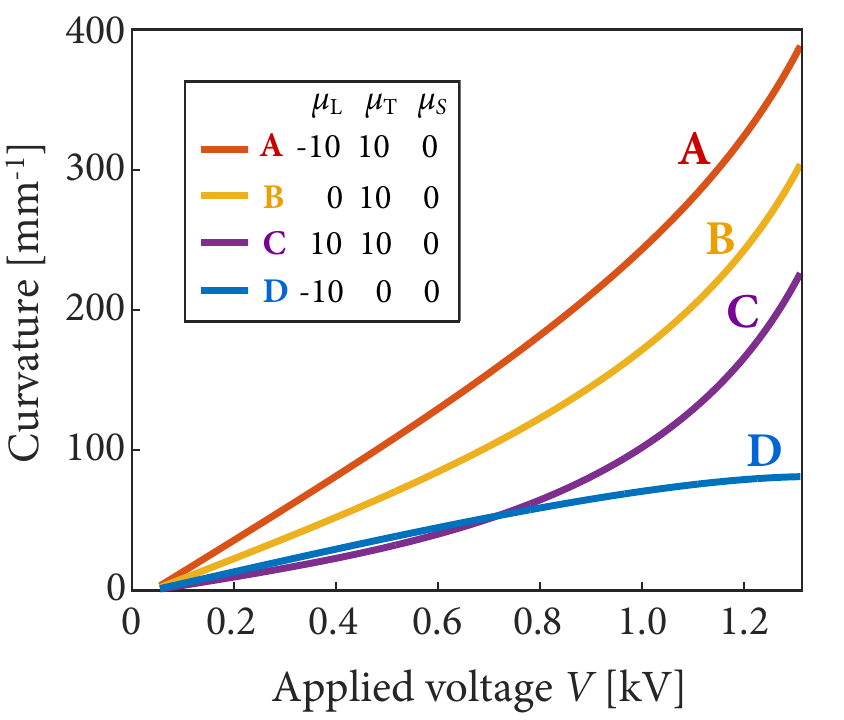}\subcaption{\label{actuator:b}Curvature $R^{-1}(V)$}
\endminipage ~
\minipage{0.33\textwidth}
\includegraphics[scale=.62]{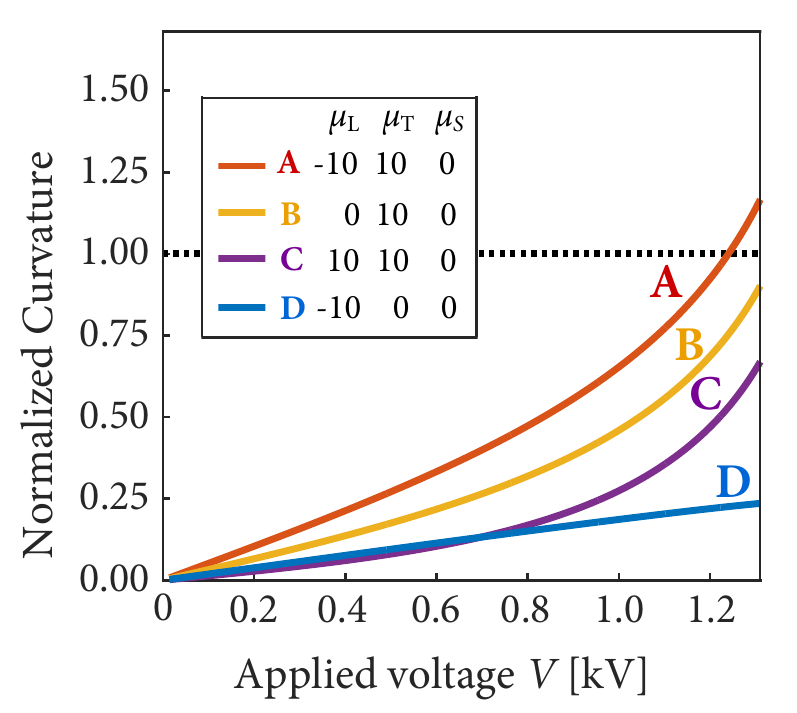}\subcaption{\label{actuator:c}Normalized curvature $\overline{R^{-1}}(V)$}
\endminipage
\\
\minipage{0.57\textwidth}
 \includegraphics[width=\textwidth]{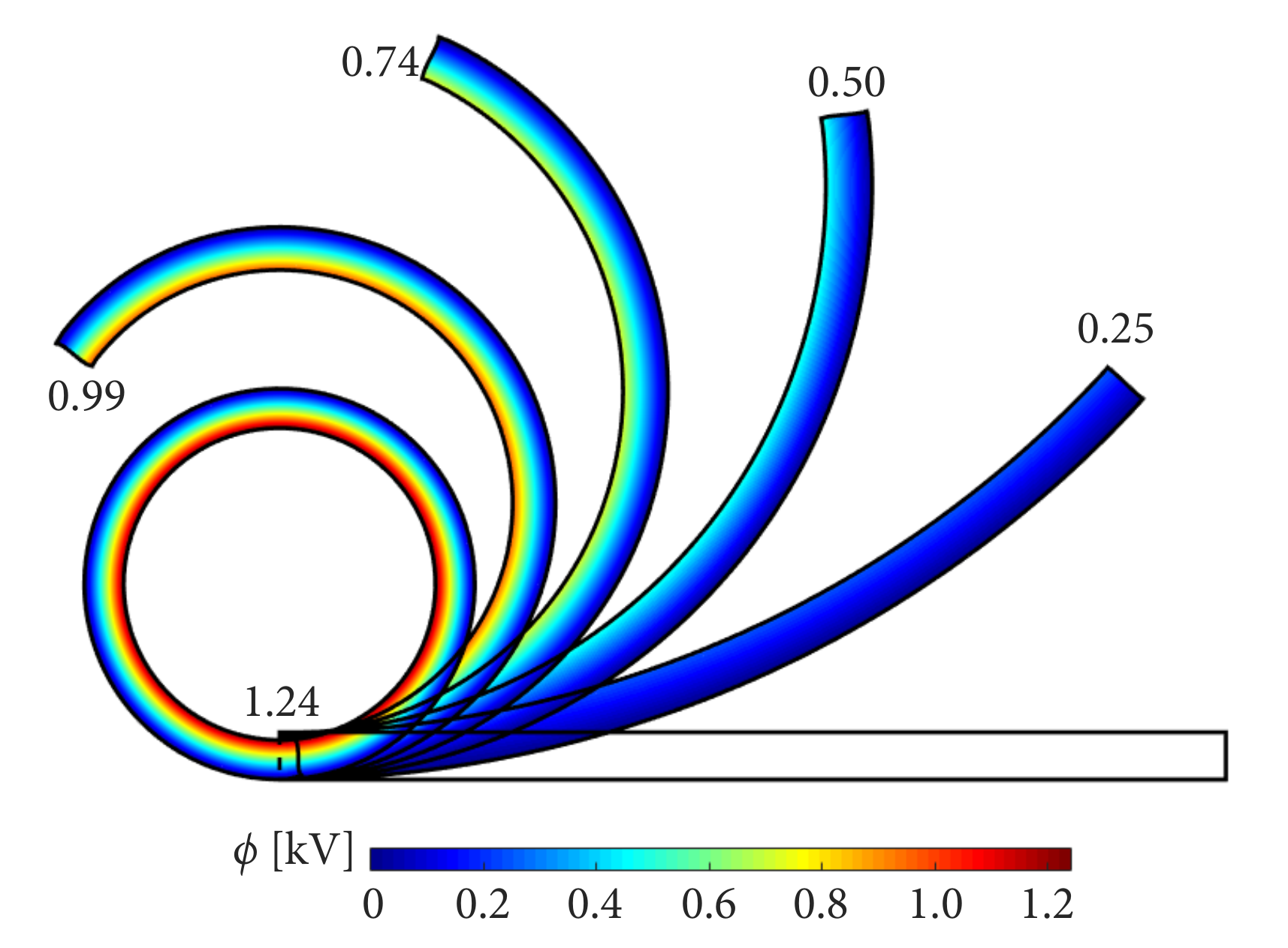}\subcaption{\label{roll-up}Deformed configuration and electric potential distribution in case A upon increasing voltage $V [\si{\kilo\volt}]$, indicated by the number at the free end.}
\endminipage \hspace{3em}
\minipage{0.25\textwidth}\vspace{6em}
  \includegraphics[width=\textwidth]{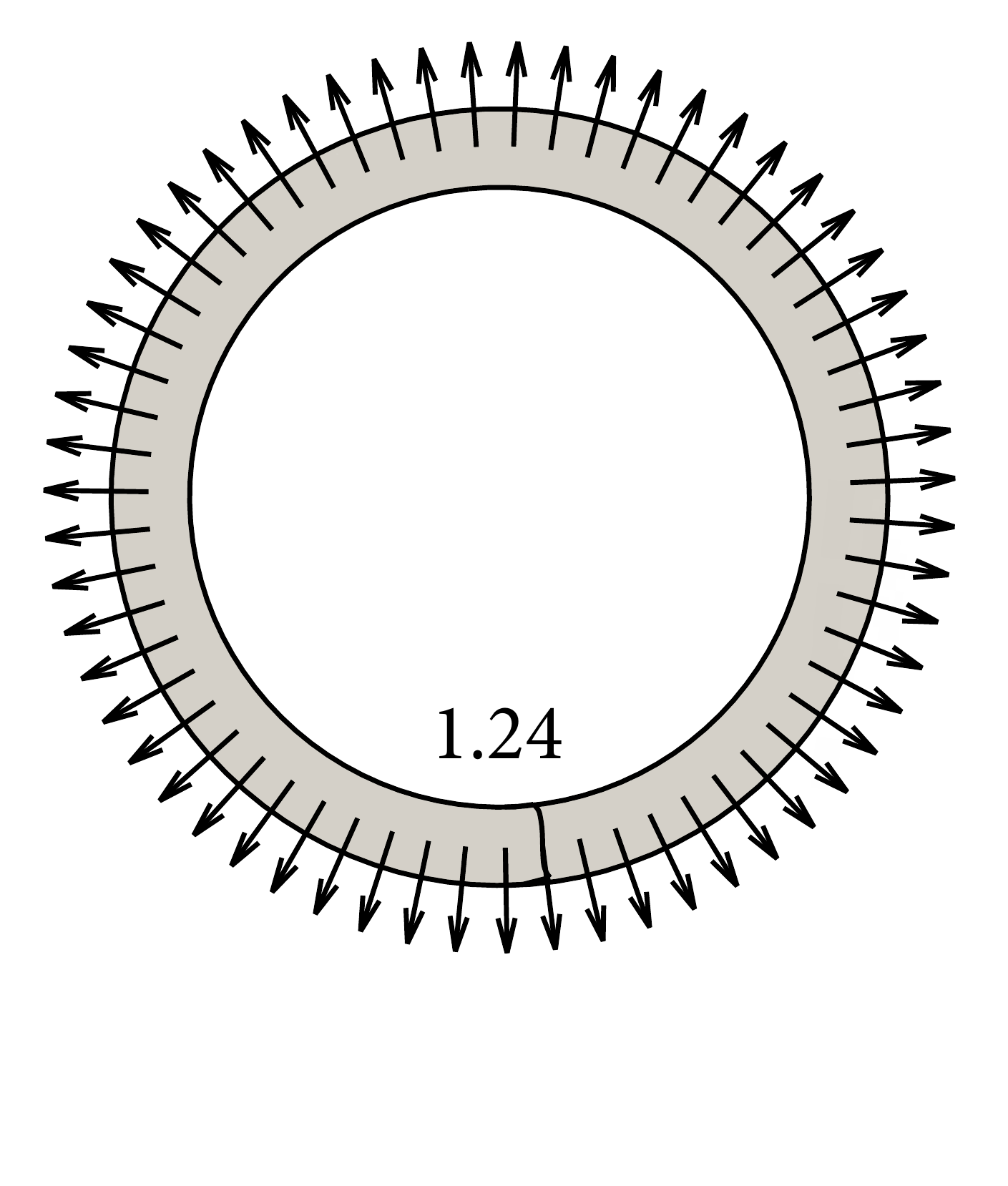}\subcaption{\label{roll-up-polarization}Distribution of polarization field in case A at $V=12.4\si{\kilo\volt}$.}
\endminipage
\caption{\label{fig:actuator}Actuation of Neo-Hookean cantilever rod with different flexoelectric tensors (expressed in  $\si{\nano\joule \per{\volt\metre}}$) }
\end{figure}

Since the curvature is found to be uniform, \cf \eq \eqref{curvature:CC}, the rod forms an arc of a circle, \cf \fig\ref{roll-up}. Thus, a natural question that arises is which set of flexoelectric parameters achieve a fully-closed circular shape more efficiently (\ie with a lower applied voltage). To address this question, we consider an isotropic Neo-Hookean elastic (see \eq\eqref{nh}) rod with 
$\nu = 0.37$, $\ell = 0.03 \si{\micro\meter}$
and varying flexoelectric constants. To quantify the curvature of the rod relative to the curvature of the closed circle, we define the normalized curvature $\overline{R^{-1}}(V)=R^{-1}(V)/R^{-1}_\circ(V)$, where $R^{-1}_\circ(V)=2\pi/\left((1+\zeta(V))L\right)$ is the curvature required to form a closed circular shape.

Figure \ref{fig:actuator} shows the evolution of $\zeta(V)$, $R^{-1}(V)$ and $\overline{R^{-1}}(V)$ for flexoelectric tensors with different combinations of longitudinal ($\flexo_\text{L}$), transversal ($\flexo_\text{T}$) and shear ($\flexo_\text{S}$) flexoelectric coefficients. The cases including a non-vanishing shear coefficient are omitted, since the results do not change significantly, even when $\flexo_\text{S}$ is one order of magnitude larger than $\flexo_\text{L}$ or $\flexo_\text{T}$. For the sake of brevity, the simulations \emph{(i)} with negative applied electric voltage $V$, and \emph{(ii)} yielding negative curvatures, are also omitted since the results are analogous to those simulations with \emph{(i)} positive applied voltage and \emph{(ii)} negative flexoelectric coefficients, respectively, as
\begin{subequations}\begin{gather}
\zeta(V)|_{\Flexo}=\zeta(-V)|_{\Flexo}=\zeta(V)|_{-\Flexo}=\zeta(-V)|_{-\Flexo};\\
R^{-1}(V)|_{\Flexo}=-R^{-1}(-V)|_{\Flexo}=-R^{-1}(V)|_{-\Flexo}=R^{-1}(-V)|_{-\Flexo};\\
\overline{R^{-1}}(V)|_{\Flexo}=-\overline{R^{-1}}(-V)|_{\Flexo}=-\overline{R^{-1}}(V)|_{-\Flexo}=\overline{R^{-1}}(-V)|_{-\Flexo};
\end{gather}\end{subequations}
in accordance with \eqs\eqref{Taylor-ActCF-FS} and \eqref{curvature:CC}. 

As expected, the axial strain of the rod (depicted in \fig\ref{actuator:a}) does not vary much with the different flexoelectric parameters, since it is mainly a consequence of electrostriction. The curvature (\fig\ref{actuator:b}), instead, varies significantly for the different combinations of flexoelectric parameters. The dominant parameter is the transversal flexoelectric coefficient $\flexo_\text{T}$ which leads to positive curvature, as shown in case B. The longitudinal flexoelectric coefficient $\flexo_\text{L}$ is also relevant and leads to negative curvature, as shown in case D. The largest response is found with positive $\flexo_\text{T}$ and negative $\flexo_\text{L}$, as shown in case A. Finally, case C corresponds to positive $\flexo_\text{L}$ and $\flexo_\text{T}$, and yields curvatures inbetween cases B (purely transversal $\Flexo$) and D (purely longitudinal $\Flexo$).

The normalized curvature is shown in \fig\ref{actuator:c}. For sufficiently large actuation, case A reaches $\overline{R^{-1}}>1$, which indicates that the actuator rolls up forming a closed circle. This process is shown in \fig\ref{roll-up}, where the deformed configuration and electric potential distribution within the rod is depicted at different applied voltages. We also show in \fig\ref{roll-up-polarization} the resulting polarization field once the circle is formed, which remains normal to the bent rod.

\subsection{\label{actuatorbuckling}Buckling of closed-circuit flexoelectric cantilever under electric actuation}

\begin{figure}[p]
\minipage{\textwidth}\vspace{-4em}\centering
\minipage{0.35\textwidth}
  \includegraphics[width=\textwidth]{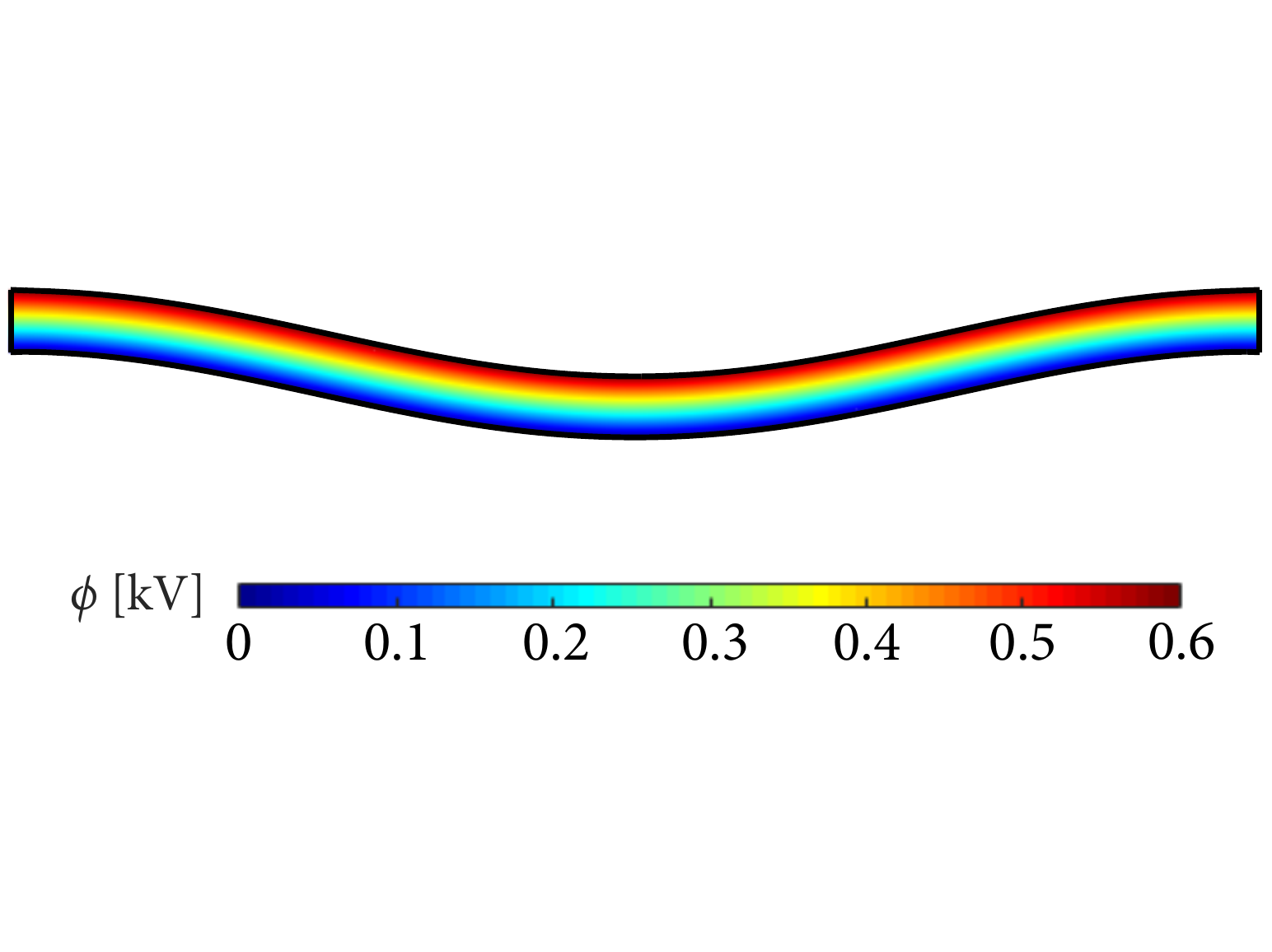}
  \subcaption{\label{actuatorbuck:def}Buckled shape and electric potential distribution of the $L/H=20$-slender rod upon electrical loading of $0.6\si{\kilo\volt}$.}
\endminipage\qquad
\minipage{0.35\textwidth}
  \includegraphics[width=\textwidth]{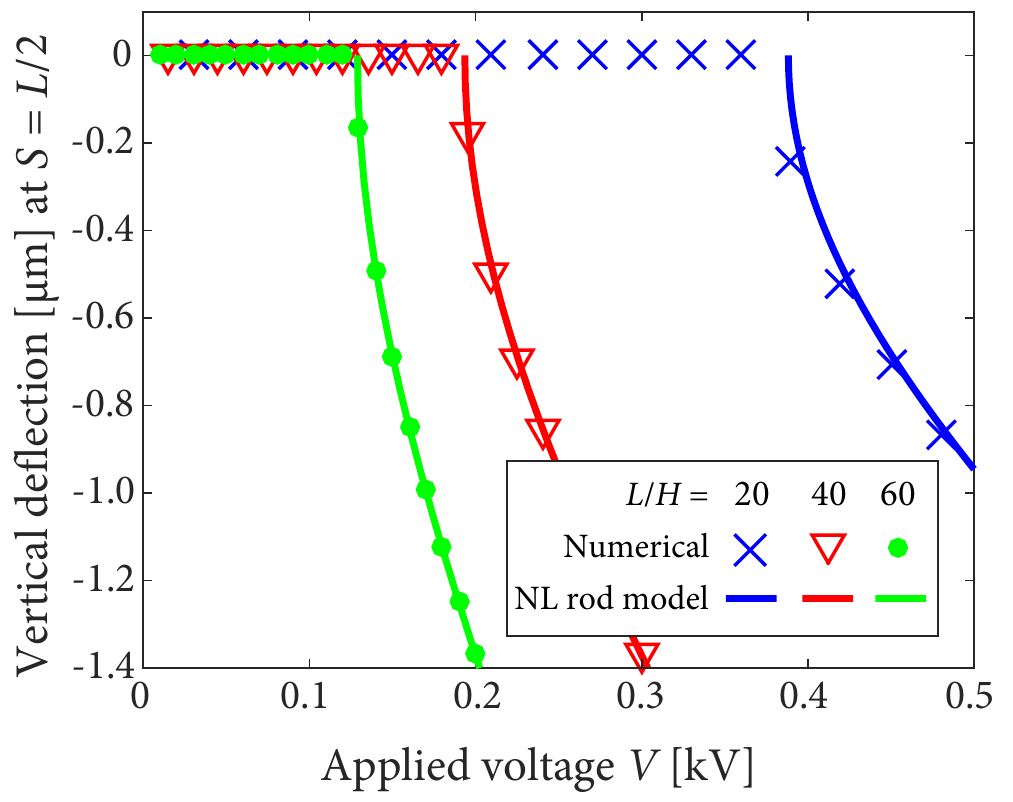}\subcaption{\label{actuatorbuck:a}Vertical deflection $r_1$ at the center of the rod.}\vspace{1.3em}
\endminipage \endminipage\\
\minipage{\textwidth}\centering
\minipage{0.35\textwidth}
\includegraphics[width=0.87\textwidth]{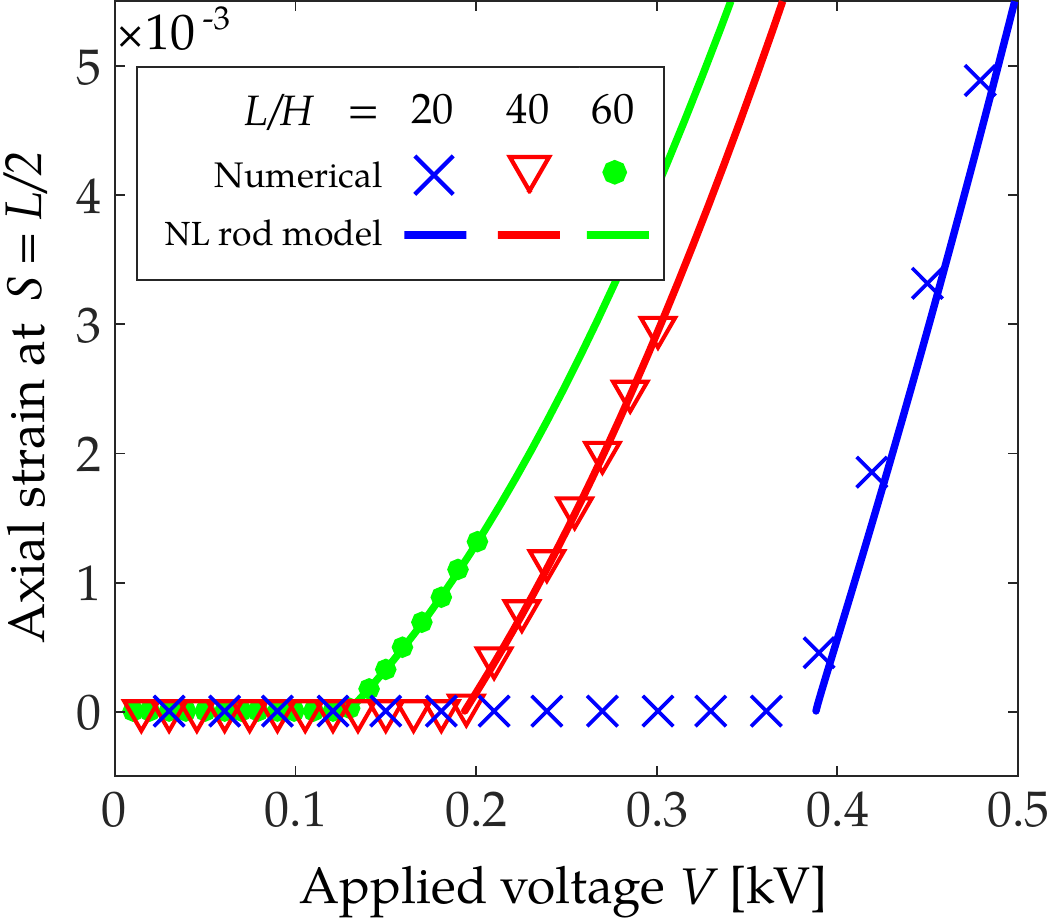}\subcaption{\label{actuatorbuck:stretch}Axial strain as a function of slenderness.}\vspace{2.4em}
\endminipage
\qquad
\minipage{0.35\textwidth}
  \includegraphics[width=\textwidth]{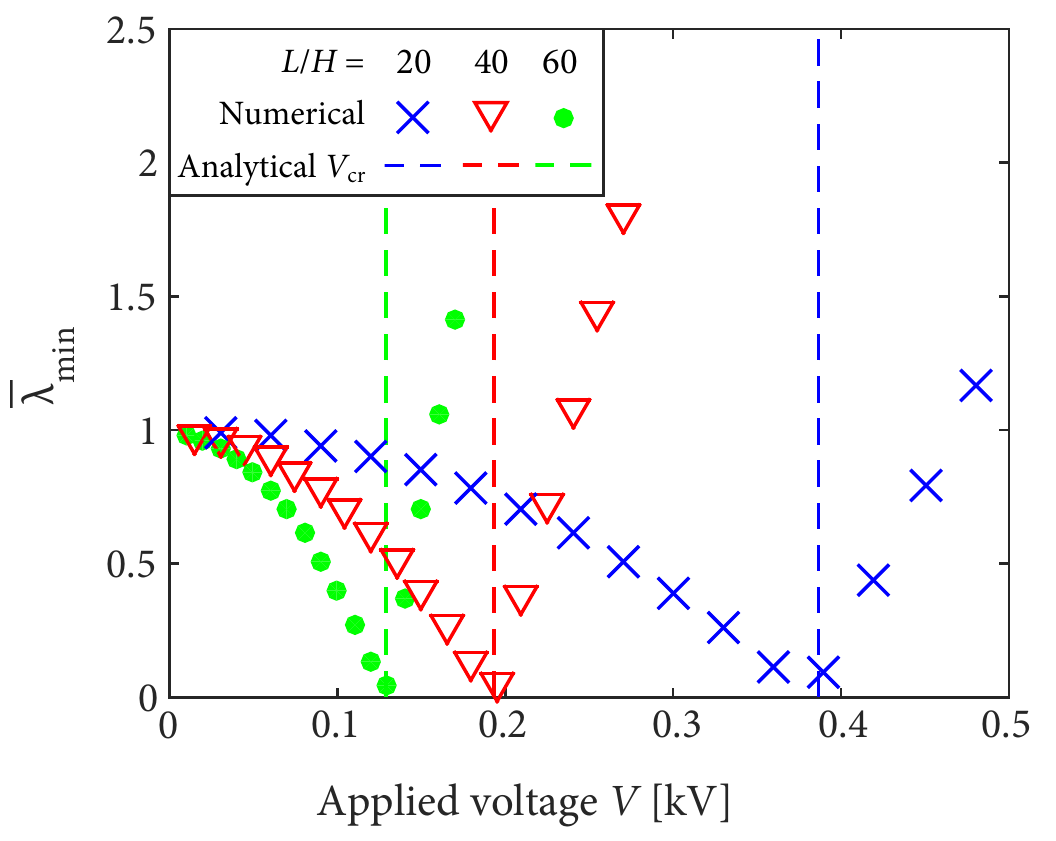}\subcaption{\label{actuatorbuck:b}Normalized minimum eigenvalue $\lambda_\text{min}\left[\left.\toMat{\widehat{H}}_{\boldsymbol{\chi}\boldsymbol{\chi}}\right.^{\!(k)}\right]$ and critical voltage of analytical 1D model.}
\endminipage\endminipage\\
\minipage{\textwidth}\vspace{1em}\centering
\minipage{0.35\textwidth}
\includegraphics[width=\textwidth]{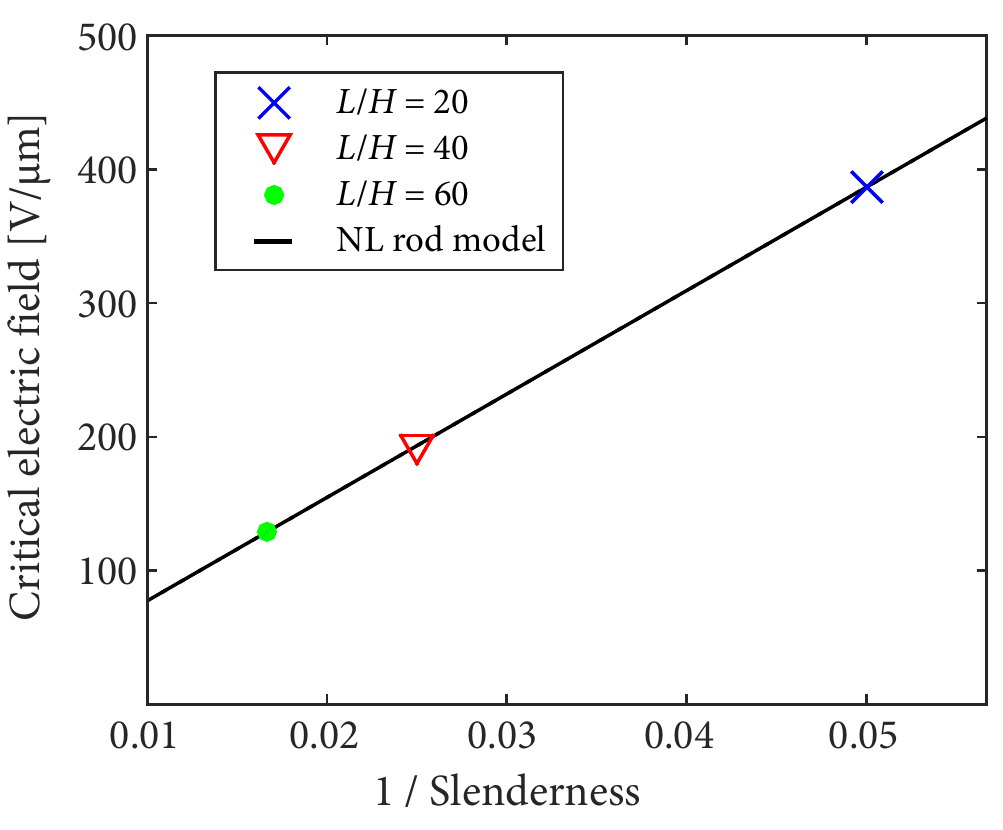}\subcaption{\label{actuatorbuck:c}Critical electric field as a function of slenderness.}
\endminipage
\endminipage
\caption{\label{fig:actuatorbuck}Actuation of Saint-Venant–Kirchhoff clamped-clamped rod with transversal flexoelectric coefficient  $\flexo_\text{T}=10\si{\nano\joule \per{\volt\metre}}$ and varying slenderness. In (d), $\bar\lambda_\text{min}=\lambda_\text{min}\left({n_\text{DOF}}/{n_0}\right)^4$ ,where $n_\text{DOF}$ is the number of degrees of freedom of each simulation, and $n_0=312$ is an arbitrary normalization constant, chosen such that $\bar\lambda_\text{min}(0)\approx 1$.} 
\end{figure}

In the previous example, the rod undergoes elongation upon electrical actuation mainly due to electrostriction. In this Section, we present a similar setup where the right tip is also clamped, as shown in \fig\ref{fig:CC-buckling}. In this case, an axial compressive force is expected at the clamped ends since the elongation of the rod is prevented. Restricting \eq\eqref{Act-CC-Force} in pre-buckling stage, the axial force grows quadratically with the applied voltage and, for a large enough applied (critical) voltage $V_\text{cr}$, \cf\eq\eqref{criticalE}, a mechanical instability is reached, inducing buckling of the rod.

Figure \ref{fig:actuatorbuck} shows numerical simulations of a flexoelectric Saint-Venant–Kirchhoff rod ($\nu = \ell = 0$) 
of dimensions $L=20\si{\micro\meter}$, $H=1\si{\micro\meter}$, with Young's modulus $Y = 1.0 \si{\giga\pascal}$, dielectric permittivity $\epsilon = 0.11 \si{\nano\joule / \square\volt\meter}$ and transversal flexoelectric coefficient $\flexo_\text{T}=10\si{\nano\joule \per{\volt\metre}}$ ($\flexo_\text{L}=\flexo_\text{S}=0$).
The postbuckling configuration and the evolution of the maximum deflection and axial strain with respect to applied voltage are depicted in \fig\ref{actuatorbuck:def}-\ref{actuatorbuck:stretch}, showing an excellent match between the numerical results and the analytical expressions in \eq\eqref{r1-CC-postbuckling}, \eqref{V-CC-postbuckling} and \eqref{eqn:strain:CC}.
The critical voltage at which the rod buckles (see \fig\ref{actuatorbuck:b}) matches also with the one predicted by the analytical 1D nonlinear model in \eq\eqref{criticalV}, and the critical electric field (\cf \fig\ref{actuatorbuck:c}) is inversely proportional to the slenderness of the rod, as predicted in \eq\eqref{criticalE_Taylor}.

\section{Conclusions and directions of future work\label{sec_06} }

We have developed the material form of the balance equations for dielectric elastomers, including the flexoelectric effect. Unlike previously considered models of the flexoelectric coupling, here we formulate our model in terms of polarization, strain gradients and flexocoupling tensor in a fully material frame. As a result, our formulation is objective by construction, and the flexocoupling tensor has the same symmetries as that used in linearized theories. After partial Legendre transform, the equations are written in terms of the electric potential and the displacement field as a fourth order unconstrained system of partial differential equations, which is convenient for finding numerical and analytical solutions.
A numerical implementation of the theory is developed using open B-spline basis of sufficient smoothness on a uniform Cartesian grid, enabling robust simulations deep into the nonlinear regime, for very large deformations, and including mechanical instabilities \citep{Yvonnet2017}.
On the other hand, analytical closed-form solutions are derived for open- and closed-circuit nonlinear extensible flexoelectric rods under bending and buckling. Direct comparison of this model with direct numerical simulations of the full model shows excellent agreement well into the nonlinear regime in conditions where the rod theory is expected to apply. The analytical rod theory serves both as a means of validation of our nonlinear simulations, and as fast and simple model to analyze and design nonlinear flexoelectric devices.  

The current model could be easily extended in several ways. For instance, rather than homogeneous electric Neumann boundary conditions on the free surfaces, it may be more realistic to directly model the surrounding medium as a dielectric when considering soft materials materials with relatively low dielectric constant \citep{Yvonnet2017,Thai2018}. Our model can be extended to account for converse flexoelectricity \citep{LandauLifshitz1951,Sharma2010,landau2013course}, for polarization gradient dielectricity \citep{Mindlin1968b}, for material incompressibility, and coupled with flexible discretization methods, e.g.~based on immersed boundaries \citep{codony2019immersed}, to model domains of general, and possibly complex, geometry that might enhance field gradients.

\section*{Acknowledgments}
This work was supported by the Generalitat de Catalunya (“ICREA Academia” award for excellence in research to I.A., and Grant No.~2017-SGR-1278), and the European Research Council (StG-679451 to I.A.). CIMNE is recipient of a Severo Ochoa Award of Excellence from the MINECO.

\appendix

\section{Material characterization}\label{sec_app01}
The material is fully characterized by specifying the elastic energy density $\Psi^\textnormal{Elast}(\C)$ and the material tensors of flexoelectricity $\Flexo$ and strain gradient elasticity $\StrGr$.

\subsection*{Isotropic Saint-Venant–Kirchhoff model.}
It corresponds to the extension of the linear isotropic elastic material model to the non-linear regime, and depends on the Lam\'{e} parameters $\uplambda=Y \nu /(1+\nu)(1-2\nu)$ and $\upmu=Y/2(1+\nu)$ as follows:
\begin{subequations}\label{sv}\begin{align}
\Psi^\textnormal{Elast}(\C)
&=
\frac{\uplambda}{2}{\left[\trace{\mathfrak{E}}\right]}^2+\upmu\trace{\mathfrak{E}^2}
,\\
\frac{\partial\Psi^\textnormal{Elast}(\C)}{\partial C_{IJ}}
&=
\frac{\uplambda}{2}\left[\trace{\mathfrak{E}}\right]\id_{IJ}+\upmu\mathfrak{E}_{IJ}
,\\
\frac{\partial^2\Psi^\textnormal{Elast}(\C)}{\partial C_{IJ}C_{KL}}
&=
\frac{\uplambda}{4}\id_{IJ}\id_{KL}+\frac{\upmu}{2}\id_{IK}\id_{JL}.
\end{align}\end{subequations}

\subsection*{Isotropic Neo-Hookean model}
The Neo-Hookean model is adequate for describing nonlinear stress-strain behavior of cross-linked polymers at moderate strains. It is mathematically defined as
\begin{subequations}\label{nh}\begin{align}
\Psi^\textnormal{Elast}(\C)
&=
\frac{\uplambda}{2}{\left[\Log{J}\right]}^2+\frac{\upmu}{2}\left[\trace{\C}-2\right]
,\\
\frac{\partial\Psi^\textnormal{Elast}(\C)}{\partial C_{IJ}}
&=
\frac{\uplambda}{2}\Log{J}C^{-1}_{IJ}+\frac{\upmu}{2}\left(\id_{IJ}-C^{-1}_{IJ}\right)
,\\
\frac{\partial^2\Psi^\textnormal{Elast}(\C)}{\partial C_{IJ}C_{KL}}
&=
\frac{\uplambda}{4}C^{-1}_{IJ}C^{-1}_{KL}+\frac{1}{4}\left[\upmu-\uplambda\Log{J}\right]\left(C^{-1}_{IK}C^{-1}_{JL}+C^{-1}_{IL}C^{-1}_{JK}\right).
\end{align}\end{subequations}

\subsection*{Flexoelectricity tensor $\Flexo$.}
The cubic flexoelectric tensor depends on the longitudinal $\flexo_\text{L}$, transversal $\flexo_\text{T}$ and shear $\flexo_\text{S}$ parameters \citep{LeQuang2011,codony2019immersed}. In the Cartesian axes, it takes the following form:
\begin{align}\label{flexotensor}
\flexo_{LIJK}=
\begin{cases}
	\flexo_\text{L},& \text{for } L=I=J=K ,\\
	\flexo_\text{T},& \text{for } I=J\neq K=L ,\\
	\flexo_\text{S},& \text{for } L=I\neq J=K \text{ or } L=J\neq I=K,\\
	0        & \text{otherwise}.
    \end{cases}
\end{align}

\subsection*{Strain gradient elasticity tensor $\StrGr$.}

We consider an isotropic simplified strain gradient elasticity tensor \citep{Altan1997}, which depends on $\uplambda$, $\upmu$ and the length scale $\ell$ in the following form:
\begin{align}\label{strgrtensor}
\strGr_{IJKLMN}=\left(\uplambda\id_{IJ}\id_{LM}+2\upmu\id_{IL}\id_{JM}\right)\ell^2\id_{KN}.
\end{align}

\section{Second variation of the enthalpy functional}\label{App02}

The second variation of the enthalpy functional, required in our solution method, is given by
\begin{equation}\label{secvar}\begin{split}
\delta^2 \Pi[\boldsymbol{\chi},\phi;\delta\boldsymbol{\chi},\delta\phi;\Delta\boldsymbol{\chi},\Delta\phi]\hspace{-7em}&
\\{}=&
\delta \left( R[\boldsymbol{\chi},\phi;\delta\boldsymbol{\chi},\delta\phi] \right) [\Delta\boldsymbol{\chi},\Delta\phi]
\\{}=&
\frac{\partial R[\boldsymbol{\chi},\phi;\delta\boldsymbol{\chi},\delta\phi]}{\partial\boldsymbol{\chi}}\Delta\boldsymbol{\chi}
+
\frac{\partial R[\boldsymbol{\chi},\phi;\delta\boldsymbol{\chi},\delta\phi]}{\partial\phi}\Delta\phi
\\{}=&\int_{\Omega_0}\Bigg\{
\delta \mathfrak{E}_{IJ} \Delta \mathfrak{E}_{KL}\left(
4\frac{\partial^2\bar\Psi^\text{Elast}(\C)}{\partial C_{IJ} \partial C_{KL}}
\right)
+\left(2\frac{\partial\Psi^\text{Elast}(\C)}{\partial C_{IJ}}\right)(\Delta\delta) \mathfrak{E}_{IJ}
\\&{}+\strGr_{IJKLMN}\delta\widetilde{\mathfrak{E}}_{IJK} \Delta \widetilde{\mathfrak{E}}_{IJK}
+\left(\strGr_{IJKLMN}\widetilde{\mathfrak{E}}_{LMN}\right)(\Delta\delta) \widetilde{\mathfrak{E}}_{IJK}
\\&{}-\epsilon JC^{-1}_{MF}\delta E_F \Delta E_M
\\&{}+\epsilon J\mathscr{C}_{MFIJ} E_F\left(\frac{1}{2}E_M(\Delta\delta) \mathfrak{E}_{IJ}+\delta \mathfrak{E}_{IJ}\Delta E_M + \delta E_M \Delta \mathfrak{E}_{IJ}\right)
\\&{}+
\epsilon J\widetilde{\mathscr{C}}_{MFIJKL}\frac{1}{2}E_ME_F \delta \mathfrak{E}_{IJ} \Delta \mathfrak{E}_{KL}
\\&{}-\flexo_{FABK}JC^{-1}_{MF}\left(E_M(\Delta\delta) \widetilde{\mathfrak{E}}_{ABK}+\delta E_M \Delta\widetilde{\mathfrak{E}}_{ABK}+\delta\widetilde{\mathfrak{E}}_{ABK}\Delta E_M\right)
\\&{}+\flexo_{FABK}J\mathscr{C}_{MFIJ}\Big(
\widetilde{\mathfrak{E}}_{ABK}\left(\delta \mathfrak{E}_{IJ}\Delta E_M + \delta E_M\Delta \mathfrak{E}_{IJ}\right)
+ E_M \left(\delta \mathfrak{E}_{IJ}\Delta\widetilde{\mathfrak{E}}_{ABK} + \delta\widetilde{\mathfrak{E}}_{ABK}\Delta \mathfrak{E}_{IJ}\right)
\\&\hphantom{{}+\flexo_{FABK}J\mathscr{C}_{MFIJ}\Big(}
+ E_M \widetilde{\mathfrak{E}}_{ABK}(\Delta\delta) \mathfrak{E}_{IJ}
\Big)
\\&{}+\flexo_{FABK} J\widetilde{\mathscr{C}}_{MFIJPQ}E_M\widetilde{\mathfrak{E}}_{ABK}\delta \mathfrak{E}_{IJ}\Delta \mathfrak{E}_{PQ}
\Bigg\}\dd \Omega_0
\\{}=&\int_{\Omega_0}\Bigg\{
\sp_{IJ}(\Delta\delta) \mathfrak{E}_{IJ}
+\ss_{IJK}(\Delta\delta) \widetilde{\mathfrak{E}}_{IJK}
\\&{}+\left(\mathbb{A}^\text{Elast}_{IJKL}+\mathbb{A}^\text{Diele}_{IJKL}+\mathbb{A}^\text{Flexo}_{IJKL}\right)
\delta \mathfrak{E}_{IJ}\Delta \mathfrak{E}_{KL}
\\&{}+\widetilde{\mathbb{A}}^\text{SGEla}_{IJKLMN}\delta\widetilde{\mathfrak{E}}_{IJK}\Delta\widetilde{\mathfrak{E}}_{LMN}
\\&{}+\widetilde{\mathbb{A}}^\text{Flexo}_{IJKLM} \left(\delta \mathfrak{E}_{IJ}\Delta\widetilde{\mathfrak{E}}_{KLM} + \delta\widetilde{\mathfrak{E}}_{KLM}\Delta \mathfrak{E}_{IJ}\right)
\\&{}+\mathbb{B}^\text{Diele}_{IJ}\left(\delta E_I \Delta E_J\right)
\\&{}+\left(\mathbb{C}^\text{Diele}_{IJK}+\mathbb{C}^\text{Flexo}_{IJK}\right)
\left(\delta \mathfrak{E}_{IJ}\Delta E_K + \delta E_K\Delta \mathfrak{E}_{IJ}\right)
\\&{}+\widetilde{\mathbb{C}}^\text{Flexo}_{IJKL}\left(\delta E_L\Delta\widetilde{\mathfrak{E}}_{IJK}+\delta\widetilde{\mathfrak{E}}_{IJK}\Delta E_L\right)
\Bigg\}\dd \Omega_0
,
\end{split}\end{equation}
where $\Delta\boldsymbol{\chi}$ and $\Delta\phi$ are variations of $\boldsymbol{\chi}$ and $\phi$, respectively, and
\begin{subequations}\begin{align}
\Delta E_L&\coloneqq -\frac{\partial(\Delta\phi)}{\partial X_L},
\\
\Delta F_{iI}&\coloneqq\frac{\partial(\Delta x_i)}{\partial X_I},
\\
\Delta \ff_{iIJ}&\coloneqq\frac{\partial^2(\Delta x_i)}{\partial X_I\partial X_J},
\\
\Delta \mathfrak{E}_{IJ}=\frac{\Delta C_{IJ}}{2}&\coloneqq \symm[IJ]{\Delta F_{kI} F_{kJ}},
\\
\Delta \widetilde{\mathfrak{E}}_{IJK}=\frac{\Delta \cc_{IJK}}{2}&\coloneqq \symm[IJ]{\Delta F_{kI} \ff_{kJK} + F_{kI} \Delta\ff_{kJK}},
\\
(\Delta\delta) \mathfrak{E}_{IJ}=\frac{(\Delta\delta) C_{IJ}}{2}&\coloneqq \symm[IJ]{\Delta F_{kI} \delta F_{kJ}},
\\
(\Delta\delta) \widetilde{\mathfrak{E}}_{IJK}=\frac{(\Delta\delta) \cc_{IJK}}{2}&\coloneqq \symm[IJ]{\Delta F_{kI} \delta\ff_{kJK} + \delta F_{kI} \Delta\ff_{kJK}}.
\end{align}\end{subequations}

The material tensors in the right hand side of \eq\eqref{secvar} are defined as follows:
\begin{subequations}\begin{align}
\mathbb{A}^\text{Elast}_{IJKL}(\C)&\coloneqq \frac{\partial^2\bar\Psi^\text{Elast}}{\partial \mathfrak{E}_{IJ} \partial \mathfrak{E}_{KL}}
\\
\mathbb{A}^\text{Diele}_{IJKL}(\C,\E)&\coloneqq \frac{\partial^2\bar\Psi^\text{Diele}}{\partial \mathfrak{E}_{IJ}\partial \mathfrak{E}_{KL}}
=\frac{1}{2}J\widetilde{\mathscr{C}}_{MFIJKL}E_M \epsilon E_F
\\
\mathbb{A}^\text{Flexo}_{IJKL}(\C,\CC,\E)&\coloneqq \frac{\partial^2\bar\Psi^\text{Flexo}}{\partial \mathfrak{E}_{IJ}\partial \mathfrak{E}_{KL}}
=J\widetilde{\mathscr{C}}_{MFIJKL}E_M\flexo_{FABC}\widetilde{\mathfrak{E}}_{ABC}
\\
\widetilde{\mathbb{A}}^\text{SGEla}_{IJKLMN}&\coloneqq \frac{\partial^2\bar\Psi^\text{SGEla}}{\partial \widetilde{\mathfrak{E}}_{IJK}\partial \widetilde{\mathfrak{E}}_{LMN}}=\strGr_{IJKLMN},
\\
\widetilde{\mathbb{A}}^\text{Flexo}_{IJKLM}(\C,\E)&\coloneqq \frac{\partial^2\bar\Psi^\text{Flexo}}{\partial \mathfrak{E}_{IJ}\partial \widetilde{\mathfrak{E}}_{KLM}}=J\mathscr{C}_{ABIJ}\flexo_{BKLM} E_A 
\\
\mathbb{B}^\text{Diele}_{IJ}(\C)&\coloneqq \frac{\partial^2\bar\Psi^\text{Diele}}{\partial E_{I}\partial E_{J}}=-\epsilon JC^{-1}_{IJ}
\\
\mathbb{C}^\text{Diele}_{IJK}(\C,\E)&\coloneqq \frac{\partial^2\bar\Psi^\text{Diele}}{\partial \mathfrak{E}_{IJ}\partial E_{K}}=\epsilon J\mathscr{C}_{KMIJ} E_M
\\
\mathbb{C}^\text{Flexo}_{IJK}(\C,\CC)&\coloneqq \frac{\partial^2\bar\Psi^\text{Flexo}}{\partial \mathfrak{E}_{IJ}\partial E_{K}}=\flexo_{MABC}J\mathscr{C}_{KMIJ}\widetilde{\mathfrak{E}}_{ABC}
\\
\widetilde{\mathbb{C}}^\text{Flexo}_{IJKL}(\C)&\coloneqq \frac{\partial^2\bar\Psi^\text{Flexo}}{\partial \widetilde{\mathfrak{E}}_{IJK}\partial E_{L}}=-\flexo_{MIJK}JC^{-1}_{ML}
\end{align}\end{subequations}

The tensor $\toVect{\widetilde{\mathscr{C}}}$ in \eq\eqref{secvar} is defined as
\begin{gather}
\widetilde{\mathscr{C}}_{ABCDEF}\coloneqq
\frac{2}{J}\frac{\partial\left(J\mathscr{C}_{ABCD}\right)}{\partial C_{EF}}=
\left(
\mathbb{D}_{ACBDEF}+\mathbb{D}_{BDACEF}+\mathbb{D}_{ADBCEF}+\mathbb{D}_{BCADEF}-\mathbb{D}_{ABCDEF}-\mathbb{D}_{CDABEF}
\right),
\end{gather}
where $\mathbb{D}_{ABCDEF}\coloneqq C^{-1}_{AB}\left(\frac{1}{2}C^{-1}_{CD}C^{-1}_{EF}-C^{-1}_{CE}C^{-1}_{DF}-C^{-1}_{CF}C^{-1}_{DE}\right).$

\section{Analytical solutions for the displacement and the electric field in flexoelectric rods under bending} \label{appendix:bending}
Following \citet{Bisshopp1945}, \eq \eqref{EL-cantilever-moment} is integrated as
\begin{equation}\label{eqn:bending}
\dfrac{1}{2}\left(\dfrac{\dd \theta}{\dd S}\right)^2 + \beta^2 \left(\sin\theta^\text{max} -\sin \theta\right) =0, 
\end{equation}
where $\theta(L)=\theta^\textnormal{max} \leq 0$ is the rotation at the free end of the rod produced by the applied load,
and equivalently
\begin{equation}\label{eqn:dX3-1}
\mathrm{d} S = - \dfrac{\mathrm{d} \theta}{\beta \sqrt{2(\sin{\theta}-\sin{\theta^\text{max}})}}, 
\end{equation}
since  $\theta \leq 0$ and $\dd \theta / \dd S \leq 0$ for a rod bending downwards. 
The integral of \eq \eqref{eqn:dX3-1} along the rod yields approximately its length, since
\begin{equation}
L = \int_{0}^{L}\mathrm{d} S = \int_{\theta(0)}^{\theta(L)}  \dfrac{\mathrm{d} S}{\mathrm{d} \theta} \dd \theta =  \int_{\theta^\text{max}}^{0} \dfrac{\mathrm{d}\theta}{\beta \sqrt{2\left(\sin{\theta} -\sin{\theta^\text{max}}\right)}},
\end{equation}
and thus
\begin{equation} \label{integral}
 \beta L = \int_{\theta^\text{max}}^{0} \dfrac{\mathrm{d}\theta}{\sqrt{2\left (\sin{\theta} -\sin{\theta^\text{max}}\right)}}.
\end{equation}
In order to evaluate this integral, let us assume
\begin{eqnarray} \label{changeVar}
\sin \theta^{\text{max}} = 1-2p^2, \quad \sin \theta = 1-2p^2\sin^2\psi, \quad \psi \in [\psi_0, \frac{\pi}{2}],
\end{eqnarray}
with
\begin{equation}\label{psi0}
\psi_0 = \sin^{-1}\left(\dfrac{1}{p\sqrt{2}}\right) = \sin^{-1}\left(\dfrac{1}{\sqrt{1-\sin{\theta^\text{max}}}}\right).
\end{equation}
Using $\cos{\theta} = \sqrt{1-\sin^2{\theta}} = 2 p \sin{\psi} \sqrt{1-p^2\sin{\psi}^2}$, we obtain
\begin{eqnarray} \label{d-theta}
\dd\theta = -\dfrac{2p\cos\psi}{\sqrt{1-p^2\sin^2\psi}}  \dd\psi,
\end{eqnarray}
and substituting in \eq \eqref{integral} yields
\begin{eqnarray}\label{beta-L}
\beta L =  \int_{\psi_0}^{\pi/2}\dfrac{\dd\psi}{\sqrt{1 - p^2\sin^2\psi}},
\end{eqnarray}
which can be written as
\begin{eqnarray} \label{eqF}
\beta L = F(p) -F(p,\psi_0),
\end{eqnarray}
where 
\begin{eqnarray}\label{elipticalInt_1stKind}
F(p) = \int_{0}^{\pi/2}\dfrac{1}{\sqrt{1 - p^2\sin^2\psi}} \dd\psi, \quad \text{and } \quad  F(p,\psi_0) = \int_{0}^{\psi_0}\dfrac{1}{\sqrt{1 - p^2\sin^2\psi}} \dd\psi.
\end{eqnarray}
are the complete and incomplete elliptical integrals of the first kind, respectively \citep{Jahnke1945}.
Hence, for a given value of $\theta^\text{max}$, $\beta$ can be determined from \eq\eqref{eqF} using \eqs \eqref{changeVar} and \eqref{psi0}, and the corresponding applied vertical load producing the rotation $\theta^\text{max}$ at the free end is then $N=\bar{Y} I^\text{eff} \beta^2 $.  For a given $N$, the problem is thus solved by the shooting method. 

Using \eq \eqref{t-vector}, the vertical displacement of the rod is
\begin{eqnarray}
r_1 (S) &=& \int_{0}^{S} (1+\zeta)\sin \theta \dd \tilde{S} \approx \int_{0}^{\theta} \dfrac{\sin \theta \dd\theta}{\beta \sqrt{2\left(\sin{\theta^{\text{max}}} -\sin \theta\right)}}  = \int_{\psi_0}^{\psi} \dfrac{2p^2\sin^2\tilde{\psi}-1}{\beta \sqrt{1 - p^2\sin^2\tilde{\psi}}} \dd \tilde{\psi} \nonumber\\
&=& \frac{1}{\beta} \left [F(p,\psi) - F(p,\psi_0)  \right ] + \frac{2}{\beta} \left [\tilde{E}(p,\psi) - \tilde{E}(p,\psi_0)  \right ],
\end{eqnarray}
where
\begin{eqnarray} \label{elipticalInt_2ndKind}
\tilde{E}(p) = \int_{0}^{\pi/2}\sqrt{1 - p^2\sin^2\psi} \dd\psi, \quad \text{and } \quad \tilde{E}(p,\psi_0) = \int_{0}^{\psi_0}\sqrt{1 - p^2\sin^2\psi} \dd\psi,
\end{eqnarray}
are the complete and incomplete elliptical integrals of the second kind, respectively \citep{Jahnke1945}. Thus, the deflection of the rod at its loaded end is 
\begin{eqnarray}
r_1 (L) &=& L+\dfrac{2}{\beta} \left[\tilde{E}(p,\psi_0) -\tilde{E}(p)\right] .
\end{eqnarray}
Finally, the vertical electric field is computed from \eq \eqref{rod-OC-EF} as
\begin{equation}
E_1 (S) =  \dfrac{\mu_\text{T}}{\epsilon} \theta'=  -\dfrac{\beta\mu_\text{T}}{\epsilon} \sqrt{2\left(\sin{\theta} -\sin{\theta^\text{max}}\right)} =  
-\dfrac{\mu_\text{T}}{\epsilon}\sqrt{\dfrac{2N}{\bar{Y}I^\text{eff}}\left(\sin{\theta} -\sin {\theta^\text{max}}\right)} .
\end{equation}
Therefore, the electric field at the fixed end is
\begin{equation}\label{rod-OC-maxEF}
E_1(0) = - \dfrac{\mu_\text{T}}{\epsilon}\sqrt{\dfrac{2N}{\bar{Y}I^\text{eff}}\sin{\vert \theta^\text{max}}\vert}.
\end{equation}

\section{Analytical solutions for displacement and electric field in flexoelectric rods under compressive axial load}\label{appendix:buckling}
Integration of \eq \eqref{EL-moment-buckling-2} yields
\begin{equation}\label{eqn:bending:CC:OC}
\dfrac{1}{2}\left(\dfrac{\dd \theta}{\dd S}\right)^2 - \beta^2 \left(\cos\theta -\cos \theta^\textnormal{max}\right) =0, 
\end{equation}
where we assume upward buckling without loss of generality, and $\theta (L/4) = \theta^\text{max} >0$. Equivalently,
\begin{equation}\label{eqn:dX3-2}
\mathrm{d} S = \dfrac{\mathrm{d} \theta}{\beta \sqrt{2(\cos{\theta}-\cos{\theta}^\text{max})}}.
\end{equation}
Since the right end of the rod is allowed to move horizontally under the action of the compressive load, the length of the rod is assumed to remain approximately unaltered after buckling. Hence, using \eq\eqref{eqn:dX3-2},
\begin{equation}
\frac{L}{4} =  \int\limits_{\theta(0)}^{\theta\left(\frac{L}{4}\right)}  \dfrac{\mathrm{d} S}{\mathrm{d} \theta} \dd \theta =  \displaystyle \int\limits_{0}^{\theta^\text{max}} \dfrac{\mathrm{d}\theta}{\beta \sqrt{2(\cos{\theta}-\cos{\theta}^\text{max})}},
\end{equation}
and thus
\begin{eqnarray}
\frac{\beta L}{4} =\int\limits_{0}^{\theta^\text{max}}\dfrac{1}{2\sqrt{\sin^2 \dfrac{\theta^\text{max}}{2} -\sin^2\dfrac{\theta}{2}}}\dd\theta .
\end{eqnarray}
To compute this integral, we define
\begin{eqnarray}\label{change-of-var-buckling}
\sin \dfrac{\theta^\text{max}}{2} = p, \quad \sin \dfrac{\theta}{2}= p \sin{\psi}, \quad \psi \in [0,\frac{\pi}{2}].
\end{eqnarray}
Hence, 
\begin{eqnarray}\label{buckling-CC-OC-solution}
\beta L = 4F(p) = 4F \left( \sin \dfrac{\theta^\text{max}}{2} \right), 
\end{eqnarray}
where again $F(p)$ is the complete elliptical integral of the first kind, see \eq\eqref{elipticalInt_1stKind}. So, for a given load $N$, $\theta^\text{max}$ is determined by the shooting method, \ie by giving values to $\theta^\text{max}$ and computing the corresponding loading parameter $\beta$ from \eq\eqref{buckling-CC-OC-solution} until the target $\beta= \sqrt{N/\bar{Y}I^\text{eff}}$ is reached.

Similarly, the change in the horizontal displacement, $\Delta r_3$, can be evaluated by the difference of actual length ($L$) and the length projected over axial direction upon buckling as
\begin{eqnarray}\label{buckling-OC-horDisp}
    \Delta r_3 &\approx& L - 4\int\limits_{\theta(0)}^{\theta\left(\frac{L}{4}\right)} \cos{\theta} \dfrac{\mathrm{d} S}{\mathrm{d} \theta} \dd \theta  = L - \int\limits_{0}^{\theta^\text{max}}\dfrac{2\cos{\theta}}{\beta\sqrt{\sin^2 \dfrac{\theta^\text{max}}{2} -\sin^2\dfrac{\theta}{2}}}\dd\theta =  
     L - \dfrac{4}{\beta}\int\limits_{0}^{\pi/2} \dfrac{1-2p^2 \sin^2 \psi}{\sqrt{1-p^2 \sin^2 \psi}} \dd\psi 
    \nonumber \\
    &=& L - \dfrac{8\tilde{E}(p) - 4F(p)}{\beta} = \dfrac{8[F(p) - \tilde{E}(p)]}{\beta}, 
\end{eqnarray}
where we have used $\sqrt{\sin^{2}{(\theta^{\text{max}}/2)} - \sin^{2}{(\theta/2})}  = p \cos{\psi}$, $\cos{\theta}/\cos{(\theta/2)} = \dfrac{1-2p^2 \sin^2 \psi}{\sqrt{1-p^2 \sin^2 \psi}}$, and \eq \eqref{buckling-CC-OC-solution}, and again $\tilde{E}(p)$ is the complete elliptical integral of the second kind, see \eq\eqref{elipticalInt_2ndKind}.

Since, the deformations  in the half-rod are antisymmetric with respect to $S=L/4$, we split the vertical deflection into two parts. Hence, assuming that the rod buckles upwards without loss of generality, 
\begin{subequations}\label{verDisp-OC-buckling}
\begin{eqnarray} 
S \in \left[0,\frac{L}{4}\right] : && r_1 (S) \approx \int_{0}^{S} \sin{\theta (u)} \dd u =\int_{0}^{\theta}  \dfrac{\sin \gamma \dd\gamma}{\beta \sqrt{2\left(\cos\gamma -\cos \theta^\text{max}\right)}} \nonumber\\ &&=  \int_{0}^{\psi}  \dfrac{2p \sin \xi\dd\xi}{\beta} = \dfrac{2p}{\beta}  (1-\cos \psi) , \quad \psi \in [0,\frac{\pi}{2}] \label{verDisp-OC-buckling-1}
\end{eqnarray}
\begin{eqnarray}
S \in \left[\frac{L}{4},\frac{L}{2}\right] : && r_1 (S) \approx \frac{2p}{\beta} - \int_{\theta^\text{max}}^{\theta} \dfrac{\sin \gamma \dd\gamma}{\beta \sqrt{2\left(\cos\gamma -\cos \theta^\text{max}\right)}} \nonumber\\ &&= \frac{2p}{\beta} + \int_{\frac{\pi}{2}}^{\psi}  \dfrac{2p \sin{\left(\frac{\pi}{2}-\xi\right)}\dd\xi}{\beta}  = \dfrac{2p}{\beta}\left(1+\cos{\left(\frac{\pi}{2}-\psi\right)}\right), \quad \psi \in [0,\frac{\pi}{2}],
\label{verDisp-OC-buckling-2}
\end{eqnarray}
\begin{eqnarray}
\end{eqnarray}
\end{subequations}
where we have used $\sqrt{\sin^{2}{(\theta^{\text{max}}/2)} - \sin^{2}{(\theta/2})}  = p \cos{\psi}$, and $\sin{\theta}/\cos{(\theta/2)} = 2 p \sin{\psi}$.  
Finally, the electric field can be evaluated as
\begin{subequations}\label{EF-CC-buckling}
\begin{align}\label{EF-CC-buckling-1}
S \in \left[0,\frac{L}{4}\right] ~~:~~ E_1 (S) &= \dfrac{\mu_\text{T}}{\epsilon} \theta'=   \dfrac{\mu_\text{T}\beta}{\epsilon} \sqrt{2\left(\cos\theta -\cos \theta^\text{max}\right)} \nonumber\\ &=\dfrac{\mu_\text{T}}{\epsilon}\sqrt{\dfrac{2N}{\bar{Y}I^\text{eff}} \left(\cos\theta -\cos \theta^\text{max}\right)}, \quad \theta \in \left[ 0,\theta^\text{max} \right ]
\\
\label{EF-CC-buckling-2}
S \in \left[\frac{L}{4},\frac{L}{2}\right] ~~:~~ E_1 (S)&= -\dfrac{\mu_\text{T}}{\epsilon}\sqrt{\dfrac{2N}{\bar{Y}I^\text{eff}}\left(\cos{\left(\theta\left(\frac{L}{2}-S\right)\right)} -\cos \theta^\text{max}\right)}, \quad \theta \in \left[ 0,\theta^\text{max} \right ] 
\end{align}
\end{subequations}
Therefore, the vertical deflection and electric field at the center of the rod are
\begin{eqnarray}
r_1\left(\dfrac{L}{2}\right) &=& \dfrac{4}{\beta} \sin{\left(\dfrac{\theta^{\text{max}}}{2}\right)},\\  
E_1 (0) = -E_1\left(\dfrac{L}{2}\right)  = E_1\left(L\right)  &=&  \dfrac{\mu_\text{T}}{\epsilon}\sqrt{\dfrac{2N}{\bar{Y}I^\text{eff}}\left(1 -\cos \theta^\text{max}\right)}.
\end{eqnarray}

\section{Analytical solutions for displacement and voltage in flexoelectric rods  under transversal voltage actuation}\label{appendix:buckling_CC}
Similarly to \ref{appendix:buckling}, integration of the moment balance \eq\eqref{momentbalance-final-cc} yields
\begin{equation}\label{eqn:buckling:CC:CC}
\dfrac{1}{2}\left(\dfrac{\dd \theta}{\dd S}\right)^2 - (1+\zeta) \ \tilde{\beta}^2 \left(\cos\theta -\cos \theta^\textnormal{max}\right) =0,
\end{equation}
where $\theta (L/4) = \theta^\text{max}$ and upon integration
\begin{equation}
 \frac{L}{4} = \int_{0}^{L/4}  \ \mathrm{d} S  = \int\limits_{\theta(0)}^{\theta\left(\frac{L}{4}\right)}  \dfrac{\mathrm{d} S}{\mathrm{d} \theta} \dd \theta =  \dfrac{1}{\sqrt{1+\zeta}} \displaystyle \int\limits_{0}^{\theta^\text{max}} \dfrac{\mathrm{d}\theta}{\tilde{\beta} \sqrt{2(\cos{\theta}-\cos{\theta}^\text{max})}} = \dfrac{F(p)}{\tilde{\beta}\sqrt{1+\zeta}}.
\end{equation}
Thus
\begin{eqnarray}\label{eqn:buckling:CC:CC:UL}
\tilde{\beta} \sqrt{1+\zeta} L = 4F(p) = 4F \left( \sin \dfrac{\theta^\text{max}}{2} \right).
\end{eqnarray}
In this case, the right end of the rod is clamped and thus the length of the rod after buckling is unknown, but its projection on the horizontal axis is the undeformed length $L$, therefore with the help of constraint \eq \eqref{constraint-CC-buckling} 
\begin{eqnarray}\label{eqn:buckling:CC:CC:PL}
\dfrac{L}{4} = \int_{0}^{L/4}\dd r_3 = \int\limits_{0}^{\theta^\text{max}} (1+\zeta)\cos{\theta} \dfrac{\dd S}{\dd \theta} \dd \theta = \dfrac{\sqrt{(1+\zeta)}\left[2\tilde{E}(p) - F(p)\right]}{\tilde{\beta}}.
\end{eqnarray}
Therefore, by using \eqs \eqref{eqn:buckling:CC:CC:UL} and \eqref{eqn:buckling:CC:CC:PL}
\begin{eqnarray}\label{eqn:strain:CC}
\zeta = \dfrac{F(p)}{2\tilde{E}(p) -F(p)} -1.
\end{eqnarray}
Once $\zeta$ is known, $\tilde{\beta}$ can be evaluated using \eq \eqref{eqn:buckling:CC:CC:UL} for any $\theta^{\text{max}}$.

Now, similar to \ref{appendix:buckling},
\begin{align}
S \in \left[0,\frac{{L}}{4}\right] ~~:~~ r_1(S) &= -\int_{0}^{\theta}  \dfrac{ (1+\zeta)\sin \theta \dd\theta}{\tilde{\beta} \sqrt{1+\zeta} \sqrt{2\left(\cos\theta -\cos \theta^\text{max}\right)}} \nonumber\\&=  -\int_{0}^{\psi}  \dfrac{2p\sqrt{1+\zeta} \sin \psi\dd\psi}{\tilde{\beta}} =-\dfrac{2p\sqrt{1+\zeta} (1-\cos \psi) }{\tilde{\beta}},\nonumber\\
S \in \left[\frac{{L}}{4},\frac{{L}}{2}\right] ~~:~~  r_1(S) &=-\dfrac{\sqrt{1+\zeta}}{\tilde{\beta} }\left(\int_{0}^{\theta^\text{max}}  \dfrac{ \sin \theta \dd\theta}{\sqrt{2\left(\cos\theta -\cos \theta^\text{max}\right)}} - \int_{\theta^\text{max}}^{\theta} \dfrac{\sin \theta \dd\theta}{ \sqrt{2\left(\cos\theta -\cos \theta^\text{max}\right)}} \right)\nonumber\\&= -\dfrac{2p\sqrt{1+\zeta} (1+\cos \psi) }{\tilde{\beta}},\nonumber\\
\end{align}
with $\sin \dfrac{\theta}{2}= p \sin{\psi}$. 
Hence, the deflection at the center of the rod  and the curvature at the left end for downward buckling are
\begin{eqnarray}
r_1\left(\dfrac{{L}}{2}\right) &=& -\dfrac{4p\sqrt{1+\zeta}}{\tilde{\beta}}, \label{verDisp-CC-buckling}\\
\theta' (0) &=& \tilde{\beta} \sqrt{2 (1+\zeta)(1-\cos{\theta^\text{max}})}. \label{curvature-CC-buckling}
\end{eqnarray}

\bibliography{References}

\begin{thebibliography}{134}
\providecommand{\natexlab}[1]{#1}
\providecommand{\url}[1]{\texttt{#1}}
\expandafter\ifx\csname urlstyle\endcsname\relax
  \providecommand{\doi}[1]{doi: #1}\else
  \providecommand{\doi}{doi: \begingroup \urlstyle{rm}\Url}\fi

\bibitem[Abdollahi and Arias(2015)]{abdollahi2015constructive}
Amir Abdollahi and Irene Arias.
\newblock Constructive and destructive interplay between piezoelectricity and
  flexoelectricity in flexural sensors and actuators.
\newblock \emph{Journal of Applied Mechanics}, 82\penalty0 (12), 2015.
\newblock URL \url{https://doi.org/10.1115/1.4031333}.

\bibitem[Abdollahi et~al.(2014)Abdollahi, Peco, Mill\'an, Arroyo, and
  Arias]{Abdollahi2014}
Amir Abdollahi, Christian Peco, Daniel Mill\'an, Marino Arroyo, and Irene
  Arias.
\newblock Computational evaluation of the flexoelectric effect in dielectric
  solids.
\newblock \emph{Journal of Applied Physics}, 116\penalty0 (9):\penalty0 093502,
  2014.
\newblock URL \url{https://doi.org/10.1063/1.4893974}.

\bibitem[Abdollahi et~al.(2015{\natexlab{a}})Abdollahi, Mill\'an, Peco, Arroyo,
  and Arias]{Abdollahi2015a}
Amir Abdollahi, Daniel Mill\'an, Christian Peco, Marino Arroyo, and Irene
  Arias.
\newblock Revisiting pyramid compression to quantify flexoelectricity: A
  three-dimensional simulation study.
\newblock \emph{Phys. Rev. B}, 91:\penalty0 104103, 2015{\natexlab{a}}.
\newblock URL \url{https://doi.org/10.1103/PhysRevB.91.104103}.

\bibitem[Abdollahi et~al.(2015{\natexlab{b}})Abdollahi, Peco, Mill\'an, Arroyo,
  Catalan, and Arias]{Abdollahi2015b}
Amir Abdollahi, Christian Peco, Daniel Mill\'an, Marino Arroyo, Gustau Catalan,
  and Irene Arias.
\newblock Fracture toughening and toughness asymmetry induced by
  flexoelectricity.
\newblock \emph{Phys. Rev. B}, 92:\penalty0 094101, 2015{\natexlab{b}}.
\newblock URL \url{https://doi.org/10.1103/PhysRevB.92.094101}.

\bibitem[Ahmadpoor and Sharma(2015)]{Ahmadpoor2015}
F.~Ahmadpoor and P.~Sharma.
\newblock Flexoelectricity in two-dimensional crystalline and biological
  membranes.
\newblock \emph{Nanoscale}, 7:\penalty0 16555, 2015.

\bibitem[Ahmadpoor et~al.(2013)Ahmadpoor, Deng, Liu, and Sharma]{Ahmadpoor2013}
F.~Ahmadpoor, Q.~Deng, L.~P. Liu, and P.~Sharma.
\newblock Apparent flexoelectricity in lipid bilayer membranes due to external
  charge and dipolar distributions.
\newblock \emph{Physical Review E}, 88:\penalty0 050701, 2013.

\bibitem[Altan and Aifantis(1997)]{Altan1997}
BS~Altan and EC~Aifantis.
\newblock On some aspects in the special theory of gradient elasticity.
\newblock \emph{Journal of the Mechanical Behavior of Materials}, 8\penalty0
  (3):\penalty0 231--282, 1997.
\newblock URL \url{https://doi.org/10.1515/JMBM.1997.8.3.231}.

\bibitem[Anqing et~al.(2015)Anqing, Shenjie, Lu, and
  Xi]{anqing2015flexoelectric}
Li~Anqing, Zhou Shenjie, Qi~Lu, and Chen Xi.
\newblock A flexoelectric theory with rotation gradient effects for elastic
  dielectrics.
\newblock \emph{Modelling and Simulation in Materials Science and Engineering},
  24\penalty0 (1):\penalty0 015009, 2015.
\newblock URL \url{https://doi.org/10.1088/0965-0393/24/1/015009}.

\bibitem[Antman(1995)]{Antman1995}
S.S. Antman.
\newblock \emph{Nonlinear Problems of Elasticity}.
\newblock Springer-Verlag, New York, 1995.
\newblock ISBN 9780387276496.
\newblock URL \url{https://doi.org/10.1007/0-387-27649-1}.

\bibitem[Barbero et~al.(1986)Barbero, Dozov, Palierne, and Durand]{Barbero1986}
G.~Barbero, I.~Dozov, J.~F. Palierne, and G.~Durand.
\newblock Order electricity and surface orientation in nematic liquid crystals.
\newblock \emph{Physical Review Letters}, 56\penalty0 (19):\penalty0
  2056--2059, 1986.

\bibitem[Baroudi and Najar(2019)]{baroudi2019dynamic}
S~Baroudi and F~Najar.
\newblock Dynamic analysis of a nonlinear nanobeam with flexoelectric
  actuation.
\newblock \emph{Journal of Applied Physics}, 125\penalty0 (4):\penalty0 044503,
  2019.
\newblock URL \url{https://doi.org/10.1063/1.5057727}.

\bibitem[Baskaran et~al.(2011)Baskaran, He, Chen, and Fu]{baskaran2011}
Sivapalan Baskaran, Xiangtong He, Qin Chen, and John~Y Fu.
\newblock Experimental studies on the direct flexoelectric effect in
  $\alpha$-phase polyvinylidene fluoride films.
\newblock \emph{Applied Physics Letters}, 98\penalty0 (24):\penalty0 242901,
  2011.
\newblock URL \url{https://doi.org/10.1063/1.3599520}.

\bibitem[Baskaran et~al.(2012)Baskaran, He, Wang, and Fu]{baskaran2012}
Sivapalan Baskaran, Xiangtong He, Yu~Wang, and John~Y Fu.
\newblock Strain gradient induced electric polarization in $\alpha$-phase
  polyvinylidene fluoride films under bending conditions.
\newblock \emph{Journal of Applied Physics}, 111\penalty0 (1):\penalty0 014109,
  2012.
\newblock URL \url{https://doi.org/10.1063/1.3673817}.

\bibitem[Bauer and Bauer(2008)]{bauer2008piezoelectric}
S~Bauer and F~Bauer.
\newblock Piezoelectric polymers and their applications.
\newblock In \emph{Piezoelectricity}, pages 157--177. Springer, 2008.
\newblock URL \url{https://doi.org/10.1007/978-3-540-68683-5_6}.

\bibitem[Bisshopp and Drucker(1945)]{Bisshopp1945}
KE~Bisshopp and DC~Drucker.
\newblock Large deflection of cantilever beams.
\newblock \emph{Quarterly of Applied Mathematics}, 3\penalty0 (3):\penalty0
  272--275, 1945.
\newblock URL \url{www.jstor.org/stable/43633516}.

\bibitem[Breger et~al.(1976)Breger, Furukawa, and Fukada]{breger1976bending}
Lance Breger, Takeo Furukawa, and Eiichi Fukada.
\newblock Bending piezoelectricity in polyvinylidene fluoride.
\newblock \emph{Japanese Journal of Applied Physics}, 15\penalty0
  (11):\penalty0 2239, 1976.
\newblock URL \url{https://doi.org/10.1143/JJAP.15.2239}.

\bibitem[Bursian and Trunov(1974)]{Bursian1974}
{\'E}.~V. Bursian and N.~N. Trunov.
\newblock Nonlocal piezoelectric effect.
\newblock \emph{Sov. Phys. Solid State}, 16\penalty0 (4):\penalty0 760 -- 762,
  1974.
\newblock URL
  \url{https://www.tib.eu/de/suchen/id/tema-archive\%3ATEMAE75020089214}.

\bibitem[Bursian and Zaikovskii(1968)]{bursian1968}
J~M Bursian and O~I Zaikovskii.
\newblock {Changes in curvature of a ferroelectric film due to polarization}.
\newblock \emph{Soviet Physics Solid State}, 10\penalty0 (5):\penalty0
  1121--1124, 1968.

\bibitem[{\v C}epi{\v c} et~al.(2000){\v C}epi{\v c}, Rov{\v s}ek, and {\v
  Z}ek{\v s}]{Cepic2000}
Mojca {\v C}epi{\v c}, Barbara Rov{\v s}ek, and Bo{\v s}tjan {\v Z}ek{\v s}.
\newblock Flexoelectrically induced polarization in polar smectic films.
\newblock \emph{Ferroelectrics}, 244\penalty0 (1):\penalty0 59--66, 2000.

\bibitem[Chu and Salem(2012)]{chu2012}
Baojin Chu and DR~Salem.
\newblock Flexoelectricity in several thermoplastic and thermosetting polymers.
\newblock \emph{Applied Physics Letters}, 101\penalty0 (10):\penalty0 103905,
  2012.
\newblock URL \url{https://doi.org/10.1063/1.4750064}.

\bibitem[Codony et~al.(2019)Codony, Marco, Fern{\'a}ndez-M{\'e}ndez, and
  Arias]{codony2019immersed}
David Codony, Onofre Marco, Sonia Fern{\'a}ndez-M{\'e}ndez, and Irene Arias.
\newblock An immersed boundary hierarchical b-spline method for
  flexoelectricity.
\newblock \emph{Computer Methods in Applied Mechanics and Engineering},
  354:\penalty0 750--782, 2019.
\newblock URL \url{https://doi.org/10.1016/j.cma.2019.05.036}.

\bibitem[de~Boor(2001)]{deBoor2001}
C.~de~Boor.
\newblock \emph{A Practical Guide to Splines}.
\newblock Applied Mathematical Sciences. Springer New York, 2001.
\newblock ISBN 9780387953663.
\newblock URL \url{10.1002/zamm.19800600129}.

\bibitem[de~Gennes and Prost(1993)]{Gennes1993}
P.~G. de~Gennes and J.~Prost.
\newblock \emph{The Physics of Liquid Crystals}.
\newblock Number~83 in International Series of Monographs on Physics. Oxford
  Science Publications, second edition edition, 1993.

\bibitem[Deng et~al.(2017)Deng, Deng, Yu, and Shen]{deng2017mixed}
Feng Deng, Qian Deng, Wenshan Yu, and Shengping Shen.
\newblock Mixed finite elements for flexoelectric solids.
\newblock \emph{Journal of Applied Mechanics}, 84\penalty0 (8), 2017.
\newblock URL \url{https://doi.org/10.1115/1.4036939}.

\bibitem[Deng et~al.(2018)Deng, Deng, and Shen]{deng2018three}
Feng Deng, Qian Deng, and Shengping Shen.
\newblock A three-dimensional mixed finite element for flexoelectricity.
\newblock \emph{Journal of Applied Mechanics}, 85\penalty0 (3), 2018.
\newblock URL \url{https://doi.org/10.1115/1.4038919}.

\bibitem[Deng et~al.(2014{\natexlab{a}})Deng, Kammoun, Erturk, and
  Sharma]{deng2014nanoscale}
Qian Deng, Mejdi Kammoun, Alper Erturk, and Pradeep Sharma.
\newblock Nanoscale flexoelectric energy harvesting.
\newblock \emph{International Journal of Solids and Structures}, 51\penalty0
  (18):\penalty0 3218--3225, 2014{\natexlab{a}}.
\newblock URL \url{https://doi.org/10.1016/j.ijsolstr.2014.05.018}.

\bibitem[Deng et~al.(2014{\natexlab{b}})Deng, Liu, and
  Sharma]{Deng2014electrets}
Qian Deng, Liping Liu, and Pradeep Sharma.
\newblock Electrets in soft materials: Nonlinearity, size effects, and giant
  electromechanical coupling.
\newblock \emph{Physical Review E}, 90\penalty0 (1):\penalty0 012603,
  2014{\natexlab{b}}.
\newblock URL \url{https://doi.org/10.1103/PhysRevE.90.012603}.

\bibitem[Deng et~al.(2014{\natexlab{c}})Deng, Liu, and
  Sharma]{Deng2014flexoelectricity}
Qian Deng, Liping Liu, and Pradeep Sharma.
\newblock Flexoelectricity in soft materials and biological membranes.
\newblock \emph{Journal of the Mechanics and Physics of Solids}, 62:\penalty0
  209--227, 2014{\natexlab{c}}.
\newblock URL \url{https://doi.org/10.1016/j.jmps.2013.09.021}.

\bibitem[Derzhanski et~al.(1990)Derzhanski, Petrov, Todorov, and
  Hristova]{Derzhanski1990}
A.~Derzhanski, A.~G. Petrov, A.~T. Todorov, and K.~Hristova.
\newblock Flexoelectricity of lipid bilayers.
\newblock \emph{Liquid Crystals}, 7\penalty0 (3):\penalty0 439--449, 1990.

\bibitem[Devonshire(1949)]{Devonshire1949}
A.~F. Devonshire.
\newblock Theory of barium titanate. part i.
\newblock \emph{Philosophical Magazine}, 40:\penalty0 1040, 1949.

\bibitem[Devonshire(1951)]{Devonshire1951}
A.~F. Devonshire.
\newblock Theory of barium titanate. part ii.
\newblock \emph{Philosophical Magazine}, 42:\penalty0 1065, 1951.

\bibitem[Devonshire(1954)]{Devonshire1954}
A.~F. Devonshire.
\newblock Theory of ferroelectrics.
\newblock \emph{A quarterly Supplement of the Philosophical Magazine},
  3\penalty0 (10):\penalty0 85--130, 1954.

\bibitem[Dorfmann and Ogden(2005)]{Dorfmann2005}
A~Dorfmann and RW~Ogden.
\newblock Nonlinear electroelasticity.
\newblock \emph{Acta Mechanica}, 174\penalty0 (3-4):\penalty0 167--183, 2005.
\newblock URL \url{https://doi.org/10.1007/s00707-004-0202-2}.

\bibitem[Dorfmann and Ogden(2014)]{dorfmann2014nonlinear}
Luis Dorfmann and Ray~W Ogden.
\newblock \emph{Nonlinear theory of electroelastic and magnetoelastic
  interactions}, volume~1.
\newblock Springer, 2014.
\newblock URL \url{https://doi.org/10.1007/978-1-4614-9596-3}.

\bibitem[Dorfmann and Ogden(2017)]{dorfmann2017nonlinear}
Luis Dorfmann and Ray~W Ogden.
\newblock Nonlinear electroelasticity: material properties, continuum theory
  and applications.
\newblock \emph{Proceedings of the Royal Society A: Mathematical, Physical and
  Engineering Sciences}, 473\penalty0 (2204):\penalty0 20170311, 2017.
\newblock URL \url{https://doi.org/10.1098/rspa.2017.0311}.

\bibitem[Fu et~al.(2006)Fu, Zhu, Li, and Cross]{fu2006experimental}
John~Y Fu, Wenyi Zhu, Nan Li, and L~Eric Cross.
\newblock Experimental studies of the converse flexoelectric effect induced by
  inhomogeneous electric field in a barium strontium titanate composition.
\newblock \emph{Journal of Applied Physics}, 100\penalty0 (2):\penalty0 024112,
  2006.
\newblock URL \url{https://doi.org/10.1063/1.2219990}.

\bibitem[Gao et~al.(2008)Gao, Feng, Yin, and Gao]{Gao2008}
L.T. Gao, X-Q Feng, Y-J Yin, and H.~Gao.
\newblock An electromechanical liquid crystal model of vesicles.
\newblock \emph{Journal of the Mechanics and Physics of Solids}, 56:\penalty0
  2844--2862, 2008.

\bibitem[Ghasemi et~al.(2017)Ghasemi, Park, and Rabczuk]{ghasemi2017level}
Hamid Ghasemi, Harold~S Park, and Timon Rabczuk.
\newblock A level-set based iga formulation for topology optimization of
  flexoelectric materials.
\newblock \emph{Computer Methods in Applied Mechanics and Engineering},
  313:\penalty0 239--258, 2017.
\newblock URL \url{https://doi.org/10.1016/j.cma.2016.09.029}.

\bibitem[Ghasemi et~al.(2018)Ghasemi, Park, and Rabczuk]{ghasemi2018multi}
Hamid Ghasemi, Harold~S Park, and Timon Rabczuk.
\newblock A multi-material level set-based topology optimization of
  flexoelectric composites.
\newblock \emph{Computer Methods in Applied Mechanics and Engineering},
  332:\penalty0 47--62, 2018.
\newblock URL \url{https://doi.org/10.1016/j.cma.2017.12.005}.

\bibitem[Gupta and Kumar(2017)]{gupta2017effect}
Prakhar Gupta and Ajeet Kumar.
\newblock Effect of material nonlinearity on spatial buckling of nanorods and
  nanotubes.
\newblock \emph{Journal of Elasticity}, 126\penalty0 (2):\penalty0 155--171,
  2017.
\newblock URL \url{https://doi.org/10.1007/s10659-016-9586-1}.

\bibitem[Hadjesfandiari(2013)]{hadjesfandiari2013size}
Ali~R Hadjesfandiari.
\newblock Size-dependent piezoelectricity.
\newblock \emph{International Journal of Solids and Structures}, 50\penalty0
  (18):\penalty0 2781--2791, 2013.
\newblock URL \url{https://doi.org/10.1016/j.ijsolstr.2013.04.020}.

\bibitem[Hamdia et~al.(2018)Hamdia, Ghasemi, Zhuang, Alajlan, and
  Rabczuk]{hamdia2018sensitivity}
Khader~M Hamdia, Hamid Ghasemi, Xiaoying Zhuang, Naif Alajlan, and Timon
  Rabczuk.
\newblock Sensitivity and uncertainty analysis for flexoelectric
  nanostructures.
\newblock \emph{Computer Methods in Applied Mechanics and Engineering},
  337:\penalty0 95--109, 2018.
\newblock URL \url{https://doi.org/10.1016/j.cma.2018.03.016}.

\bibitem[Harden et~al.(2006)Harden, Mbanga, {\'E}ber, Fodor-Csorba, Sprunt,
  Gleeson, and Jakli]{harden2006giant}
John Harden, Badel Mbanga, Nandor {\'E}ber, Katalin Fodor-Csorba, Samuel
  Sprunt, James~T Gleeson, and Antal Jakli.
\newblock Giant flexoelectricity of bent-core nematic liquid crystals.
\newblock \emph{Physical review letters}, 97\penalty0 (15):\penalty0 157802,
  2006.
\newblock URL \url{https://doi.org/10.1103/PhysRevLett.97.157802}.

\bibitem[Hong and Vanderbilt(2011)]{hong2011first}
Jiawang Hong and David Vanderbilt.
\newblock First-principles theory of frozen-ion flexoelectricity.
\newblock \emph{Physical Review B}, 84\penalty0 (18):\penalty0 180101, 2011.
\newblock URL \url{https://doi.org/10.1103/PhysRevB.84.180101}.

\bibitem[Hu and Shen(2010)]{hu2010}
ShuLing Hu and ShengPing Shen.
\newblock Variational principles and governing equations in nano-dielectrics
  with the flexoelectric effect.
\newblock \emph{Science China Physics, Mechanics and Astronomy}, 53\penalty0
  (8):\penalty0 1497--1504, 2010.
\newblock URL \url{https://doi.org/10.1007/s11433-010-4039-5}.

\bibitem[Huang et~al.(2018)Huang, Qi, Huang, Shu, Zhou, and
  Jiang]{huang2018flexoelectricity}
Shujin Huang, Lu~Qi, Wenbin Huang, Longlong Shu, Shenjie Zhou, and Xiaoning
  Jiang.
\newblock Flexoelectricity in dielectrics: Materials, structures and
  characterizations.
\newblock \emph{Journal of Advanced Dielectrics}, 8\penalty0 (02):\penalty0
  1830002, 2018.
\newblock URL \url{https://doi.org/10.1142/S2010135X18300025}.

\bibitem[Indenbom et~al.(1981{\natexlab{a}})Indenbom, Loginov, and
  Osipov]{indenbom1981}
VL~Indenbom, EB~Loginov, and MA~Osipov.
\newblock Flexoelectric effect and structure of crystals.
\newblock \emph{Kristallografiya}, 28:\penalty0 1157--1162, 1981{\natexlab{a}}.

\bibitem[Indenbom et~al.(1981{\natexlab{b}})Indenbom, Loginov, and
  Osipov]{indenbom1981_}
VL~Indenbom, EB~Loginov, and MA~Osipov.
\newblock Flexoelectric effect and crystal-structure.
\newblock \emph{Kristallografiya}, 26\penalty0 (6):\penalty0 1157--1162,
  1981{\natexlab{b}}.

\bibitem[Jahnke(1945)]{Jahnke1945}
Eugene Jahnke.
\newblock Tables of functions with formulae and curves.
\newblock \emph{New York: Dover Publications,| c1945, 4th ed.}, 1945.
\newblock URL \url{https://ui.adsabs.harvard.edu/abs/1945tfwf.book.....J}.

\bibitem[Jewell(2011)]{Jewell2011}
S.~A. Jewell.
\newblock Living systems and liquid crystals.
\newblock \emph{Liquid Crystals}, 38\penalty0 (11-12):\penalty0 1699--1714,
  2011.

\bibitem[Jiang et~al.(2013)Jiang, Huang, and Zhang]{jiang2013flexoelectric}
Xiaoning Jiang, Wenbin Huang, and Shujun Zhang.
\newblock Flexoelectric nano-generator: Materials, structures and devices.
\newblock \emph{Nano Energy}, 2\penalty0 (6):\penalty0 1079--1092, 2013.
\newblock URL \url{https://doi.org/10.1016/j.nanoen.2013.09.001}.

\bibitem[Kogan(1964)]{kogan1964piezoelectric}
Sh~M Kogan.
\newblock Piezoelectric effect during inhomogeneous deformation and acoustic
  scattering of carriers in crystals.
\newblock \emph{Soviet Physics-Solid State}, 5\penalty0 (10):\penalty0
  2069--2070, 1964.

\bibitem[Krichen and Sharma(2016)]{Krichen2016}
Sana Krichen and Pradeep Sharma.
\newblock Flexoelectricity: a perspective on an unusual electromechanical
  coupling.
\newblock \emph{Journal of Applied Mechanics}, 83\penalty0 (3), 2016.
\newblock URL \url{https://doi.org/10.1115/1.4032378}.

\bibitem[Kuczynski and Hoffmann(2005)]{Kuczynski2005}
W.~Kuczynski and J.~Hoffmann.
\newblock Determination of piezoelectric and flexoelectric polarization in
  ferroelectric liquid crystals.
\newblock \emph{Physical Review E}, 72\penalty0 (4):\penalty0 041701, 2005.

\bibitem[Lagerwall and Dahl(1984)]{Lagerwall1984}
S.T. Lagerwall and I.~Dahl.
\newblock Ferroelectric liquid crystals.
\newblock \emph{Molecular Crystals and Liquid Crystals}, 114\penalty0
  (1-3):\penalty0 151--187, 1984.

\bibitem[Landau and Lifshitz(2013)]{landau2013course}
Lev~Davidovich Landau and Evgenii~Mikhailovich Lifshitz.
\newblock \emph{Course of theoretical physics}.
\newblock Elsevier, 2013.
\newblock URL \url{https://books.google.es/books?id=LuBbAwAAQBAJ}.

\bibitem[Lax and Nelson(1976)]{Lax1976}
M~Lax and DF~Nelson.
\newblock Maxwell equations in material form.
\newblock \emph{Physical Review B}, 13\penalty0 (4):\penalty0 1777, 1976.
\newblock URL \url{https://doi.org/10.1103/PhysRevB.13.1777}.

\bibitem[Le~Quang and He(2011)]{LeQuang2011}
H.~Le~Quang and Q.-C. He.
\newblock The number and types of all possible rotational symmetries for
  flexoelectric tensors.
\newblock \emph{Proceedings of the Royal Society of London A: Mathematical,
  Physical and Engineering Sciences}, 467\penalty0 (2132):\penalty0 2369--2386,
  2011.
\newblock ISSN 1364-5021.
\newblock URL \url{https://doi.org/10.1098/rspa.2010.0521}.

\bibitem[Liang et~al.(2014)Liang, Hu, and Shen]{liang2014effects}
Xu~Liang, Shuling Hu, and Shengping Shen.
\newblock Effects of surface and flexoelectricity on a piezoelectric nanobeam.
\newblock \emph{Smart materials and structures}, 23\penalty0 (3):\penalty0
  035020, 2014.
\newblock URL \url{https://doi.org/10.1088/0964-1726/23/3/035020}.

\bibitem[Lifshitz and Landau(1951)]{LandauLifshitz1951}
EM~Lifshitz and LD~Landau.
\newblock Statistical physics (course of theoretical physics, volume 5), 1951.

\bibitem[Lin and Chiao(1998)]{Lin1998}
Liwei Lin and Mu~Chiao.
\newblock Electro, thermal and elastic characterizations of suspended micro
  beams.
\newblock \emph{Microelectronics journal}, 29\penalty0 (4-5):\penalty0
  269--276, 1998.
\newblock URL \url{https://doi.org/10.1016/S0026-2692(97)00066-9}.

\bibitem[Lines and Glass(1979)]{Lines1979}
M.E. Lines and A.M. Glass.
\newblock \emph{Principles and applications of ferroelectrics and related
  materials}.
\newblock Oxford University Press, 1979.

\bibitem[Liu(2014)]{Liu2014}
Liping Liu.
\newblock An energy formulation of continuum magneto-electro-elasticity with
  applications.
\newblock \emph{Journal of the Mechanics and Physics of Solids}, 63:\penalty0
  451--480, 2014.
\newblock URL \url{https://doi.org/10.1016/j.jmps.2013.08.001}.

\bibitem[Liu and Sharma(2013)]{Liu2013}
L.P. Liu and P.~Sharma.
\newblock Flexoelectricity and thermal fluctuations of lipid bilayer membranes:
  renormalization of flexoelectric, dielectric, and elastic properties.
\newblock \emph{Physical Review E}, 87:\penalty0 032715, 2013.

\bibitem[Ma and Cross(2001{\natexlab{a}})]{ma2001large}
Wenhui Ma and L~Eric Cross.
\newblock Large flexoelectric polarization in ceramic lead magnesium niobate.
\newblock \emph{Applied Physics Letters}, 79\penalty0 (26):\penalty0
  4420--4422, 2001{\natexlab{a}}.
\newblock URL \url{https://doi.org/10.1063/1.1426690}.

\bibitem[Ma and Cross(2001{\natexlab{b}})]{ma2001observation}
Wenhui Ma and L~Eric Cross.
\newblock Observation of the flexoelectric effect in relaxor {Pb (Mg ${}_{1/3}$
  Nb ${}_{2/3}$) O${}_3$} ceramics.
\newblock \emph{Applied Physics Letters}, 78\penalty0 (19):\penalty0
  2920--2921, 2001{\natexlab{b}}.
\newblock URL \url{https://doi.org/10.1063/1.1356444}.

\bibitem[Ma and Cross(2002)]{ma2002flexoelectric}
Wenhui Ma and L~Eric Cross.
\newblock Flexoelectric polarization of barium strontium titanate in the
  paraelectric state.
\newblock \emph{Applied Physics Letters}, 81\penalty0 (18):\penalty0
  3440--3442, 2002.
\newblock URL \url{https://doi.org/10.1063/1.1518559}.

\bibitem[Ma and Cross(2003)]{ma2003strain}
Wenhui Ma and L~Eric Cross.
\newblock Strain-gradient-induced electric polarization in lead zirconate
  titanate ceramics.
\newblock \emph{Applied Physics Letters}, 82\penalty0 (19):\penalty0
  3293--3295, 2003.
\newblock URL \url{https://doi.org/10.1063/1.1570517}.

\bibitem[Ma and Cross(2005)]{ma2005flexoelectric}
Wenhui Ma and L~Eric Cross.
\newblock Flexoelectric effect in ceramic lead zirconate titanate.
\newblock \emph{Applied Physics Letters}, 86\penalty0 (7):\penalty0 072905,
  2005.
\newblock URL \url{https://doi.org/10.1063/1.1868078}.

\bibitem[Ma and Cross(2006)]{ma2006flexoelectricity}
Wenhui Ma and L~Eric Cross.
\newblock Flexoelectricity of barium titanate.
\newblock \emph{Applied Physics Letters}, 88\penalty0 (23):\penalty0 232902,
  2006.
\newblock URL \url{https://doi.org/10.1063/1.2211309}.

\bibitem[Majdoub et~al.(2009)Majdoub, Sharma, and \ifmmode \mbox{\c{C}}\else
  \c{C}\fi{}a\ifmmode~\breve{g}\else \u{g}\fi{}in]{Majdoub2009}
M.~S. Majdoub, P.~Sharma, and T.~\ifmmode \mbox{\c{C}}\else
  \c{C}\fi{}a\ifmmode~\breve{g}\else \u{g}\fi{}in.
\newblock Erratum: Enhanced size-dependent piezoelectricity and elasticity in
  nanostructures due to the flexoelectric effect [phys. rev. b 77, 125424
  (2008)].
\newblock \emph{Phys. Rev. B}, 79:\penalty0 119904, 2009.
\newblock \doi{10.1103/PhysRevB.79.119904}.
\newblock URL \url{https://doi.org/10.1103/PhysRevB.79.119904}.

\bibitem[Majdoub et~al.(2008)Majdoub, Sharma, and \ifmmode \mbox{\c{C}}\else
  \c{C}\fi{}a\ifmmode~\breve{g}\else \u{g}\fi{}in]{Majdoub2008}
MS~Majdoub, P~Sharma, and T~\ifmmode \mbox{\c{C}}\else
  \c{C}\fi{}a\ifmmode~\breve{g}\else \u{g}\fi{}in.
\newblock Enhanced size-dependent piezoelectricity and elasticity in
  nanostructures due to the flexoelectric effect.
\newblock \emph{Physical Review B}, 77\penalty0 (12):\penalty0 125424, 2008.
\newblock URL \url{https://doi.org/10.1103/PhysRevB.77.125424}.

\bibitem[Mao and Purohit(2014)]{Mao2014}
Sheng Mao and Prashant~K. Purohit.
\newblock Insights into flexoelectric solids from strain-gradient elasticity.
\newblock \emph{ASME Journal of Applied Mechanics}, 81\penalty0 (8):\penalty0
  1--10, 2014.
\newblock URL \url{https://doi.org/10.1115/1.4027451}.

\bibitem[Mao et~al.(2016)Mao, Purohit, and Aravas]{mao2016mixed}
Sheng Mao, Prashant~K Purohit, and Nikolaos Aravas.
\newblock Mixed finite-element formulations in piezoelectricity and
  flexoelectricity.
\newblock \emph{Proceedings of the Royal Society A: Mathematical, Physical and
  Engineering Sciences}, 472\penalty0 (2190):\penalty0 20150879, 2016.
\newblock URL \url{https://doi.org/10.1098/rspa.2015.0879}.

\bibitem[Maranganti et~al.(2006)Maranganti, Sharma, and
  Sharma]{maranganti2006electromechanical}
R~Maranganti, ND~Sharma, and P~Sharma.
\newblock Electromechanical coupling in nonpiezoelectric materials due to
  nanoscale nonlocal size effects: Green's function solutions and embedded
  inclusions.
\newblock \emph{Physical Review B}, 74\penalty0 (1):\penalty0 014110, 2006.
\newblock URL \url{https://doi.org/10.1103/PhysRevB.74.014110}.

\bibitem[Marcerou and Prost(1980)]{Marcerou1990}
J.~P. Marcerou and J.~Prost.
\newblock The different aspects of flexoelectricity in nematics.
\newblock \emph{Molecular Crystals and Liquid Crystals}, 58\penalty0
  (3-4):\penalty0 259--284, 1980.

\bibitem[Marvan and Havr{\'a}nek(1998)]{marvan1998}
M~Marvan and A~Havr{\'a}nek.
\newblock Flexoelectric effect in elastomers.
\newblock In \emph{Relationships of Polymeric Structure and Properties}, pages
  33--36. Springer, 1998.
\newblock URL \url{https://doi.org/10.1007/BFb0114342}.

\bibitem[Mashkevich and Tolpygo(1957)]{Mashkevich1957}
V.S. Mashkevich and K.B. Tolpygo.
\newblock Electrical, optical and elastic properties of diamond type crystals.
  1.
\newblock \emph{Soviet Physics JETP-USSR}, 5\penalty0 (3):\penalty0 435--439,
  1957.
\newblock URL \url{http://www.jetp.ac.ru/cgi-bin/e/index/e/5/3/p435?a=list}.

\bibitem[McBride et~al.(2019)McBride, Davydov, and
  Steinmann]{mcbride2019modelling}
Andrew McBride, Denis Davydov, and Paul Steinmann.
\newblock Modelling the flexoelectric effect in solids: a micromorphic
  approach.
\newblock \emph{arXiv preprint}, 2019.
\newblock URL \url{https://arxiv.org/abs/1909.08695}.

\bibitem[Meyer(1969)]{meyer1969piezoelectric}
Robert~B Meyer.
\newblock Piezoelectric effects in liquid crystals.
\newblock \emph{Physical Review Letters}, 22\penalty0 (18):\penalty0 918, 1969.
\newblock URL \url{https://doi.org/10.1103/PhysRevLett.22.918}.

\bibitem[Mindlin(1964)]{Mindlin1964}
Raymond~David Mindlin.
\newblock Micro-structure in linear elasticity.
\newblock \emph{Archive for Rational Mechanics and Analysis}, 16\penalty0
  (1):\penalty0 51--78, 1964.

\bibitem[Mindlin(1968)]{Mindlin1968b}
Raymond~David Mindlin.
\newblock Polarization gradient in elastic dielectrics.
\newblock \emph{International Journal of Solids and Structures}, 4\penalty0
  (6):\penalty0 637--642, 1968.
\newblock URL \url{https://doi.org/10.1016/0020-7683(68)90079-6}.

\bibitem[Mindlin and Eshel(1968)]{Mindlin1968a}
R.D. Mindlin and N.N. Eshel.
\newblock On first strain-gradient theories in linear elasticity.
\newblock \emph{International Journal of Solids and Structures}, 4\penalty0
  (1):\penalty0 109--124, 1968.
\newblock URL \url{https://doi.org/10.1016/0020-7683(68)90036-X}.

\bibitem[Mohammadi et~al.(2014)Mohammadi, Liu, and Sharma]{Mohammadi2014}
P.~Mohammadi, L.P. Liu, and P.~Sharma.
\newblock A theory of flexoelectric membranes and effective properties of
  heterogeneous membranes.
\newblock \emph{Journal of Applied Mechanics}, 81:\penalty0 011007, 2014.

\bibitem[Morozovska et~al.(2016)Morozovska, Eliseev, Scherbakov, and
  Vysochanskii]{Morozovska2016}
Anna~N Morozovska, Eugene~A Eliseev, Christian~M Scherbakov, and Yulian~M
  Vysochanskii.
\newblock Influence of elastic strain gradient on the upper limit of
  flexocoupling strength, spatially modulated phases, and soft phonon
  dispersion in ferroics.
\newblock \emph{Physical Review B}, 94\penalty0 (17):\penalty0 174112, 2016.
\newblock URL \url{https://doi.org/10.1103/PhysRevB.94.174112}.

\bibitem[Morozovska et~al.(2018)Morozovska, Khist, Glinchuk, Scherbakov,
  Silibin, Karpinsky, and Eliseev]{Mozorovska2018}
Anna~N. Morozovska, Victoria~V. Khist, Maya~D. Glinchuk, Christian~M.
  Scherbakov, Maxim~V. Silibin, Dmitry~V. Karpinsky, and Eugene~A. Eliseev.
\newblock Flexoelectricity induced spatially modulated phases in ferroics and
  liquid crystals.
\newblock \emph{Journal of Molecular Liquids}, 267:\penalty0 550--559, 2018.

\bibitem[Nanthakumar et~al.(2017)Nanthakumar, Zhuang, Park, and
  Rabczuk]{nanthakumar2017topology}
SS~Nanthakumar, Xiaoying Zhuang, Harold~S Park, and Timon Rabczuk.
\newblock Topology optimization of flexoelectric structures.
\newblock \emph{Journal of the Mechanics and Physics of Solids}, 105:\penalty0
  217--234, 2017.
\newblock URL \url{https://doi.org/10.1016/j.jmps.2017.05.010}.

\bibitem[Nguyen et~al.(2019)Nguyen, Zhuang, and Rabczuk]{nguyen2019nurbs}
BH~Nguyen, X~Zhuang, and Timon Rabczuk.
\newblock Nurbs-based formulation for nonlinear electro-gradient elasticity in
  semiconductors.
\newblock \emph{Computer Methods in Applied Mechanics and Engineering},
  346:\penalty0 1074--1095, 2019.
\newblock URL \url{https://doi.org/10.1016/j.cma.2018.08.026}.

\bibitem[Nguyen et~al.(2013)Nguyen, Mao, Yeh, Purohit, and
  McAlpine]{Nguyen2013}
Thanh~D. Nguyen, Sheng Mao, Yao-Wen Yeh, Prashant~K. Purohit, and Michael~C.
  McAlpine.
\newblock Nanoscale flexoelectricity.
\newblock \emph{Advanced Materials}, 25\penalty0 (7):\penalty0 946--974, 2013.
\newblock URL \url{https://doi.org/10.1002/adma.201203852}.

\bibitem[O'Halloran et~al.(2008)O'Halloran, O'malley, and McHugh]{o2008review}
Ailish O'Halloran, Fergal O'malley, and Peter McHugh.
\newblock A review on dielectric elastomer actuators, technology, applications,
  and challenges.
\newblock \emph{Journal of Applied Physics}, 104\penalty0 (7):\penalty0 9,
  2008.
\newblock URL \url{https://doi.org/10.1063/1.2981642}.

\bibitem[Osipov and Pikin(1995)]{Osipov1995}
M.~A. Osipov and S.~A. Pikin.
\newblock Dipolar and quadrupolar ordering in ferroelectric iquid crystals.
\newblock \emph{J. Phys. II France}, 5:\penalty0 1223--1240, 1995.

\bibitem[Pelrine et~al.(1998)Pelrine, Kornbluh, and
  Joseph]{pelrine1998electrostriction}
Ronald~E Pelrine, Roy~D Kornbluh, and Jose~P Joseph.
\newblock Electrostriction of polymer dielectrics with compliant electrodes as
  a means of actuation.
\newblock \emph{Sensors and Actuators A: Physical}, 64\penalty0 (1):\penalty0
  77--85, 1998.
\newblock URL \url{https://doi.org/10.1016/S0924-4247(97)01657-9}.

\bibitem[Petrov et~al.(1989)Petrov, Ramsey, and Usherwood]{petrov1989curvature}
AG~Petrov, RL~Ramsey, and PNR Usherwood.
\newblock Curvature-electric effects in artificial and natural membranes
  studied using patch-clamp techniques.
\newblock \emph{European Biophysics Journal}, 17\penalty0 (1):\penalty0 13--17,
  1989.
\newblock URL \url{https://doi.org/10.1007/BF00257141}.

\bibitem[Petrov(1975)]{petrov1975flexoelectric}
Alexander~G Petrov.
\newblock Flexoelectric model for active transport.
\newblock In \emph{Physical and Chemical Bases of Biological Information
  Transfer}, pages 111--125. Springer, 1975.
\newblock URL \url{https://doi.org/10.1007/978-1-4684-2181-1_9}.

\bibitem[Petrov(1999)]{petrov1999}
Alexander~G. Petrov.
\newblock Liquid crystal physics and the physics of living matter.
\newblock \emph{Molecular Crystals and Liquid Crystals Science and Technology.
  Section A. Molecular Crystals and Liquid Crystals}, 332\penalty0
  (1):\penalty0 577--584, 1999.

\bibitem[Petrov(2002)]{petrov2002flexoelectricity}
Alexander~G Petrov.
\newblock Flexoelectricity of model and living membranes.
\newblock \emph{Biochimica et Biophysica Acta (BBA)-Biomembranes},
  1561\penalty0 (1):\penalty0 1--25, 2002.
\newblock URL \url{https://doi.org/10.1016/S0304-4157(01)00007-7}.

\bibitem[Piegl and Tiller(2012)]{Piegl2012}
L.~Piegl and W.~Tiller.
\newblock \emph{The NURBS Book}.
\newblock Monographs in Visual Communication. Springer Berlin Heidelberg, 2012.
\newblock ISBN 9783642973857.
\newblock \doi{10.1007/978-3-642-97385-7}.
\newblock URL \url{https://books.google.es/books?id=58KqCAAAQBAJ}.

\bibitem[Pikin and indenbom(1978)]{Pikin1978}
S.A. Pikin and V.L. indenbom.
\newblock piezoeffects and ferroelectric phenomena in smectic liquid crystals.
\newblock \emph{Ferroelectrics}, 20:\penalty0 151--153, 1978.

\bibitem[Poya et~al.(2019)Poya, Gil, Ortigosa, and Palma]{poya2019family}
Roman Poya, Antonio~J Gil, Rogelio Ortigosa, and Roberto Palma.
\newblock On a family of numerical models for couple stress based
  flexoelectricity for continua and beams.
\newblock \emph{Journal of the Mechanics and Physics of Solids}, 125:\penalty0
  613--652, 2019.
\newblock URL \url{https://doi.org/10.1016/j.jmps.2019.01.013}.

\bibitem[Prost and Marcerou(1977)]{prost1977microscopic}
Jacques Prost and JP~Marcerou.
\newblock On the microscopic interpretation of flexoelectricity.
\newblock \emph{Journal de Physique}, 38\penalty0 (3):\penalty0 315--324, 1977.
\newblock URL \url{https://doi.org/10.1051/jphys:01977003803031500}.

\bibitem[Resta(2010)]{resta2010towards}
Raffaele Resta.
\newblock Towards a bulk theory of flexoelectricity.
\newblock \emph{Physical review letters}, 105\penalty0 (12):\penalty0 127601,
  2010.
\newblock URL \url{https://doi.org/10.1103/PhysRevLett.105.127601}.

\bibitem[Rey(2006)]{Rey2006}
A.~D. Rey.
\newblock Liquid crystal model of membrane flexoelectricity.
\newblock \emph{Physical Reviw E}, 74\penalty0 (1):\penalty0 011710, 2006.

\bibitem[Rogers(2001)]{Rogers2001}
D.F. Rogers.
\newblock \emph{An Introduction to NURBS: With Historical Perspective}.
\newblock Morgan Kaufmann Series in Computer Graphics and Geometric Modeling.
  Morgan Kaufmann Publishers, 2001.
\newblock ISBN 9781558606692.
\newblock URL \url{https://doi.org/10.1016/B978-1-55860-669-2.X5000-3}.

\bibitem[Rosset and Shea(2016)]{rosset2016}
Samuel Rosset and Herbert~R Shea.
\newblock Small, fast, and tough: Shrinking down integrated elastomer
  transducers.
\newblock \emph{Applied Physics Reviews}, 3\penalty0 (3):\penalty0 031105,
  2016.
\newblock URL \url{https://doi.org/10.1063/1.4963164}.

\bibitem[Sahin and Dost(1988)]{sahin1988strain}
E~Sahin and S~Dost.
\newblock A strain-gradients theory of elastic dielectrics with spatial
  dispersion.
\newblock \emph{International Journal of Engineering Science}, 26\penalty0
  (12):\penalty0 1231--1245, 1988.
\newblock URL \url{https://doi.org/10.1016/0020-7225(88)90043-2}.

\bibitem[Schiaffino et~al.(2019)Schiaffino, Dreyer, Vanderbilt, and
  Stengel]{schiaffino2019metric}
Andrea Schiaffino, Cyrus~E Dreyer, David Vanderbilt, and Massimiliano Stengel.
\newblock Metric wave approach to flexoelectricity within density functional
  perturbation theory.
\newblock \emph{Physical Review B}, 99\penalty0 (8):\penalty0 085107, 2019.
\newblock URL \url{https://doi.org/10.1103/PhysRevB.99.085107}.

\bibitem[Sharma et~al.(2010)Sharma, Landis, and Sharma]{Sharma2010}
N.D. Sharma, C.M. Landis, and P.~Sharma.
\newblock Piezoelectric thin-film superlattices without using piezoelectric
  materials.
\newblock \emph{Journal of Applied Physics}, 108\penalty0 (2):\penalty0 1--25,
  2010.
\newblock \doi{10.1063/1.3443404}.
\newblock URL \url{http://dx.doi.org/10.1063/1.3443404}.

\bibitem[Shen and Hu(2010)]{Shen2010}
Shengping Shen and Shuling Hu.
\newblock A theory of flexoelectricity with surface effect for elastic
  dielectrics.
\newblock \emph{Journal of the Mechanics and Physics of Solids}, 58\penalty0
  (5):\penalty0 665 -- 677, 2010.
\newblock ISSN 0022-5096.
\newblock URL \url{https://doi.org/10.1016/j.jmps.2010.03.001}.

\bibitem[Steinmann and Vu(2017)]{Steinmann2017}
Paul Steinmann and Duc~Khoi Vu.
\newblock Computational challenges in the simulation of nonlinear
  electroelasticity.
\newblock \emph{Computer Assisted Methods in Engineering and Science},
  19\penalty0 (3):\penalty0 199--212, 2017.
\newblock URL \url{https://cames.ippt.pan.pl/index.php/cames/article/view/90}.

\bibitem[Sun(1997)]{sun1997toward}
Kai Sun.
\newblock Toward molecular mechanoelectric sensors: Flexoelectric sensitivity
  of lipid bilayers to structure, location, and orientation of bound
  amphiphilic ions.
\newblock \emph{The Journal of Physical Chemistry B}, 101\penalty0
  (33):\penalty0 6327--6330, 1997.

\bibitem[Tagantsev(1986)]{Tagantsev1986}
A.~K. Tagantsev.
\newblock Piezoelectricity and flexoelectricity in crystalline dielectrics.
\newblock \emph{Phys. Rev. B}, 34:\penalty0 5883--5889, 1986.
\newblock URL \url{https://doi.org/10.1103/PhysRevB.34.5883}.

\bibitem[Tagantsev(1985)]{Tagantsev1985}
AK~Tagantsev.
\newblock Theory of flexoelectric effect in crystals.
\newblock \emph{Zhurnal Eksperimental'noi i Teoreticheskoi Fiziki}, 88\penalty0
  (6):\penalty0 2108--22, 1985.
\newblock URL \url{http://www.jetp.ac.ru/cgi-bin/e/index/e/61/6/p1246?a=list}.

\bibitem[Tagantsev(1991)]{tagantsev1991electric}
Alexander~K Tagantsev.
\newblock Electric polarization in crystals and its response to thermal and
  elastic perturbations.
\newblock \emph{Phase Transitions: A Multinational Journal}, 35\penalty0
  (3-4):\penalty0 119--203, 1991.
\newblock URL \url{https://doi.org/10.1080/01411599108213201}.

\bibitem[Thai et~al.(2018)Thai, Rabczuk, and Zhuang]{Thai2018}
Tran~Quoc Thai, Timon Rabczuk, and Xiaoying Zhuang.
\newblock A large deformation isogeometric approach for flexoelectricity and
  soft materials.
\newblock \emph{Computer Methods in Applied Mechanics and Engineering},
  341:\penalty0 718--739, 2018.
\newblock URL \url{https://doi.org/10.1016/j.cma.2018.05.019}.

\bibitem[Timoshenko and Gere(2009)]{timoshenko2009theory}
Stephen~P Timoshenko and James~M Gere.
\newblock \emph{Theory of elastic stability}.
\newblock Courier Corporation, 2009.
\newblock URL \url{https://books.google.es/books?id=98B6JOW2HiUC}.

\bibitem[Todorov et~al.(1991)Todorov, Petrov, Brandt, and
  Fendler]{todorov1991electrical}
A~Todorov, A~Petrov, Michael~O Brandt, and Janos~H Fendler.
\newblock Electrical and real-time stroboscopic interferometric measurements of
  bilayer lipid membrane flexoelectricity.
\newblock \emph{Langmuir}, 7\penalty0 (12):\penalty0 3127--3137, 1991.
\newblock URL \url{https://doi.org/10.1021/la00060a036}.

\bibitem[Todorov et~al.(1994)Todorov, Petrov, and Fendler]{todorov1994first}
AT~Todorov, AG~Petrov, and JH~Fendler.
\newblock First observation of the converse flexoelectric effect in bilayer
  lipid membranes.
\newblock \emph{The Journal of Physical Chemistry}, 98\penalty0 (12):\penalty0
  3076--3079, 1994.
\newblock URL \url{https://doi.org/10.1021/j100063a004}.

\bibitem[Tolpygo(1963)]{tolpygo1963long}
KB~Tolpygo.
\newblock Long wavelength oscillations of diamond-type crystals including long
  range forces.
\newblock \emph{Soviet Physics-Solid State}, 4\penalty0 (7):\penalty0
  1297--1305, 1963.

\bibitem[Toupin(1956)]{Toupin1956}
Richard Toupin.
\newblock The elastic dielectric.
\newblock \emph{Journal of Rational Mechanics and Analysis}, 5\penalty0
  (6):\penalty0 849--915, 1956.

\bibitem[Trabi et~al.(2008)Trabi, Brown, Smith, and
  Mottram]{trabi2008interferometric}
CL~Trabi, CV~Brown, AAT Smith, and NJ~Mottram.
\newblock Interferometric method for determining the sum of the flexoelectric
  coefficients (e${}_1$+ e${}_3$) in an ionic nematic material.
\newblock \emph{Applied Physics Letters}, 92\penalty0 (22):\penalty0 223509,
  2008.
\newblock URL \url{https://doi.org/10.1063/1.2938722}.

\bibitem[Vu et~al.(2007)Vu, Steinmann, and Possart]{Vu2007}
DK~Vu, P~Steinmann, and G~Possart.
\newblock Numerical modelling of non-linear electroelasticity.
\newblock \emph{International Journal for Numerical Methods in Engineering},
  70\penalty0 (6):\penalty0 685--704, 2007.
\newblock URL \url{https://doi.org/10.1002/nme.1902}.

\bibitem[Wang et~al.(2019)Wang, Gu, Zhang, and Chen]{wang2019flexoelectricity}
Bo~Wang, Yijia Gu, Shujun Zhang, and Long-Qing Chen.
\newblock Flexoelectricity in solids: Progress, challenges, and perspectives.
\newblock \emph{Progress in Materials Science}, 2019.
\newblock URL \url{https://doi.org/10.1016/j.pmatsci.2019.05.003}.

\bibitem[Yudin and Tagantsev(2013)]{Yudin2013}
P.V. Yudin and A.K. Tagantsev.
\newblock Fundamentals of flexoelectricity in solids.
\newblock \emph{Nanotechnology}, 24\penalty0 (43):\penalty0 1--36, 2013.
\newblock URL \url{https://doi.org/10.1088/0957-4484/24/43/432001}.

\bibitem[Yudin et~al.(2014)Yudin, Ahluwalia, and Tagantsev]{Yudin2014}
PV~Yudin, R~Ahluwalia, and AK~Tagantsev.
\newblock Upper bounds for flexoelectric coefficients in ferroelectrics.
\newblock \emph{Applied Physics Letters}, 104\penalty0 (8):\penalty0 082913,
  2014.
\newblock URL \url{https://doi.org/10.1063/1.4865208}.

\bibitem[Yudin et~al.(2015)Yudin, Ahluwalia, and Tagantsev]{Yudin2015}
PV~Yudin, R~Ahluwalia, and AK~Tagantsev.
\newblock Erratum:``upper bounds for flexoelectric coefficients in
  ferroelectrics''[appl. phys. lett. 104, 082913 (2014)].
\newblock \emph{Applied Physics Letters}, 106\penalty0 (18):\penalty0 189902,
  2015.
\newblock URL \url{https://doi.org/10.1063/1.4919883}.

\bibitem[Yvonnet and Liu(2017)]{Yvonnet2017}
Julien Yvonnet and LP~Liu.
\newblock A numerical framework for modeling flexoelectricity and {Maxwell}
  stress in soft dielectrics at finite strains.
\newblock \emph{Computer Methods in Applied Mechanics and Engineering},
  313:\penalty0 450--482, 2017.
\newblock URL \url{https://doi.org/10.1016/j.cma.2016.09.007}.

\bibitem[Zhang et~al.(2016{\natexlab{a}})Zhang, Liang, and
  Shen]{zhang2016timoshenko}
Runzhi Zhang, Xu~Liang, and Shengping Shen.
\newblock A timoshenko dielectric beam model with flexoelectric effect.
\newblock \emph{Meccanica}, 51\penalty0 (5):\penalty0 1181--1188,
  2016{\natexlab{a}}.
\newblock URL \url{https://doi.org/10.1007/s11012-015-0290-1}.

\bibitem[Zhang et~al.(2015)Zhang, Xu, Liang, and Shen]{zhang2015}
Shuwen Zhang, Minglong Xu, Xu~Liang, and Shengping Shen.
\newblock Shear flexoelectric coefficient $\mu$1211 in polyvinylidene fluoride.
\newblock \emph{Journal of Applied Physics}, 117\penalty0 (20):\penalty0
  204102, 2015.
\newblock URL \url{https://doi.org/10.1063/1.4921444}.

\bibitem[Zhang et~al.(2016{\natexlab{b}})Zhang, Xu, Ma, Liang, and
  Shen]{zhang2016experimental}
Shuwen Zhang, Minglong Xu, Guoliang Ma, Xu~Liang, and Shengping Shen.
\newblock Experimental method research on transverse flexoelectric response of
  poly (vinylidene fluoride).
\newblock \emph{Japanese Journal of Applied Physics}, 55\penalty0 (7):\penalty0
  071601, 2016{\natexlab{b}}.
\newblock URL \url{https://doi.org/10.7567/JJAP.55.071601}.

\bibitem[Zhou et~al.(2017)Zhou, Liu, Hu, Chu, Chen, and
  Salem]{zhou2017flexoelectric}
Yang Zhou, Jie Liu, Xinping Hu, Baojin Chu, Shutao Chen, and David Salem.
\newblock Flexoelectric effect in {PVDF}-based polymers.
\newblock \emph{IEEE Transactions on Dielectrics and Electrical Insulation},
  24\penalty0 (2):\penalty0 727--731, 2017.
\newblock URL \url{https://doi.org/10.1109/TDEI.2017.006273}.

\bibitem[Zhuang et~al.(2019)Zhuang, Nanthakumar, and
  Rabczuk]{zhuang2019meshfree}
Xiaoying Zhuang, SS~Nanthakumar, and Timon Rabczuk.
\newblock A meshfree formulation for large deformation analysis of
  flexoelectric structures accounting for the surface effects.
\newblock \emph{arXiv preprint}, 2019.
\newblock URL \url{https://arxiv.org/abs/1911.06553}.

\bibitem[Zhuang et~al.(2020)Zhuang, Nguyen, Nanthakumar, Tran, Alajlan, and
  Rabczuk]{zhuang2020computational}
Xiaoying Zhuang, Binh~Huy Nguyen, Subbiah~Srivilliputtur Nanthakumar, Thai~Quoc
  Tran, Naif Alajlan, and Timon Rabczuk.
\newblock Computational modeling of flexoelectricity---a review.
\newblock \emph{Energies}, 13\penalty0 (6):\penalty0 1326, 2020.
\newblock URL \url{https://doi.org/10.3390/en13061326}.

\bibitem[Zubko et~al.(2007)Zubko, Catalan, Buckley, Welche, and
  Scott]{zubko2007strain}
P~Zubko, G~Catalan, A~Buckley, PRL Welche, and JF~Scott.
\newblock Strain-gradient-induced polarization in {SrTiO${}_3$} single
  crystals.
\newblock \emph{Physical Review Letters}, 99\penalty0 (16):\penalty0 167601,
  2007.
\newblock URL \url{https://doi.org/10.1103/PhysRevLett.99.167601}.

\bibitem[Zubko et~al.(2013)Zubko, Catalan, and Tagantsev]{Zubko2013}
Pavlo Zubko, Gustau Catalan, and Alexander~K. Tagantsev.
\newblock Flexoelectric effect in solids.
\newblock \emph{Annual Review of Materials Research}, 24\penalty0
  (43):\penalty0 387--421, 2013.
\newblock URL \url{https://doi.org/10.1146/annurev-matsci-071312-121634}.

\end{thebibliography}
\end{document}